%% file: main.tex
\documentclass[twoside,11pt]{article}

\usepackage[utf8]{inputenc} 
\usepackage[T1]{fontenc}    

\usepackage{url}            
\usepackage{booktabs}       
\usepackage{amsfonts}       
\usepackage{nicefrac}       
\usepackage{microtype}      
\usepackage{xcolor}         
\definecolor{darkblue}{RGB}{0,0,128}
\definecolor{darkgreen}{RGB}{0, 200, 0}
\definecolor{forestlikegreen}{RGB}{0, 160, 0}
\definecolor{darkred}{RGB}{128, 0, 0}
\definecolor{black}{RGB}{0, 0, 0}
\definecolor{errorcolor}{HTML}{8b0000}
\definecolor{viridisgreen}{HTML}{55C667}
\definecolor{observedcolor}{HTML}{eb5760}
\definecolor{networkcolor}{HTML}{2A0593}

\usepackage{amsmath}

\usepackage{accents}
\newcommand{\observed}[1]{\protect\accentset{\text{o}}{#1}}

\usepackage{array}

\newcolumntype{L}{>{\scriptstyle}l}
\newcolumntype{C}{>{\scriptstyle}c}
\newcolumntype{R}{>{\scriptstyle}r}
\newenvironment{mysubarray}{%
  \scriptstyle
  \setlength\arraycolsep{0pt}%
  \setlength\extrarowheight{-1ex}
  \renewcommand\arraystretch{0}
  \begin{array}{RCL}}{\end{array}}

\usepackage{tikz}
\usepackage{tkz-euclide}
\usetikzlibrary{positioning, shapes, arrows, fit, shapes.multipart}
\usepackage{adjustbox}
\usepackage{algorithm}
\usepackage{algpseudocode}

\usepackage{chngcntr}

\usepackage{comment}
\usepackage{enumitem}

\usepackage{subcaption}

\graphicspath{{plots/}}
\usepackage{placeins}

\usepackage{chngcntr}

\usepackage{booktabs} 
\usepackage{multirow}
\usepackage{makecell}

\newcommand{\numberGaussianMeans}{1}
\newcommand{\numberGaussianMeansCov}{2}
\newcommand{\numberCS}{3}
\newcommand{\numberDDM}{4}
\newcommand{\numberCovid}{5}

\newcommand{\x}{\boldsymbol{x}}
\newcommand{\z}{\boldsymbol{z}}

\newcommand{\mub}{\boldsymbol{\mu}}
\newcommand{\xib}{\boldsymbol{\xi}}

\newcommand{\Sigmab}{\boldsymbol{\Sigma}}
\newcommand{\thetab}{\boldsymbol{\theta}}
\newcommand{\phib}{\boldsymbol{\phi}}
\newcommand{\psib}{\boldsymbol{\psi}}
\newcommand{\Psib}{\boldsymbol{\Psi}}

\newcommand{\0}{\boldsymbol{0}}

\newcommand{\given}{\,|\,}

\newcommand{\priorm}{p(\thetab \given \mathcal{M})}
\newcommand{\likm}{p(\x \given \thetab, \mathcal{M})}

\newcommand{\postm}{p(\thetab \given \x, \mathcal{M})}
\newcommand{\NIW}{\text{N-}\mathcal{W}^{-1}}

\newcommand{\prior}{p(\thetab)}

\newcommand{\jointm}{p(\thetab, \x \given \mathcal{M})}

\newcommand{\noised}{p(\xib \given \thetab)}
\newcommand{\model}{g(\thetab, \xib)}
\newcommand{\diff}{\mathrm{d}}
\newcommand{\rMMD}{\widehat{\text{rMMD}}}
\newcommand{\M}{\mathcal{M}}

\newcommand{\ie}{i.\,e.}
\newcommand{\eg}{e.\,g.}

\newcommand{\colsquare}[1]{\fcolorbox{#1}{#1}{\rule{0pt}{3pt}\rule{3pt}{0pt}}\,}
\DeclareMathOperator*{\argmin}{argmin}

\usepackage[preprint]{stylefile}
\jmlrheading{?}{????}{?--?}{?/??}{??/??}{?????}{?????}

\renewcommand{\cite}[1]{\citep{#1}}

\ShortHeadings{Detecting Model Misspecification in Amortized SBI}{Schmitt, B\"{u}rkner, K\"{o}the, and Radev}
\firstpageno{1}

\begin{document}

\title{Detecting Model Misspecification in Amortized Bayesian Inference with Neural Networks}

\author{
    Marvin Schmitt\\
    Cluster of Excellence SimTech\\
    University of Stuttgart\\
    \texttt{mail.marvinschmitt@gmail.com}\\
    \And
    Paul-Christian Bürkner\\
    Cluster of Excellence SimTech\\
    University of Stuttgart\\
    \texttt{paul.buerkner@gmail.com}\\
    \And
    Ullrich Köthe\\
    Computer Vision and Learning Lab, IWR\\
    Heidelberg University\\
    \texttt{ullrich.koethe@iwr.uni-heidelberg.de}\\
    \And
    Stefan T. Radev\\
    Cluster of Excellence STRUCTURES\\
    Heidelberg University\\
    \texttt{stefan.radev93@gmail.com}
}

\maketitle

\begin{abstract}%
Recent advances in probabilistic deep learning enable efficient amortized Bayesian inference in settings where the likelihood function is only implicitly defined by a simulation program (simulation-based inference; SBI). But how faithful is such inference if the simulation represents reality somewhat inaccurately---that is, if the true system behavior at test time deviates from the one seen during training? We conceptualize the types of model misspecification arising in SBI and systematically investigate how the performance of neural posterior approximators gradually deteriorates under these misspecifications, making inference results less and less trustworthy. To notify users about this problem, we propose a new misspecification measure that can be trained in an unsupervised fashion (\ie, without training data from the true distribution) and reliably detects model misspecification at test time. Our experiments clearly demonstrate the utility of our new measure both on toy examples with an analytical ground-truth and on representative scientific tasks in cell biology, cognitive decision making, and disease outbreak dynamics. We show how the proposed misspecification test warns users about suspicious outputs, raises an alarm when predictions are not trustworthy, and guides model designers in their search for better simulators.
\end{abstract}


\section{Introduction}
\begin{figure*}
    \centering
    \begin{adjustbox}{width=1.0\textwidth}
    \input{plots/simulation_gap}
    \end{adjustbox}
    \caption{Conceptual overview of our neural approach.
    The summary network $h_\psi$ maps observations $\x$ to summary statistics $h_\psi(\x)$, and the inference network $f_\phi$ estimates the parameter posterior $p(\thetab\given\x,\mathcal{M})$ from the summary statistics.
    The generative model $\mathcal{M}$ creates training data $\x$ in the green region, and the networks learn to map these data to well-defined summary statistics and posteriors (green regions/dot/box).
    If the generative model $\mathcal{M}$ is misspecificed, real observations $\observed{\x}$ fall outside the training region and are therefore mapped to outlying summary statistics and potentially incorrect posteriors (red dots/box).
    Since learning enforces the inlier summary distribution to be known (\eg, a unit Gaussian), misspecification can be detected by a distribution mismatch in summary space, as signaled by a high maximum mean discrepancy \citep[MMD;][]{Gretton2012} score.}
    \label{fig:conceptual}
\end{figure*}
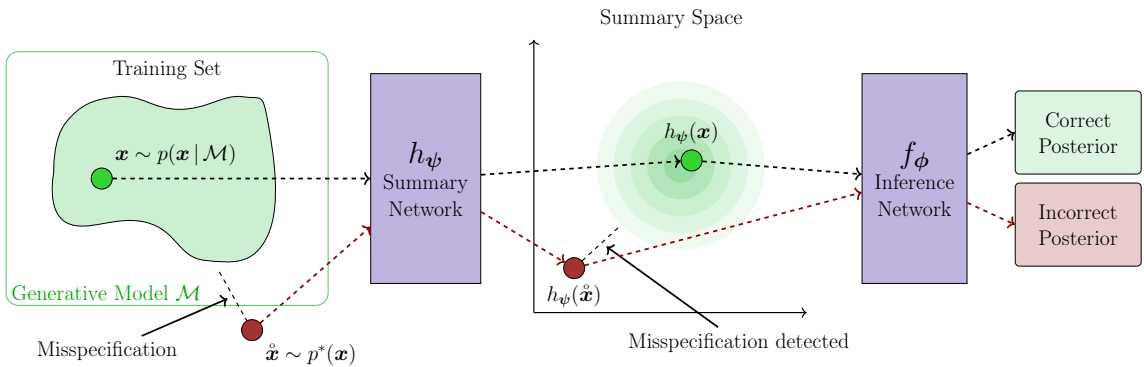

Computer simulations play a fundamental role in many fields of science. 
However, the associated {\em inverse} problems of finding simulation parameters that accurately reproduce or predict real-world behavior are generally difficult and analytically intractable.
Here, we consider \emph{simulation-based inference} \citep[SBI;][]{frontier} as a general approach to overcome this difficulty within a Bayesian inference framework.
That is, given an assumed generative model $\mathcal{M}$ (as represented by the simulation program, see Section~\ref{sec:defining-model-misspecification} for details) and observations $\x$ (real or simulated outcomes), we estimate the posterior distribution $p(\thetab\given\x,\mathcal{M})$ of the simulation parameters $\thetab$ that would reproduce the observed $\x$.
Distributional estimates are preferable over point estimates because $\thetab$ is typically not uniquely determined by $\x$ and $\mathcal{M}$.
The recent introduction of efficient neural network approximators for this task---specifically SNPE-C \citep[APT;][]{apt} and BayesFlow \cite{bayesflow}---has inspired a rapidly growing literature on SBI solutions for various application domains \citep[e.g.,][]{butter2022machine, lueckmann2021benchmarking, gonccalves2020training,  bayesflow_agent, bayesflow_qcd, von_krause_mental_2022, ghaderi-kangavari_general_2022}.
These empirical successes call for a systematic investigation of the trustworthiness of SBI, see \autoref{fig:conceptual}.

In this paper, we conduct an extensive error analysis of SNPE-C and BayesFlow, two major deep learning algorithms for {\em amortized} approximation of $p(\thetab\given\x,\mathcal{M})$.
We specifically study their accuracy under model misspecification, where the generative model $\mathcal{M}^*$ at test time (the ``true data generating process'') deviates from the one used during training (\ie, $\mathcal{M}^*\ne\mathcal{M}$), a situation commonly known as a {\em simulation gap}. 
Our investigations complement existing work on deep amortized SBI, whose main focus has been on network architectures and training algorithms achieving high accuracy in the well-specified case $\mathcal{M}^*=\mathcal{M}$ \cite{ramesh2022gatsbi, pacchiardi2022likelihood, contrastive, bayesflow, apt, bayes_lstm, papamakarios2016fast}.

\begin{figure}
\centering
    \begin{minipage}{.40\linewidth}
        \includegraphics[width=\linewidth]{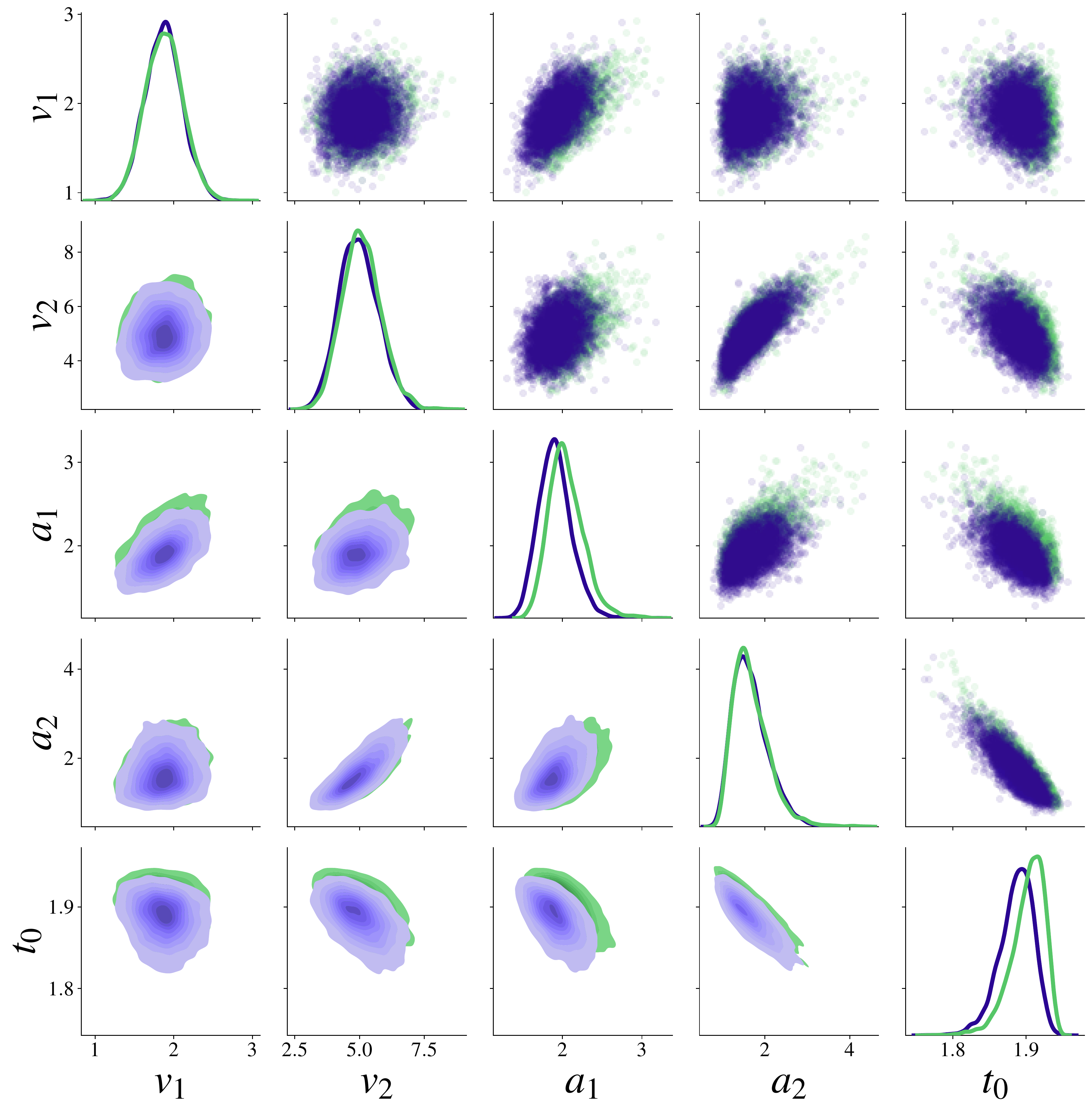}
    \end{minipage}
    \hspace*{2cm}
    \begin{minipage}{.40\linewidth}
        \includegraphics[width=\linewidth]{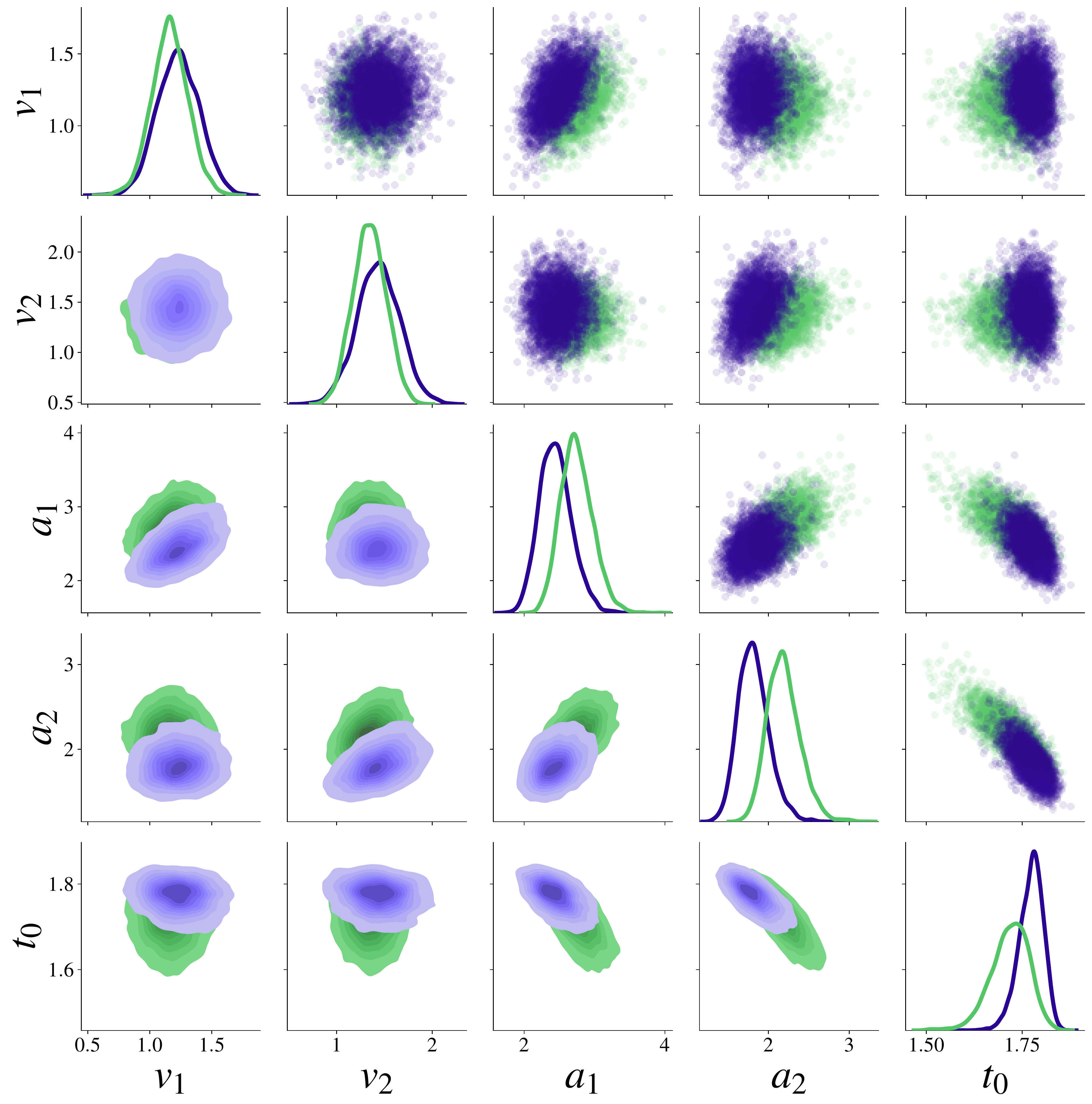}   
    \end{minipage}\\
    \begin{minipage}{.40\linewidth}
        \includegraphics[width=\linewidth]{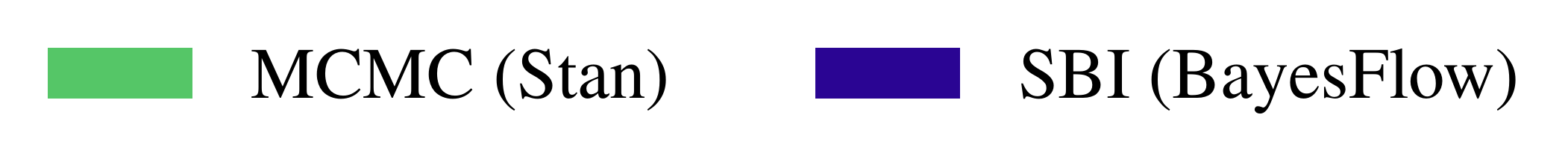}
    \end{minipage}\\
    \begin{minipage}{.45\linewidth}
        \begin{subfigure}[t]{\linewidth}
            \caption{Well-specified: Similar posteriors.}
            \label{fig:exp:ddm:stan-bf:clean}
        \end{subfigure}
    \end{minipage}
    \hfill
    \begin{minipage}{.45\linewidth}
        \begin{subfigure}[t]{\linewidth}
            \caption{Misspecified: Dissimilar posteriors.}
            \label{fig:exp:ddm:stan-bf:slow}
        \end{subfigure}
    \end{minipage}
    \caption{Preview of \textbf{Experiment \numberDDM} on reaction time modeling in psychological experiments (Section~\ref{sec:ddm-experiment}). Posteriors obtained via MCMC (Stan) and simulation-based inference (SBI; BayesFlow) are very similar when the model is well-specified (left).
    However, a simulation gap (here: not accounting for occasional slow responses due to mind wandering) leads to considerable disagreement between these methods (right).
    }
    \label{fig:exp:ddm:stan-bf}
\end{figure}

\autoref{fig:exp:ddm:stan-bf} illustrates the difference between the two situations:
When the model is well-specified, the posterior estimates of BayesFlow and a classical MCMC sampler \citep[implemented in Stan;][]{Stan2022} are essentially equal, whereas both approaches disagree considerably under model misspecification.
In order to avoid drawing incorrect conclusions from misspecified models and incorrect posteriors, it is of crucial importance to detect whether $\mathcal{M}^*\ne\mathcal{M}$ and to quantify the severity of the mismatch.
However, this is difficult in practice because the true data generating process $\mathcal{M}^*$ is generally unknown (except in controlled validation settings).

In this work, we propose a new misspecification measure that can be trained in an unsupervised fashion (\ie, without knowledge of $\mathcal{M}^*$ or training data from the true data distribution $p^*(\x)$) and reliably quantifies by how much $\mathcal{M}^*$ deviates from $\mathcal{M}$ at test time.
\autoref{fig:conceptual} illustrates how we achieve this by splitting the task between two neural networks, the summary network $h_\psi$ and the inference network $f_\phi$.
This allows us to measure the misspecification in summary space, where it amounts to a distribution mismatch test that can be elegantly implemented by maximum mean discrepancy \citep[MMD;][]{Gretton2012}.

In principle, we could detect misspecification directly in observation space by measuring the discrepancy between the true distribution $p^*(\x)$ and the model marginal $p(\x\given\mathcal{M})$.
However, this is a much harder learning problem, because $p^*(\x)$ and $p(\x\given\mathcal{M})$ are usually complex high-dimensional distributions.
Moreover, $\x$ often has variable dimension in practice (\eg, due to a varying number of observations for different subjects or a varying number of time steps in different time series), which further complicates learning in observation space.
In contrast, the summary space is designed to have fixed and typically low dimension, and end-to-end training of the two networks ensures that the learned summary statistics are maximally informative about $\x$ and follow a simple known distribution (a unit Gaussian) with meaningful MMD scores for misspecification detection.
At the same time, summary-based detectors avoid an important shortcoming of traditional Bayesian model checking methods \citep[e.g.,][]{bayes_ppc}, which base their diagnostics on the posterior distributions $p(\thetab\given\x,\mathcal{M})$ and therefore become unreliable if these posteriors get distorted in unpredictable ways under model misspecification (cf.\ \autoref{fig:exp:ddm:stan-bf:slow}).

Our experiments clearly demonstrate the power of our new measure both on toy examples with an analytical ground-truth, and on representative scientific tasks in cell biology, cognitive decision making, and disease outbreak dynamics.
We show how amortized posterior inference gradually deteriorates as the simulation gap widens and how the proposed misspecification test warns users about suspicious outputs, raises an alarm when predictions are not trustworthy, and guides model designers in their search for better simulators.

In particular, our paper makes the following key contributions:
\begin{enumerate}[label=(\roman*)]
    \item We systematically conceptualize different sources of model misspecification in amortized Bayesian inference with neural networks and propose a new detection criterion that is widely applicable to different model structures, inputs, and outputs.
    \item We incorporate this criterion into existing amortized neural posterior estimation methods, both with hand-crafted and learned summary statistics, and extend the associated learning algorithms in a largely non-intrusive manner.
    \item We conduct a systematic empirical evaluation of our detection criterion, the influence of the summary space dimension, and the relationship between summary outliers and posterior distortion under various types and strengths of model misspecification.
\end{enumerate}

\section{Related Work}\label{sec:related-work}
Model misspecification has been studied both in the context of standard Bayesian inference and generalizations thereof \citep[i.e., generalized Bayesian inference, see][]{knoblauch2019generalized, schmon_generalized_2021}.
To alleviate model misspecification in generalized Bayesian inference, researchers have investigated probabilistic classifiers \citep{mms_genbayes}, second-order PAC-Bayes bounds \citep{masegosa2020learning}, scoring rules \citep{giummole2019objective}, priors over a class of predictive models \citep{loaiza2021focused}, or Stein discrepancy as a loss function \citep{matsubara_robust_2022}.
Notably, most of these approaches deviate from the standard Bayesian formulation and investigate alternative schemes for belief updating and learning (e.g., replacing the likelihood function with a generic loss function).
In contrast, our method remains grounded in the standard Bayesian framework embodying an implicit likelihood principle \cite{berger_likelihood_1988}.
Differently, power scaling methods incorporate a modified likelihood (raised to a power $0 < \alpha < 1)$ in order to prevent potentially overconfident Bayesian updating \cite{bayesian_miss, holmes2017assigning}.
However, the SBI setting assumes that the likelihood function is not available in closed-form, which makes an explicit modification of the implicitly defined likelihood less obvious.

Neural approaches to amortized SBI can be categorized as either targeting the posterior \cite{bayesflow, apt}, the likelihood \cite{snle, ratios}, or both \cite{snpla}.
These methods employ simulations for training amortized neural approximators which can either generate samples from the posterior directly \cite{bayesflow, apt, snpla} or in tandem with Markov chain Monte Carlo (MCMC) sampling algorithms \cite{snle, ratios}.
Since the behavior of these methods depends on the fidelity of the simulations used as training data, we hypothesize that their estimation quality will be, in general, unpredictable, when faced with atypical real-world data.
Indeed, the critical impact of model misspecification in neural SBI has been commonly acknowledged in the scientific research community \cite{cannon_investigating_2022, alquier_concentration_2019, zhang_convergence_2020, frazier_model_2020, frazier_robust_2021, pacchiardi_score_2022}.

Recent approaches to detect model misspecification in simulation-based inference are usually based on the obtained approximate posterior distribution \citep[e.g.,][]{hermans2021averting, dellaporta_robust_2022, leclercq_simulation-based_2022}.
However, we show in \textbf{Experiment \numberGaussianMeans} and \textbf{Experiment \numberDDM} that the approximate posteriors in simulation-based inference tend to show pathological behavior under misspecified models. 
Posteriors from misspecified models may erroneously look legitimate, rendering diagnostic methods on their basis unreliable.
Moreover, the same applies for approaches based on the \textit{posterior predictive distribution} \citep{burkner_approximate_2020, gabry_visualization_2019, vehtari_survey_2012} since these also rely on the fidelity of the posterior distribution and can therefore only serve as an indirect measure of misspecification.

A few novel techniques aim to \emph{mitigate} model misspecification in simulation-based inference to achieve robust inference.
\citet{delaunoy_towards_2022} equip neural ratio estimation \cite{ratios} with a balancing condition which tends to produce more conservative posterior approximations.
\citet{ward_robust_2022} explore a way to alleviate model misspecification with two neural approximators and subsequent MCMC.
While both approaches are appealing in theory, the computational burden of MCMC sampling contradicts the idea of amortized inference and prohibits their use in complex applications with learned summary statistics and large amounts of data.
In fact, \citet{von_krause_mental_2022} recently used amortized neural SBI on more than a million data sets and demonstrated that an alternative inference method involving non-amortized MCMC would have taken years of sampling per model.

For robust non-amortized ABC samplers, the possibility of utilizing hand-crafted summary statistics as an important element of misspecification analysis has already been explored \cite{frazier_model_2020, frazier_robust_2021}.
Our work parallels these ideas and extends them to the case of \textit{learnable summary statistics} in amortized SBI on potentially massive data sets, where ABC becomes infeasible.
However, we show in \textbf{Experiment \numberCS} that our method also works with hand-crafted summary statistics.
Furthermore, we offer a conceptual discussion on the sources of misspecifications (and their consequences) arising in amortized SBI and explore the connection between posterior errors and model misspecification as a function of the number of summary statistics.

In the field of variational Bayes, recent work has studied the convergence and concentration rates with and without misspecification \cite{alquier_concentration_2019, zhang_convergence_2020}.
These approaches are not directly transferable to amortized models because the methods used in variational Bayes do not naturally provide a practical ``misspecification alarm'', which is needed because neural approximators remain unchanged after the training phase in SBI.

Framing Bayesian model selection as classification \cite{ruff_deep_2018, pudlo2016reliable}, diverse outlier detection techniques seem viable for uncovering simulation gaps in model comparison settings.
\citet{amortized_bmc} propose to train regularized evidential networks which learn a higher-order distribution over posterior model probabilities. 
This way, conclusions about the ``absolute misfit'' of all models in a set of candidate models can be drawn.
However, this approach is not suitable for parameter estimation and currently requires a loss function which does not guarantee a correct approximation of posterior model probabilities. 

Finally, from the perspective of deep anomaly detection, our approach for learning informative summary statistics can be viewed as a special case of \emph{generic normality feature learning} \cite{pang_deep_2022}. 
Standard learned summary statistics are optimized with a generic feature learning objective which is not primarily designed for anomaly detection \citep{bayesflow}.
However, since learned summary statistics are also optimized to be maximally informative for posterior inference, they are forced to capture underlying data regularities \cite{pang_deep_2022}.
This makes the summary statistics equipped with a tractable \textit{latent summary space} appropriate targets for misspecification detection.

\section{Method}\label{sec:methods}
This section defines model misspecification as a mismatch between marginal distributions and explains the building blocks and theoretical implications of our proposed detection method in the context of amortized SBI.
To shorten notation throughout the manuscript, we use an abbreviated set notation $\smash{\{\boldsymbol{a}^{(n)}\}\equiv\{\boldsymbol{a}^{(n)}\}_{n=1}^N}$, since we use the same lowercase and uppercase letter for referencing a set member and denoting set size respectively (e.g., $n \in \{1,\dots,N\})$.
\subsection{Defining Model Misspecification}\label{sec:defining-model-misspecification}
For the purpose of simulation-based inference, we define a generative model as a triple $\mathcal{M} = \big( \model, \noised, \prior \big)$.
Such a model generates data $\x \in \mathcal{X}$ according to the system
\begin{equation}
    \x = \model \quad\textrm{with}\quad \xib \sim \noised,\; \thetab \sim \prior,
\end{equation}
where $g$ denotes a (randomized) simulator, $\xib \in \Xi$ is a source of randomness (\ie, noise) with density function $\noised$, and $\prior$ encodes prior knowledge about plausible simulation parameters $\thetab \in \Theta$.
Intuitively, $\x$ represents quantities that we can observe and measure.
We use the decorated symbol $\observed{\x}$ to mark data that was in fact \emph{observed} in the real world and not merely simulated by the assumed model $\M$.
The parameters $\thetab$ consist of hidden properties whose role in $g$ we explicitly understand and model, and $\xib$ takes care of nuisance effects that we only treat statistically.
The abstract spaces $\mathcal{X}, \Xi$, and $\Theta$ denote the domain of possible output data (possible worlds), the scope of noise, and the set of admissible model parameters, respectively.
The distinction between hidden properties $\thetab$ and noise $\xib$ is not entirely clear-cut, but depends on our modeling goals and may vary across applications.
Moreover, our understanding of the world is constantly evolving, and yesterday's noise might become tomorrow's signal.

Whenever we employ simulations to investigate some real world phenomenon, a close correspondence between model and reality is necessary. 
Unacceptably large discrepancies between the two realms are known as a \emph{simulation gap}, and the corresponding model is said to be \emph{misspecified}.
Model misspecification can arise from any of the three model components in isolation or simultaneously.
A few illustrative examples show what can go wrong in practice:

\begin{enumerate}[label=(\roman*)]
    \item \textbf{Misspecified Simulator.} In a model for the hydraulic conductivity of a medium, the spatial composition of the material is essential: A simulator relying on the assumption of homogeneity will wrongly predict the behavior of heterogeneous materials, biasing inference results in complex and arbitrary ways \citep{Schoniger2015, Nowak2016}.
    \item \textbf{Unexpected Contamination.} During an ongoing pandemic, data collection may be severely distorted, for example by noisy measurements, systematic underreporting, and delayed data transfer \cite{covid_germany}, to name just a few.
    An epidemiological model disregarding these factors in $\noised$ will produce erroneous inferences about key disease parameters, even if the underlying theory was otherwise a good approximation of the disease dynamics.
    \item \textbf{Misspecified Prior.} When the admissible region of the prior $\prior$ is specified too large---for example, allows negative mass---physically impossible simulations may arise.
    On the other hand, when the prior is too narrow, the typical generative set of a model may leave out an important subset of observables or lead to unreasonable uncertainty estimates \citep[e.g., in satellite retrievals;][]{nguyen2019sensitivity}.
\end{enumerate}

Our generative model formulation is equivalent to the standard factorization of the Bayesian joint distribution into likelihood and prior, $\jointm = \likm\,\priorm$, where $\mathcal{M}$ expresses the prior knowledge and assumptions embodied in the model.
The likelihood is obtained by marginalizing the joint distribution $p(\xib, \x \given \thetab, \mathcal{M})$ over all possible values of the nuisance parameters $\xib$, that is, over all possible execution paths of the simulation program, for fixed $\thetab$:
\begin{equation}
    \likm = \int_{\Xi} p(\xib, \x \given \thetab, \mathcal{M})\,\diff \xib.
\end{equation}
This integral is typically intractable \cite{frontier}, but we assume that it exists and is non-degenerate, that is, it defines a proper density over the constrained manifold $\left(g(\thetab, \xib), \xib\right)$.
Whenever we model a real-world complex system, we assume an unknown (true) generator $\M^*$, which yields an unknown (true) distribution $\observed{\x} \sim p^*(\x)$ and is available to the data analyst only via a finite realization (\ie, actually observed data $\observed{\x}$).
Then, using the Bayesian formulation, we say that a generative model $\mathcal{M}$ is \emph{strictly} well-specified if
\begin{equation}\label{eq:spec}
    p^*(\x) = p(\x \given \mathcal{M}) \equiv \int_{\Theta} \likm\,\priorm\,\diff\thetab 
\end{equation}
for every $\x \in \mathcal{X}$.
Conversely, a generative model is misspecified if an observable $\x \in \mathcal{X}$ exists for which the above equality is violated. 

Note that our definition of model misspecification does not assume the existence of a \textit{true} parameter vector $\thetab^*$, as required by some definitions relying on asymptotic guarantees \cite{vandervaart2000asymptotic}. 
That is, we do not require that the (conditional) likelihood itself matches the assumed data-generating distribution, which would mean $p^*(\x) = p(\x \given \thetab^*, \mathcal{M})$ for some ground-truth $\thetab^*\in\Theta$.
Instead, we focus on the \textit{marginal likelihood} $p(\x \given \mathcal{M})$ which represents the entire prior predictive distribution of a model and does not commit to a single most representative parameter vector \cite{lotfi2022bayesian, masegosa2020learning}.
In this way, multiple models whose marginal distributions are representative of $p^*(\x)$ can be considered well-specified without any reference to some hypothetical ground-truth $\thetab^*$, which may not even exist for opaque systems with unknown parameters.

Since models necessarily simplify reality, the above strict criterion for well-specified models in Eq.~\ref{eq:spec} is often unattainable in practice.
We therefore relax the requirement by quantifying a model's degree of misspecification in terms of the information loss incurred by the following simplification:
For an acceptable upper bound $\vartheta$ on the information loss, a model is well-specified if $\mathbb{D}\big[p^*(\x)\,||\,p(\x \given \mathcal{M})\big] < \vartheta$
and misspecified otherwise. 
The symbol $\mathbb{D}$ denotes a divergence metric quantifying the ``distance'' between the data distributions implied by reality and by the model (the marginal likelihood). 
Notably, equality in Eq.~\ref{eq:spec} implies no information loss by modeling $p^*(\x)$ with $p(\x \given \mathcal{M})$ and thus leads to a divergence of zero.
A natural choice for $\mathbb{D}$ would be a metric from the family of $\mathcal{F}$-divergences, such as the Kullback-Leibler (KL) divergence. 
However, since the practical computation of $\mathcal{F}$-divergences requires closed-form densities, and thus $p^*(\x)$ to be analytically tractable, we prefer a probability integral metric, such as the Maximum Mean Discrepancy \citep[MMD;][]{Gretton2012}. Using the kernel trick, the MMD can be expressed as
\begin{equation}\label{eq:MMD:MMD-kernel-trick}
    \mathbb{MMD}^2\big[p^*(\x)\,||\,p(\x \given \mathcal{M})\big] =
    \mathbb{E}_{
    p^*(\x)}\big[\kappa(\x, \x')\big]
    + \mathbb{E}_{
    p(\x \given \mathcal{M})}\big[\kappa(\x, \x')\big]
    - 2 \mathbb{E}_{
    \begin{mysubarray}
      \x{\ }&\sim&p^*(\x) 
      \\ 
      \x'&\sim&p(\x \given \mathcal{M})
    \end{mysubarray}
    }\big[\kappa(\x, \x')\big].
\end{equation}
with any positive definite kernel $\kappa(\cdot,\cdot)$.
Crucially, this metric is practically tractable because it can be efficiently estimated via finite samples from $p^*(\x)$ and $p(\x \given \mathcal{M})$, and it equals zero if and only if the two densities are equal \cite{Gretton2012}.

We use sums of Gaussian kernels with different widths $\sigma_i$ as an established and flexible universal kernel \cite{Muandet2017}.
However, \citet{Ardizzone2018} argue that kernels with heavier tails may improve performance by yielding superior gradients for outliers.
Thus, we repeated all experiments with a sum of inverse multiquadratic kernels \citep[as proposed by][]{Tolstikhin2017}, and find that the results are essentially equal.

\subsection{Neural Architectures for Amortized Inference}

Our proposed method can be applied to any framework that uses summary statistics as an input to an amortized posterior approximator \cite{burkner2022some}.
We will exemplarily outline the seamless integration of our method into the BayesFlow \cite{bayesflow} and the SNPE-C \citep[aka APT;][]{apt} frameworks, but our method should also apply to the broader class of scoring-based minimization \cite{pacchiardi2022likelihood} or adversarial \cite{ramesh2022gatsbi} approaches.
BayesFlow and SNPE-C, as implemented in the respective software toolkits \cite{tejero2020sbi}, use different neural architectures and training regimes to minimize the expected KL divergence between approximate and correct simulation posterior
\begin{equation}
    \psib^*,\phib^* = \argmin_{\psib, \phib}\mathbb{E}_{p(\x \given \mathcal{M})}\left[\int_{\Theta} p(\thetab \given \x, \mathcal{M}) \log \frac{p(\thetab \given \x, \mathcal{M})}{q_{\phib}\big(\thetab \given h_{\psib}(\x), \mathcal{M}\big)}\, \diff\thetab\right], \label{eq:kl_bf_m}
\end{equation}
where the expectation is taken with respect to the prior predictive distribution $p(\x \given \mathcal{M})$. 
This criterion reduces to
\begin{equation}
  \psib^*, \phib^* =
  \argmin_{\psib, \phib}  
    \mathbb{E}_{\jointm}\Big[-\log q_{\phib}\big(\thetab\given h_{\psib}(\x), \mathcal{M}\big)\Big], \label{eq:kl_bf}
\end{equation}
since the correct posterior $\postm$ does not depend on the trainable neural network parameters ($\psib$, $\phib$).
The above criterion optimizes a summary (aka embedding or conditioning) network with parameters $\psib$ and an inference network with parameters $\phib$ which jointly amortize a generative model $\M$.
The summary network transforms input data $\x$ of variable size and structure to a fixed-length representation $\z = h_{\psib}(\x)$.
The inference network $f_{\phib}$ generates random draws from an approximate posterior $q_{\phib}$ via a normalizing flow, for instance, realized by a conditional invertible neural network \citep[cINN,][]{Ardizzone2019} or a conditional masked autoregressive flow \citep[cMAF,][]{papamakarios2017masked}.
Accordingly, we can compute the exact posterior density via the change of variable formula:
\begin{equation}
    q_{\phib}(\thetab \given h_{\psib}(\x), \mathcal{M}) = p\Big(f_{\phib}\big(\thetab; h_{\psib}(\x)\big)\Big)\left|\det \left(\frac{\partial f_{\phib}(\thetab;h_{\psib}(\x))}{\partial \thetab}\right)\right| \label{eq:cinn}
\end{equation}
We approximate the expectation in Eq.~\ref{eq:kl_bf} via simulations from the generative model $\M$ and repeat the process until convergence, which enables us to perform fully amortized inference (i.e., the posterior functional can be evaluated for any number of observed data sets $\x$).
Moreover, this objective is self-consistent and results in correct amortized inference under optimal convergence \cite{apt, bayesflow}.
However, simulation-based training (cf.\ Eq.~\ref{eq:kl_bf_m}) takes the expectation with respect to the model-implied prior predictive distribution~$p(\x\given\M)$, not necessarily the ''true`` real-world distribution $p^*(\x)$.
Thus, optimal convergence does not imply correct amortized inference or faithful prediction in the real world when there is a simulation gap, that is, when the assumed training model $\M$ deviates critically from the unknown true generative model $\M^*$.

At this point, we may ask whether the same problem also exists for sequential inference, as realized by multi-round SNPE methods \cite{contrastive, apt}. 
Whenever we require inference for a single observation $\observed{\x}$, we can sequentially transform the prior distribution $\prior$ into the posterior $p(\thetab \given \observed{\x}, \mathcal{M})$ through a series of simulation-based training rounds \citep{contrastive, apt}.
Typically, we will use the approximate posterior after a given round as the proposal prior for the next round, resulting in a \textit{semi-amortized} optimization criterion
\begin{equation}
  \psib^*, \phib^* =
  \argmin_{\psib, \phib}  
    \mathbb{E}_{p(\x \given \thetab, \mathcal{M}) \widehat{p}(\thetab \given \mathcal{M})}\Big[-\log \widehat{q}_{\phib}\big(\thetab \given h_{\psib}(\x), \mathcal{M}\big)\Big], \label{eq:kl_apt}
\end{equation}
where $\widehat{p}(\thetab \given \mathcal{M})$ is the current proposal distribution and the approximate posterior $\widehat{q}_{\phib}$ is represented as a categorical distribution over a discrete set of \textit{atomic proposals} in order to be tractable \cite{apt}.
In this way, the neural approximator specializes for estimating solely the parameters of $\observed{\x}$, since information from the observed data is used to narrow down the prior $\prior$ to the typical set of $\postm$.
Still, the first round in sequential inference depends only on the simulator outputs, so simulation gaps are likely to remain problematic, potentially propagating posterior inference errors through further training rounds. 
Indeed, we demonstrate this behavior in \textbf{Experiment \numberGaussianMeans}.

\subsection{Structured Summary Statistics}\label{sec:structured-summary-statistics}
In simulation-based inference, summary statistics have a dual purpose, because (i) they are fixed-length vectors, even if the input data $\x$ have variable length; and (ii) they usually contain crucial features of the data, which drastically simplifies neural posterior inference.
However, in complex real-world scenarios (such as decision making or COVID-19 modeling, cf.\ \textbf{Experiments \numberDDM} and \textbf{\numberCovid}), it is not feasible to rely on hand-crafted summary statistics.
Thus, combining neural posterior inference with \emph{learned summary statistics} leverages the benefits of summary statistics (i.e., compression to fixed-length vectors) while avoiding the virtually impossible task of designing hand-crafted summary statistics for complex models.

\begin{algorithm}[t]
\caption{Misspecification-aware amortized Bayesian inference. The algorithm illustrates online learning for simplicity. The training phase can utilize any other learning paradigm as well (e.g., round-based). Thus, details like batch size or round indices are omitted in the training phase of the algorithm.}
\label{alg:mms}
\begin{algorithmic}[1]
\State \textbf{Training phase}:
\Repeat
\State Sample parameters and data $(\thetab, \x)$ from the specified generative model $\M$.
\State Pass the data $\x$ through the summary network: $\z = h_{\psib}(\x)$.
\State Compute $-\log q_{\phib}\big(\thetab\given \z, \mathcal{M}\big)$ using Eq.~\ref{eq:cinn}.
\State Compute Monte Carlo estimate of Eq.~\ref{eq:bf_kl_mmd} as a loss function.
\State Update neural network parameters $\psib, \phib$  via backpropagation.
\Until{convergence to $\widehat{\psib}, \widehat{\phib}$}
\\
\State \textbf{Inference phase} \textit{(given $N$ observed or test data sets $\{\observed{\x}^{(n)}\}$, query $L$ draws from the amortized posterior, use $M$ draws from the validation summary distribution)}:
\For{$m=1,\ldots, M$}
\State Re-use data set $\x^{(m)}$ from the training phase or simulate a new one from $\mathcal{M}$.
\State Pass the data set $\x^{(m)}$ through the converged summary network: $\z^{(m)} = h_{\widehat{\psib}}(\x^{(m)})$.
\EndFor
\For{$n=1,\ldots, N$}
\State Pass the observed data set $\observed{\x}^{(n)}$ through the summary network: $\observed{\z}^{(n)} = h_{\widehat{\psib}}(\observed{\x}^{(n)})$.
\State Pass $\observed{\z}^{(n)}$ through the posterior approximator for $L$ draws $\{\thetab^{(n, l)}\}$ from $q_{\widehat{\phib}}(\thetab \given \observed{\z}^{(n)})$
\EndFor
\State Estimate the MMD distance of the inference data $\{\observed{\z}^{(n)}\}$ from the validation summary space under the generative model $\M$ from training: $\widehat{\mathrm{MMD}}^2\left(\left\{\observed{\z}^{(n)} \right\} \,||\,\left\{\z^{(m)} \right\}\right)$.
\State \Return Draws from the approximate posterior distribution for each queried data set $\observed{\x}^{(n)}$ $\{\thetab^{(n, l)}\}$ and $\widehat{\mathrm{MMD}}^2\left(\left\{\observed{\z}^{(n)} \right\} \,||\,\left\{\z^{(m)} \right\}\right)$.
\end{algorithmic}
\end{algorithm}

In simulation-based inference, the summary network $h_{\psib}$ acts as an interface between the data $\x$ and the inference network $f_{\phib}$.
Its role is to learn maximally informative summary vectors of fixed size $S$ from complex and structured observations (\eg, sets of $i.\-i.\-d.\-$ measurements or multivariate time series).
Since the learned summary statistics are optimized to be
maximally informative for posterior inference, they are forced to capture underlying data regularities (cf.\ Section~\ref{sec:related-work}).
Therefore, we deem the summary network's representation $\z = h_{\psib}(\x)$ as an adequate target to detect simulation gaps.%

Specifically, we propose to prescribe an $S$-dimensional multivariate unit Gaussian distribution to the summary space, $p\big(\z=h_{\psib}(\x) \given \mathcal{M}\big) \approx \mathcal{N}(\z \given\0, \mathbb{I})$, by minimizing the MMD between summary network outputs and random draws from a unit Gaussian distribution. 
To ensure that the summary vectors comply with the support of the Gaussian density, we use a linear (bottleneck) output layer with $S$ units in the summary network.
Thus, a random vector in summary space takes the form $h_{\psib}(\x) \equiv \z \equiv  (z_1, \ldots, z_S)\in\mathbb{R}^S$. 
The extended optimization objective then becomes
\begin{equation}\label{eq:bf_kl_mmd}
    \psib^*,\phib^* = \argmin_{\psib, \phib}  
    \mathbb{E}_{\jointm}\Big[-\log q_{\phib}\big(\thetab\given h_{\psib}(\x), \mathcal{M}\big)
    \Big]
    + \gamma\,\mathbb{MMD}^2\big[p\big(h_{\psib}(\x)\given \mathcal{M}\big)\,||\,\mathcal{N}\big(\0, \mathbb{I}\big)\big]
\end{equation}
with a hyperparameter $\gamma$ to control the relative weight of the MMD term.
Intuitively, this objective encourages the approximate posterior $q_{\phib}\big(\thetab\given h_{\psib}(\x), \mathcal{M}\big)$ to match the correct posterior and the summary distribution $\smash{p\big(h_{\psib}(\x) \given \mathcal{M}\big)}$ to match a unit Gaussian. %
The extended objective does not constitute a trade-off between the two terms, since the MMD term merely reshapes the summary distribution in an information preserving manner. 
Indeed, our experiments confirm that the extended objective does not impose restrictions on learnable posteriors or other limits on amortized simulation-based inference.

It is worth noting that this method is also directly applicable to hand-crafted summary statistics.
Hand-crafted summary statistics already have fixed length and usually contain rich information for posterior inference.
Thus, the task of the summary network $h_{\psi}$ simplifies to transforming the hand-crafted summary statistics to a unit Gaussian (Eq.~\ref{eq:bf_kl_mmd}) to enable model misspecification via distribution matching during test time (see below).
We apply our method to hand-crafted summary statistics in \textbf{Experiment \numberCS}.

\subsection{Theoretical Implications}
\label{sec:theoretical-implications}

Attaining the global minimum of Eq.~\ref{eq:bf_kl_mmd} with an arbitrarily expressive neural architecture $\{h_{\psib^*}, f_{\phib^*}, \mathcal{M}\}$ implies that (i) the inference and summary network jointly amortize the analytic posterior $p(\thetab \given \x, \mathcal{M})$; and (ii) the summary network transforms the data $p(\x \given \mathcal{M})$ into a unit Gaussian in summary space: $p\big(\z = h_{\psib^*}(\x)\big) = \mathcal{N}(\z \given \mathbf{0}, \mathbb{I})$.
According to (i), the set of inference network parameters $\phib^*$ is a minimizer of the (expected) negative log posterior learned by the inference network,
\begin{equation}
    \phib^* = \argmin_{\phib}  
    \mathbb{E}_{p(\z)} \mathbb{E}_{p(\thetab \given \z)}\Big[-\log q_{\phib}\big(\thetab\given \z\big)\Big].
\end{equation}
At the same time, (ii) ensures that deviances in the summary space (according to MMD) imply differences in the data generating processes,
\begin{equation}\label{eq:implications:mmd-greater-zero}
    \underbrace{\mathbb{MMD}^2\big[p^*(\z)\,||\,p(\z\given\M)\big] > 0}_{\text{summary space difference}} \implies \underbrace{\mathbb{MMD}^2\big[p^*(\x)\,||\,p(\x \given \mathcal{M})\big] > 0}_{\text{data space difference}},
\end{equation}
since a deviation of $p^*(\z=h_{\psib^*}(\x))$ from a unit Gaussian means that the summary network is no longer transforming samples from $p(\x \given \mathcal{M})$. 

Accordingly, the LHS of Eq.~\ref{eq:implications:mmd-greater-zero} no longer guarantees that the inference network parameters $\phib^*$ are maximally informative for posterior inference. 
The preceding argumentation also motivates our augmented objective, since a divergence of summary statistics for observed data $\observed{\z}=h_{\psib}(\observed{\x})$ from a unit Gaussian signalizes a deficiency in the assumed generative model $\M$ and a need to revise the latter. 
We also hypothesize and show empirically that we can successfully detect simulation gaps in practice even when the summary network outputs have not exactly converged to a unit Gaussian (\eg, in the presence of correlations in summary space, cf.\ \textbf{Experiment \numberDDM} and \textbf{Experiment \numberCovid}).

However, the converse of Eq.~\ref{eq:implications:mmd-greater-zero} is not true in general.
In other words, a discrepancy in data space (non-zero MMD on the RHS of Eq.~\ref{eq:implications:mmd-greater-zero}) does \emph{not} generally imply a difference in summary space (non-zero MMD on the LHS of Eq.~\ref{eq:implications:mmd-greater-zero}).
To illustrate this via a counter-example, consider the assumed Gaussian generative model $\M$ defined by
\begin{equation}\label{eq:implications:process-1}
    \begin{aligned}
        \mu &\sim p(\mu), \\
        x_1,x_2  &\sim \mathcal{N}(\mu, \sigma^2 = 2),
    \end{aligned}
\end{equation}
for $N = 2$ observations and a summary network with a single-output ($S = 1$).
Since the variance is fixed, the only inference target is the mean $\mu$.

Then, an optimal summary network $\psib^*$ outputs the minimal sufficient summary statistic for recovering the mean, namely the empirical average: $h_{\psib^*}(x_1, x_2) = \bar{x} \equiv (x_1 + x_2) / 2$. 
Consequently, the distribution in summary space is given as $p(\bar{x}) = \mathcal{N}(0, 1)$.\footnote{This follows from the property $\text{Var}(\bar{x}) = \text{Var}((x_1 + x_2)/ 2) = (\text{Var}(x_1) + \text{Var}(x_2)) / 2^2 = 1$.}
In terms of the MMD criterion, we see that $\mathbb{MMD}^2\big[\underbrace{p(h_{\psib^*}(x_1, x_2))}_{=\mathcal{N}(0, 1)}\,||\,\mathcal{N}(0, 1)\big] = 0$.

Now, suppose that the real data are actually generated by a different model $\M^*$ with $\observed{x}\sim p^*(\x)$ given by 
\begin{equation}\label{eq:implications:process-2}
    \begin{aligned}
        \mu &\sim p(\mu),\\
        \observed{x}_1 &\sim \mathcal{N}(\mu, \sigma^2 = 1), \\
        \observed{x}_2 &\sim \mathcal{N}(\mu, \sigma^2 = 3).
    \end{aligned}
\end{equation}
Clearly, this process $p^*(\x)$ differs from $p(\x\given\M)$ (Eq.~\ref{eq:implications:process-1}) on the data domain according to the MMD metric: 
$\mathbb{MMD}^2\big[p^*(\x)\,||\,p(\x \given \mathcal{M})\big] > 0$. 
However, using the same calculations as above, we find that the summary space for the process $p^*(\x)$ also follows a unit Normal distribution: $p^*(\bar{x}) = \mathcal{N}(0, 1)$.
Thus, the processes $p^*(\x)$ and $p(\x\given\M)$ are indistinguishable in the summary space despite the fact that the first generative model $\M$ is clearly misspecified.

The above example shows that learning \emph{minimal sufficient summary statistics} for solving the inference task (i.e., the mean in this example) might not be optimal for detecting simulation gaps. 
In fact, increasing the output dimensions $S$ of the summary network $h_{\psib}$ would enable the network to learn structurally richer (overcomplete) sufficient summary statistics.
The latter would be invariant to fewer misspecifications and thus more useful for uncovering simulation gaps. 
In the above example, an \emph{overcomplete} summary network with $S = 2$ which simply copies and scales the two variables by their corresponding variances is able to detect the misspecification. 
\textbf{Experiment \numberGaussianMeans} studies the influence of the number of summary statistics in a controlled setting and provides empirical illustrations.
\textbf{Experiments \numberDDM} and \textbf{\numberCovid} further address the choice of the number of summary statistics in more complex models of decision making and disease outbreak.
Next, we describe how to practically detect simulation gaps during inference using only \emph{finite realizations} from $\M$ and $\mathcal{M}^*$.

\subsection{Detecting Model Misspecification with Finite Data}
Once the simulation-based training phase is completed, we can generate $M$ validation samples $\smash{\{\thetab^{(m)}, \x^{(m)}\}}$ of from our generative model $\M$ and pass them through the summary network to obtain a sample of latent summary vectors $\smash{\{\z^{(m)}\}}$, where $\smash{\z=h_{\psib}(\x)}$ denotes the output of the summary network.
The properties of this sample contain important convergence information: If $\smash{\z}$ is approximately unit Gaussian, we can assume a structured summary space given the training model $\M$.
This enables model misspecification diagnostics via distribution checking during inference on real data (see \autoref{alg:mms}).

Let $\smash{\{\observed{\x}^{(n)}\}}$ be an \emph{observed} sample, either simulated from a different generative model, or arising from real-world observations with an unknown generator. 
Before invoking the inference network, we pass this sample through the summary network to obtain the summary statistics for the sample: $\smash{\{\observed{\z}^{(n)}\}}$.
We then compare the validation summary distribution $\{\z^{(m)}\}$ with the summary statistics of the observed data $\smash{\{\observed{\z}^{(n)}\}}$ according to the sample-based MMD estimate $\smash{\widehat{\text{MMD}}^2\left(p(\z)\,||\, p(\observed{\z})\right)}$ \citep[cf.][]{Gretton2012}. %
Importantly, we are not limited to pre-determined sizes of simulated or real-world data sets, as the MMD estimator is defined for arbitrary $M$ and $N$.%
\footnote{To allow MMD estimation for data sets with single instances ($N=1$ or $M=1$), we do not use the unbiased MMD version from \citet{Gretton2012}. 
Singleton data sets are an important use case for our method in practice, and potential advantages of unbiased estimators do not justify exclusion of such data.}
To enhance visibility, the figures in the experimental section will depict the square root $\rMMD(\cdot, \cdot)$ of the originally squared MMD estimate.

Whenever we estimate the MMD from finite data, its estimates vary according to a sampling distribution and we can resort to a frequentist hypothesis test to determine the probability of observed MMD values under well-specified models.
Although this sampling distribution under the null hypothesis is unknown, we can estimate it from multiple sets of simulations from the generative model, $\smash{\{\z^{(m)}\}}$ and $\smash{\{\observed{\z}^{(n)}\}}$, with $M$ large and $N$ equal to the number of real data sets.
Based on the estimated sampling distribution, we can obtain a critical MMD value for a fixed Type I error probability $(\alpha)$ and compare it to the one estimated with the observed data. 
In general, a larger $\alpha$ level corresponds to a more conservative modeling approach: A larger type I error implies that more tests reject the null hypothesis, which corresponds to more frequent model misspecification alarms and a higher chance that incorrect models will be recognised.
Note that the Type II error probability ($\beta$) of this test will generally be high (\ie, the \emph{power} of the test will be low) whenever the number of real data sets $N$ is very small.
However, we show in \textbf{Experiment \numberCovid} that even as few as $5$ real data sets suffice to achieve $\beta \approx 0$ for a complex model on COVID-19 time series.

\subsection{Posterior Inference Errors}\label{sec:posterior-inference-errors}

Given a generative model $\mathcal{M}$, the \emph{analytic posterior under the potentially misspecified model} $p(\thetab \given \x,\mathcal{M})$ always exists, even if $\mathcal{M}$ is misspecified for the data $\x$.
Obtaining a trustworthy approximation of the analytic posterior is the fundamental basis for any follow-up inference (\eg, parameter estimation or model comparison) and must be at least an intermediate goal in real world applications.
Assuming optimal convergence under a misspecified model $\mathcal{M}$, the amortized posterior $q_{\phib}\big(\thetab \given \z = h_{\psib}(\x), \mathcal{M}\big)$ still corresponds to the analytic posterior $p(\thetab \given \x,\mathcal{M})$, as any transformed $\observed{\x}$ arising from $p^*(\x)$ has non-zero density in the latent Gaussian summary space.\footnote{We assume that we have no hard limits in the prior or simulator in $\M$.}
Thus, the posterior approximator should still be able to obtain the correct pushforward density under $\mathcal{M}$ for any query $\observed{\x}$.
However, optimal convergence can never be achieved after finite training time, so we need to address its implications for the validity of amortized simulation-based posterior inference in practice.

Given finite training data, the summary and inference networks will mostly see simulations from the \emph{typical set} $\mathcal{T}(\M) \subset \mathcal{X}$ of the generative model $\M$, that is, training instances whose self-information $-\log p(\x \given \mathcal{M})$ is close to the entropy $\mathbb{E}\big[-\log p(\x \given \mathcal{M})\big]$.
In high dimensional problems, the typical set will comprise a rather small subset of the possible outcome space, 
determined by a complex interaction between the components of $\M$ \citep{betancourt2017}.
Accordingly, good convergence in practice may mean that i) only observations from $\mathcal{T}(\M)$ actually follow the approximate Gaussian in latent summary space and ii) the inference network has only seen enough training examples in $\mathcal{T}(\M)$ to learn accurate posteriors for observables $\x \in \mathcal{T}(\M)$, but remains inaccurate for well-specified, but rare $\x\not\in\mathcal{T}(\M)$.

Since atypical or improbable outcomes occur rarely during simulation-based training, they have negligible influence on the loss in Eq.~\ref{eq:bf_kl_mmd}.
Consequently, posterior approximation errors for observations outside of $\mathcal{T}(\M)$ can be large, simply because the networks have not yet converged in these unusual regions, and the highly non-linear mapping of the inference network still deviates considerably from the true solution.
Better training methods might resolve this problem in the future, but for now our proposed MMD criterion reliably signals low fidelity posterior estimates by quantifying the ``distance from the typical generative set'' $\mathcal{T}(\M)$ in the structured summary space.

Moreover, we hypothesize and demonstrate empirically in the following experiments that the difference between the correct posterior $p(\thetab \given \x, \mathcal{M})$ and the approximate posterior $q_{\phib}(\thetab \given h_{\psib}(\x), \mathcal{M})$ for misspecified models increases as a function of MMD, and thus the latter also measures the amount of misspecification.
Therefore, our MMD criterion serves a dual purpose in practice: It can uncover potential simulation gaps and, simultaneously, signal errors in posterior estimation of rare (but valid) events.

\section{Experiments}
In the following, we illustrate our method to detect model misspecification in 5 experiments, namely (i) a 2D Gaussian conjugate model for the mean with various simulation gaps in the prior, simulator, and noise; (ii) a 5D Gaussian model with fully estimated covariance matrix and 20 inference parameters; (iii) a point process model of cancer and stromal cell development with hand-crafted summary statistics; (iv) a cognitive model of decision making with tractable likelihood to allow for a comparison with \texttt{Stan}; and (v) a complex epidemiological model with an application to real-world COVID-19 time series and 192 learned summary statistics.

\begin{figure}[t]
    \centering
    \begin{subfigure}[t]{.49\textwidth}
     \includegraphics[width=\linewidth]{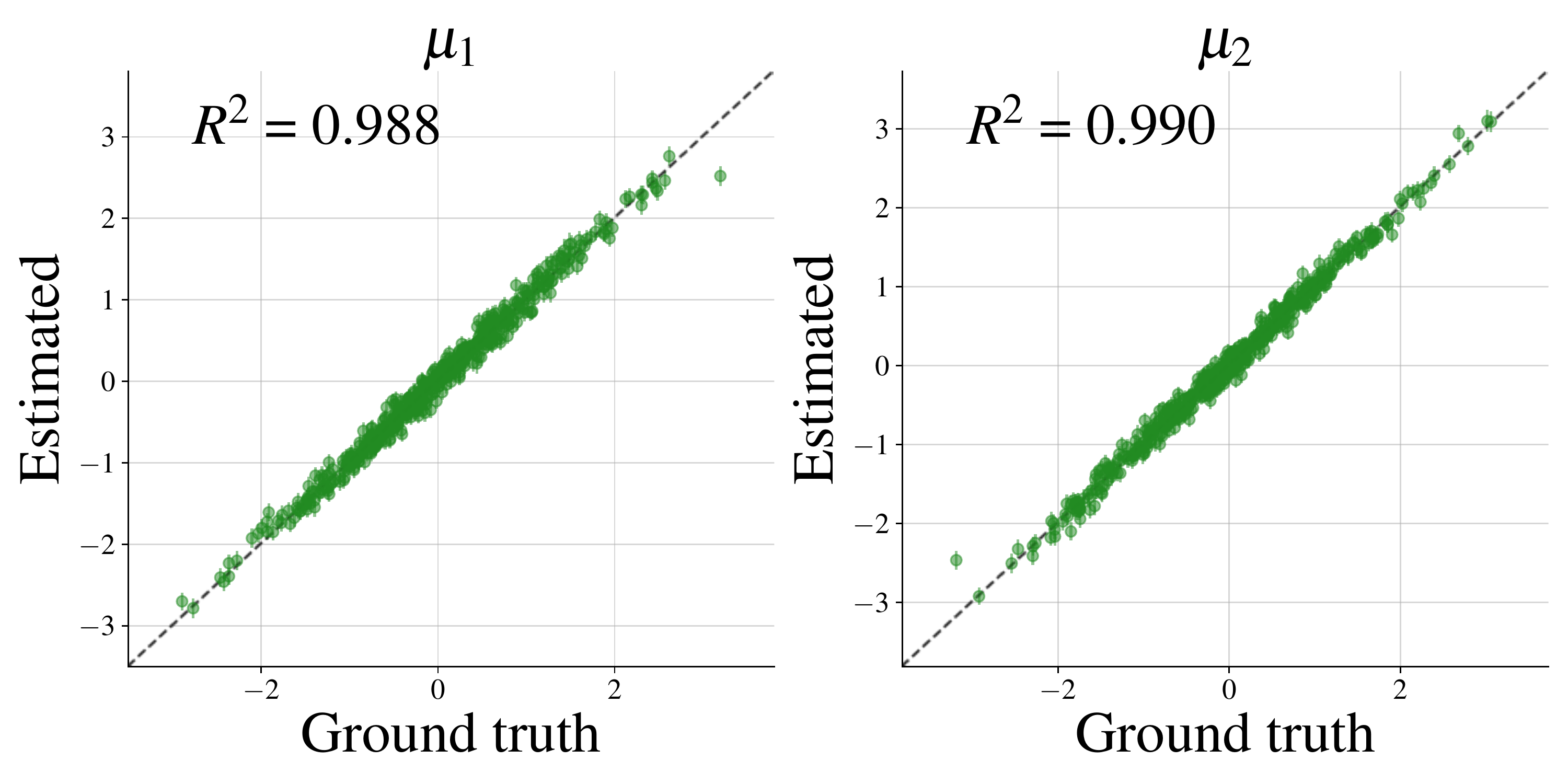}
     \caption{Excellent parameter recovery.}
     \label{fig:mvn:well-specified-performance:recovery}
    \end{subfigure}
    \hfill
    \begin{subfigure}[t]{.49\textwidth}
     \includegraphics[width=\linewidth]{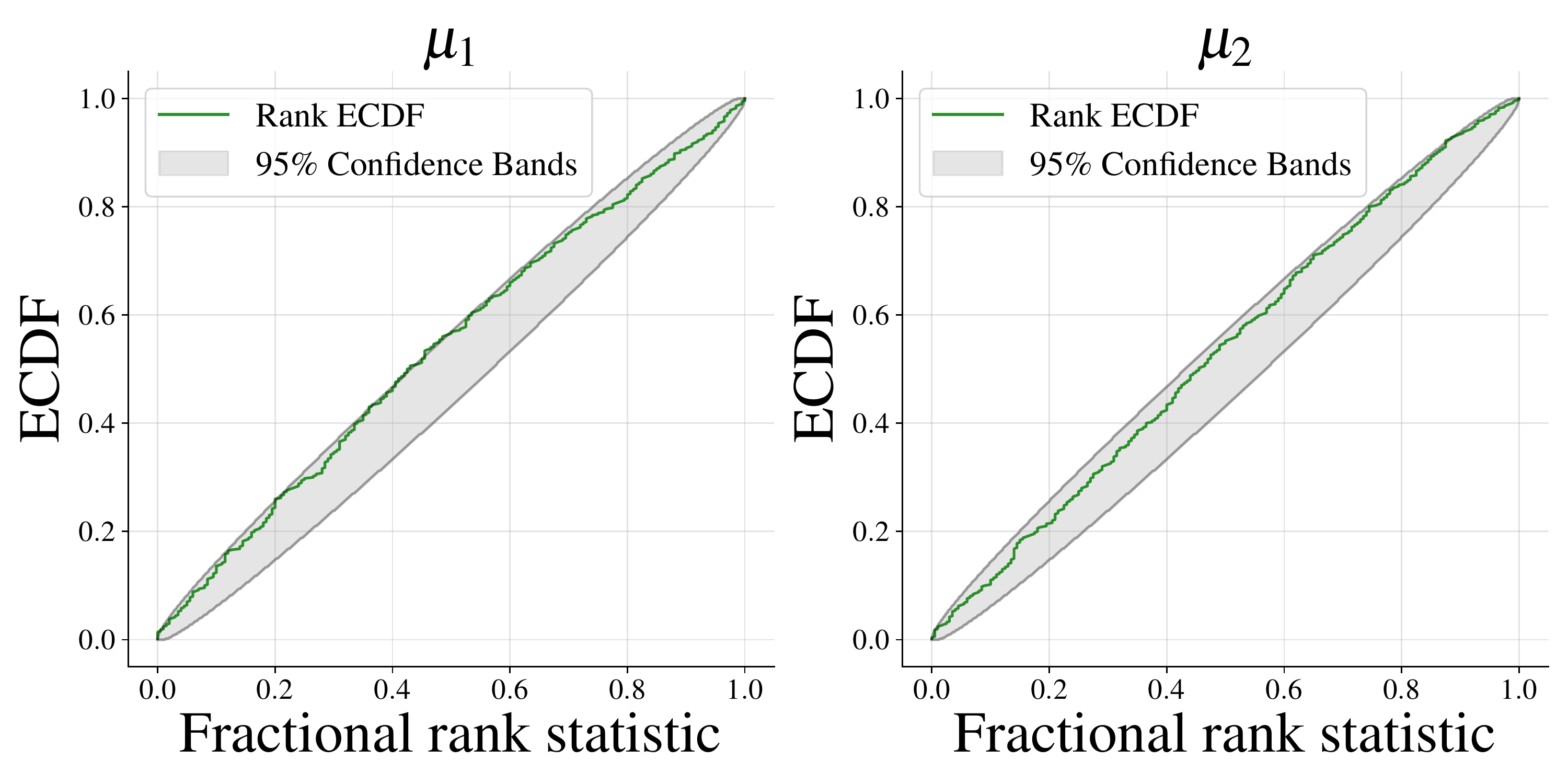}
     \caption{Excellent calibration.}
     \label{fig:mvn:well-specified-performance:calibration}
    \end{subfigure}
    \caption{\textbf{Experiment \numberGaussianMeans}. When the assumed model is well-specified for the observed data ($\M=\M^*$), both posterior recovery (\subref{fig:mvn:well-specified-performance:recovery}) and calibration (\subref{fig:mvn:well-specified-performance:calibration}) remain excellent with our adjusted optimization objective.}
    \label{fig:mvn:well-specified-performance}
\end{figure}
\subsection{Experiment \numberGaussianMeans: 2D Gaussian Means}\label{sec:experiment-toy-conjugate}
\begin{table}[b]
\small
    \begin{tabular}{l|ll}
    \centering
        \bfseries Model (MMS) &\bfseries Prior &\bfseries Likelihood\\
        \hline
        $\mathcal{M}$ (No MMS) &
        $\mub\sim\mathcal{N}(\mub_0=\0, \Sigmab_0=\mathbb{I})$&
        $\x_k\sim\mathcal{N}(\mub, \Sigmab=\mathbb{I})$
        \\
        $\mathcal{M}_P$ (Prior) &
        $\mub\sim\mathcal{N}(\mub_0 \neq \0, \Sigmab_0=\tau_0\mathbb{I}), \tau_0\in\mathbb{R}^+$&
        $\x_k\sim\mathcal{N}(\mub, \Sigmab=\mathbb{I})$\\
        $\mathcal{M}_S$ (Simulator) &
        $\mub\sim\mathcal{N}(\mub_0=\0, \Sigmab_0=\mathbb{I})$&
        $\x_k\sim\mathcal{N}(\mub, \Sigmab=\tau\mathbb{I}), \tau\in\mathbb{R}^+$
        \\
        $\mathcal{M}_N$ (Noise)&
        $\mub\sim\mathcal{N}(\mub_0=\0, \Sigmab_0=\mathbb{I})$ & 
        $\x_k\sim\lambda\cdot\mathrm{Beta}(2, 5)+(1-\lambda)\cdot\mathcal{N}(\mub, \Sigmab=\mathbb{I})
        $
    \end{tabular}%
    \caption{\textbf{Experiment \numberGaussianMeans.} Investigated model misspecifications.}
    \label{tab:mvn-mean-misspecifications}
\end{table}
We set the stage by estimating the mean of a $2$-dimensional conjugate Gaussian model with $K=100$ observations per data set and a known analytic posterior in order to illustrate our method.
This experiment contains the Gaussian examples from \citet{frazier_model_2020} and \citet{ward_robust_2022}, and extends them by (i) studying misspecifications beyond the likelihood variance (see below); and (ii) implementing continuously widening simulation gaps, as opposed to a single discrete misspecification.
The data generating process is defined as
\begin{equation}
    \x_k \sim \mathcal{N}(\x\given\mub, \Sigmab) \quad\text{for } k = 1,...,K\qquad \text{with }\;\;
    \mub \sim \mathcal{N}(\mub\given\mub_0, \Sigmab_0).
\end{equation}
We use a permutation invariant summary network \cite{invariant} with $S=2$ output dimensions, which equal the number of minimal sufficient statistics\footnote{The terms ``minimal'', ``sufficient'', and ``overcomplete'' refer to the inference task and \emph{not} to the data. Thus, $S=2$ summary statistics are \emph{sufficient} to solve the inference task, namely recover two means.} implied by the analytic posterior.
For training the posterior approximator, we set the prior of the generative model $\M$ to a unit Gaussian and the likelihood covariance $\Sigmab$ to an identity matrix.
The induced misspecifications during test time are outlined in \autoref{tab:mvn-mean-misspecifications}.\footnote{Note that Experiment \numberGaussianMeans~from \citet{ward_robust_2022} is represented by the scenario $\mathcal{M}_S$ with $\tau=2$.
In addition, we study model misspecification across the entire plausible parameter space of the likelihood variance, as well as prior ($\M_P$) and noise ($\M_N$) misspecification.
}
We conduct the experiment with BayesFlow and SNPE-C, both equipped with our adjusted optimization objective.%
\begin{figure}[t]
    \centering
    \begin{subfigure}[t]{0.45\linewidth}
        \includegraphics[width=\linewidth]{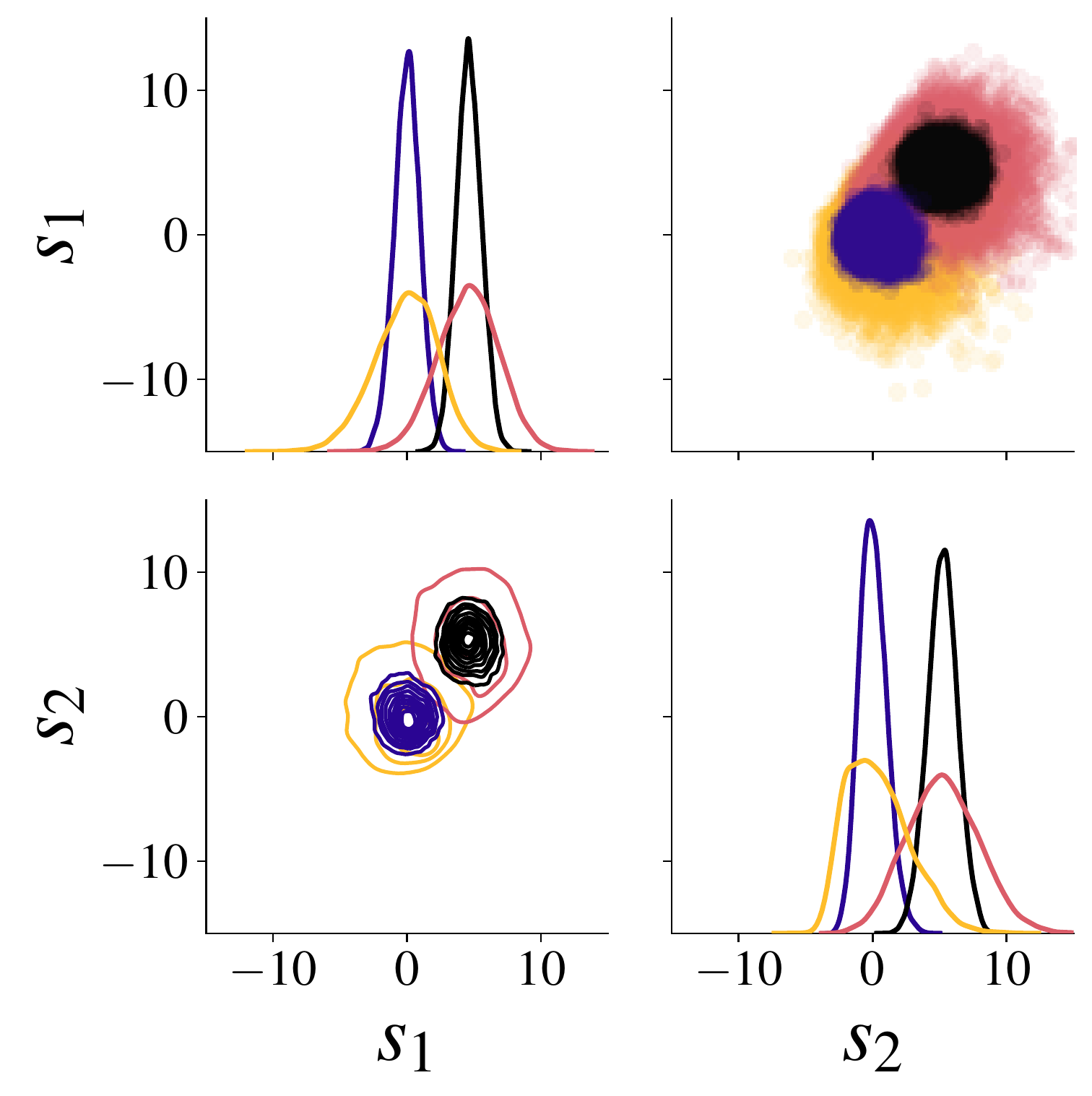}
    \end{subfigure}\hfill
    \begin{subfigure}[t]{0.45\linewidth}
        \includegraphics[width=\linewidth]{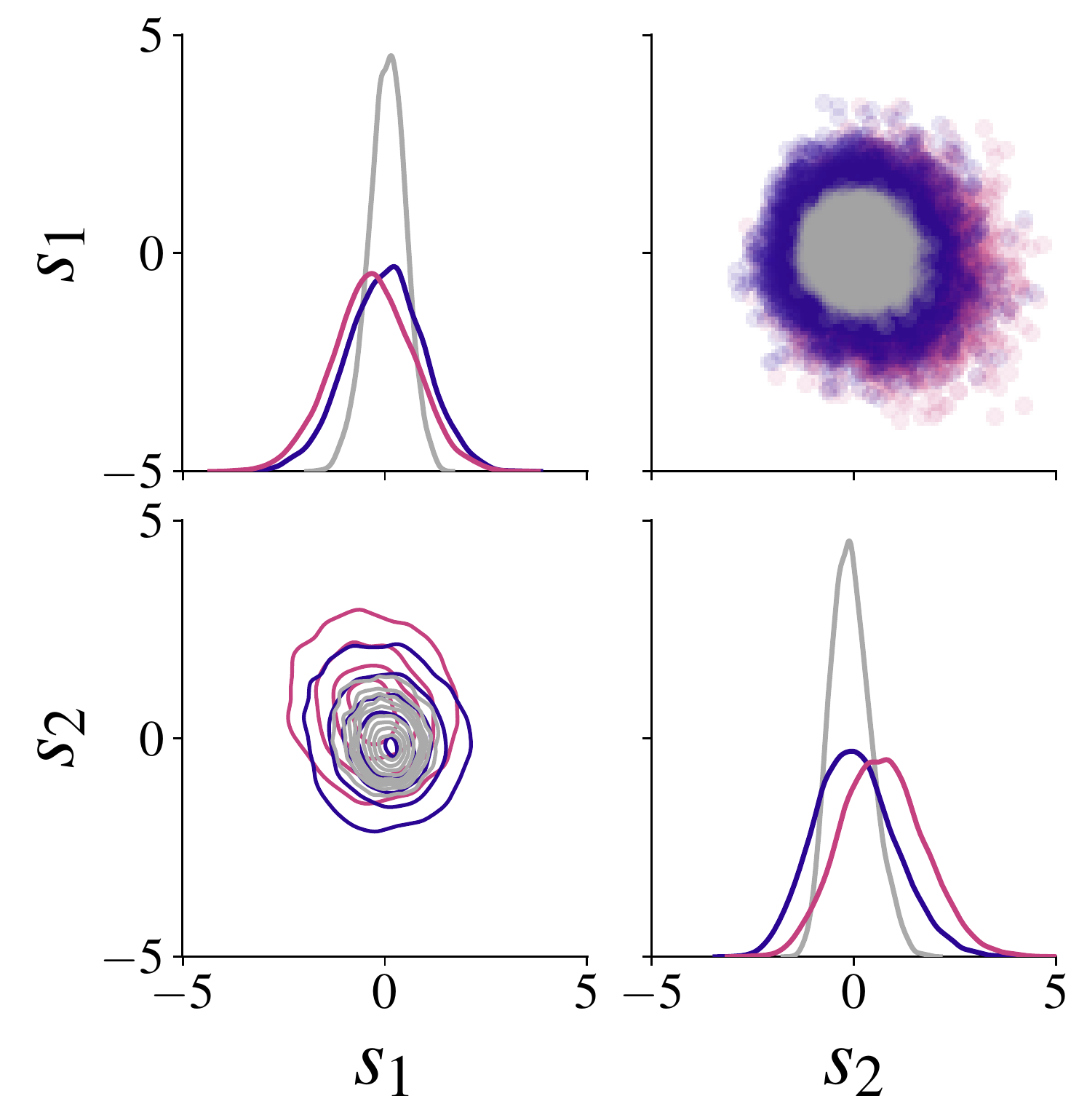}
    \end{subfigure}\\
    \includegraphics[width=0.95\linewidth]{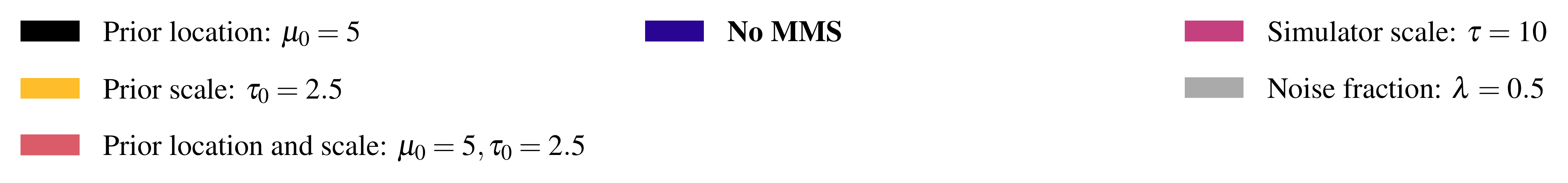}
    \caption{\textbf{Experiment \numberGaussianMeans.} Summary space samples for the minimal sufficient summary network ($S=2$) from a well-specified model $\M$ (blue) and several misspecified configurations. \textbf{Left:} Prior misspecification can be detected. \textbf{Right:} Noise misspecification can be detected, while simulator scale misspecification is indistinguishable from the validation summary statistics (but see results for $S=4$).}
    \label{fig:mvn:pairplot}
\end{figure}

\begin{figure}[t]
    \centering
    \begin{subfigure}[c]{0.9\linewidth}%
    \setlength\tabcolsep{2pt}%
    \begin{tabular}{ccccc}
    && \multicolumn{2}{c}{\textbf{Model Misspecification}} & \\
        & &
        \textbf{Prior} ($\M_P$) &
        \textbf{Simulator} ($\M_S$) \textbf{\& noise} ($\M_N$)
        \\
        \multirow{2}{*}{\hspace*{-0.1cm}\rotatebox[origin=c]{90}{\textbf{Summary Network}}} &
        \rotatebox[origin=c]{90}{\textbf{minimal}} &
        \raisebox{-0.48\height}{\includegraphics[width=0.45\linewidth]{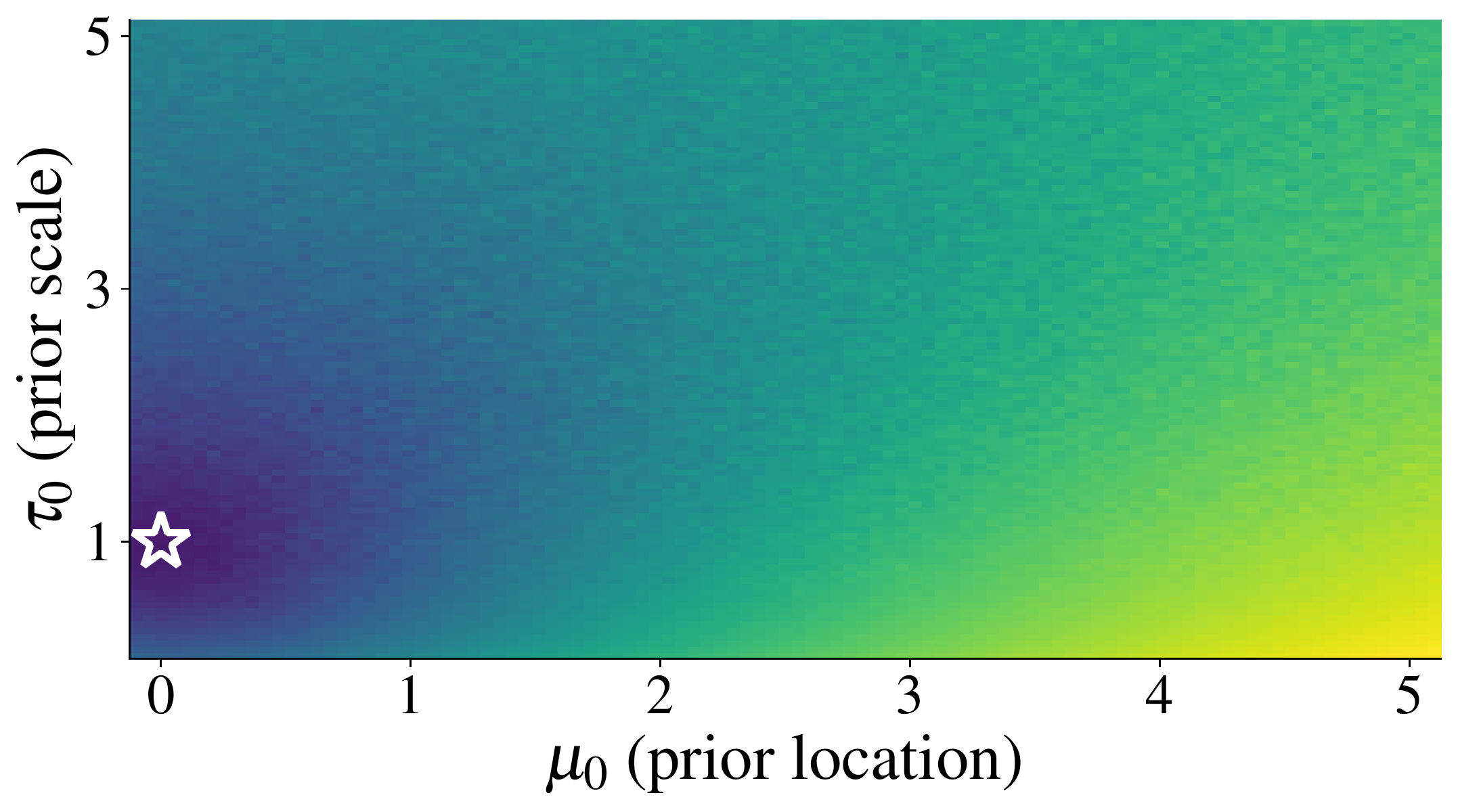}} &
        \raisebox{-0.48\height}{\includegraphics[width=0.46\linewidth]{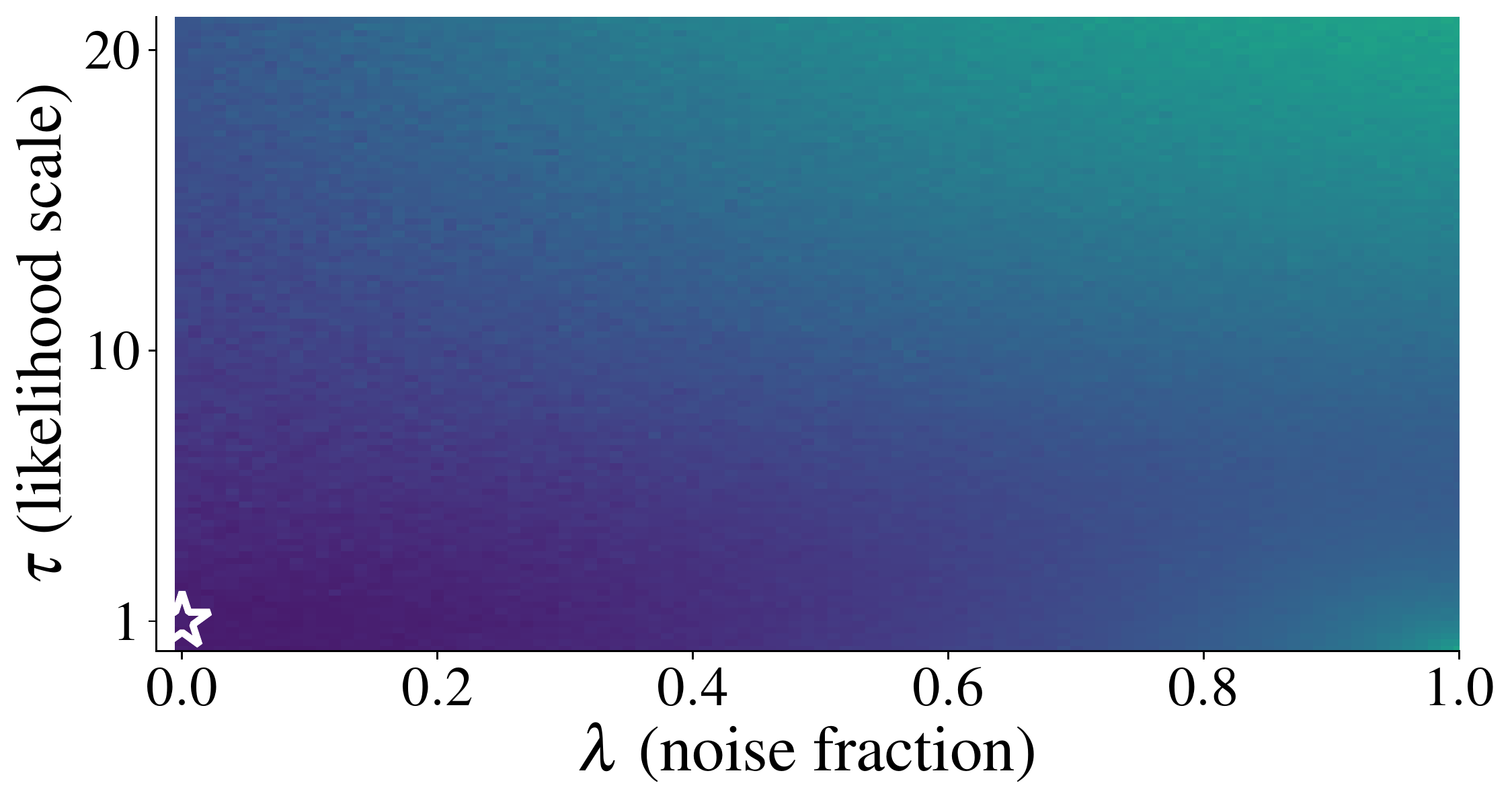}}
        \\
        &\rotatebox[origin=c]{90}{\textbf{overcomplete}} &
        \raisebox{-0.5\height}{\includegraphics[width=0.45\linewidth]{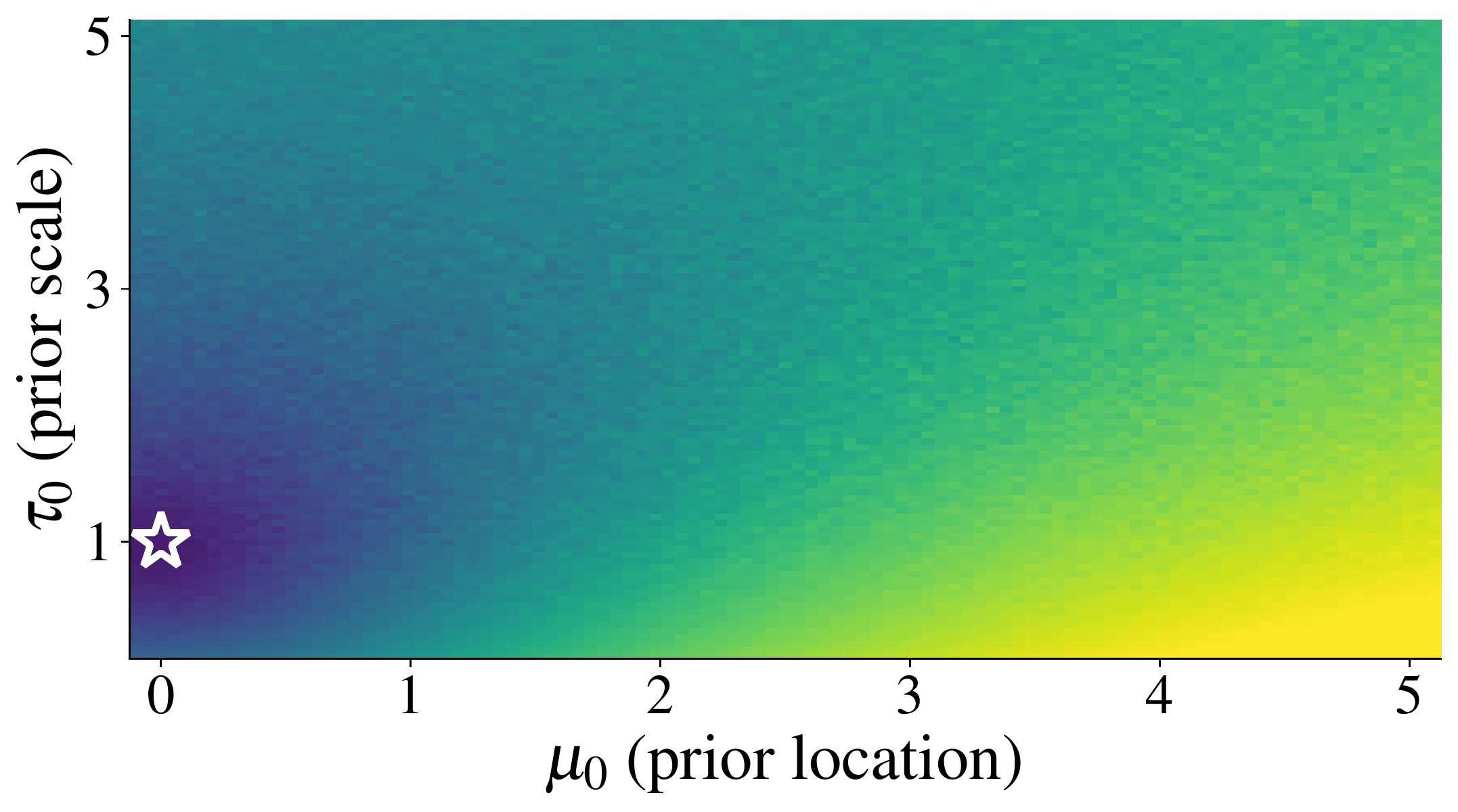}} &
        \raisebox{-0.5\height}{\includegraphics[width=0.46\linewidth]{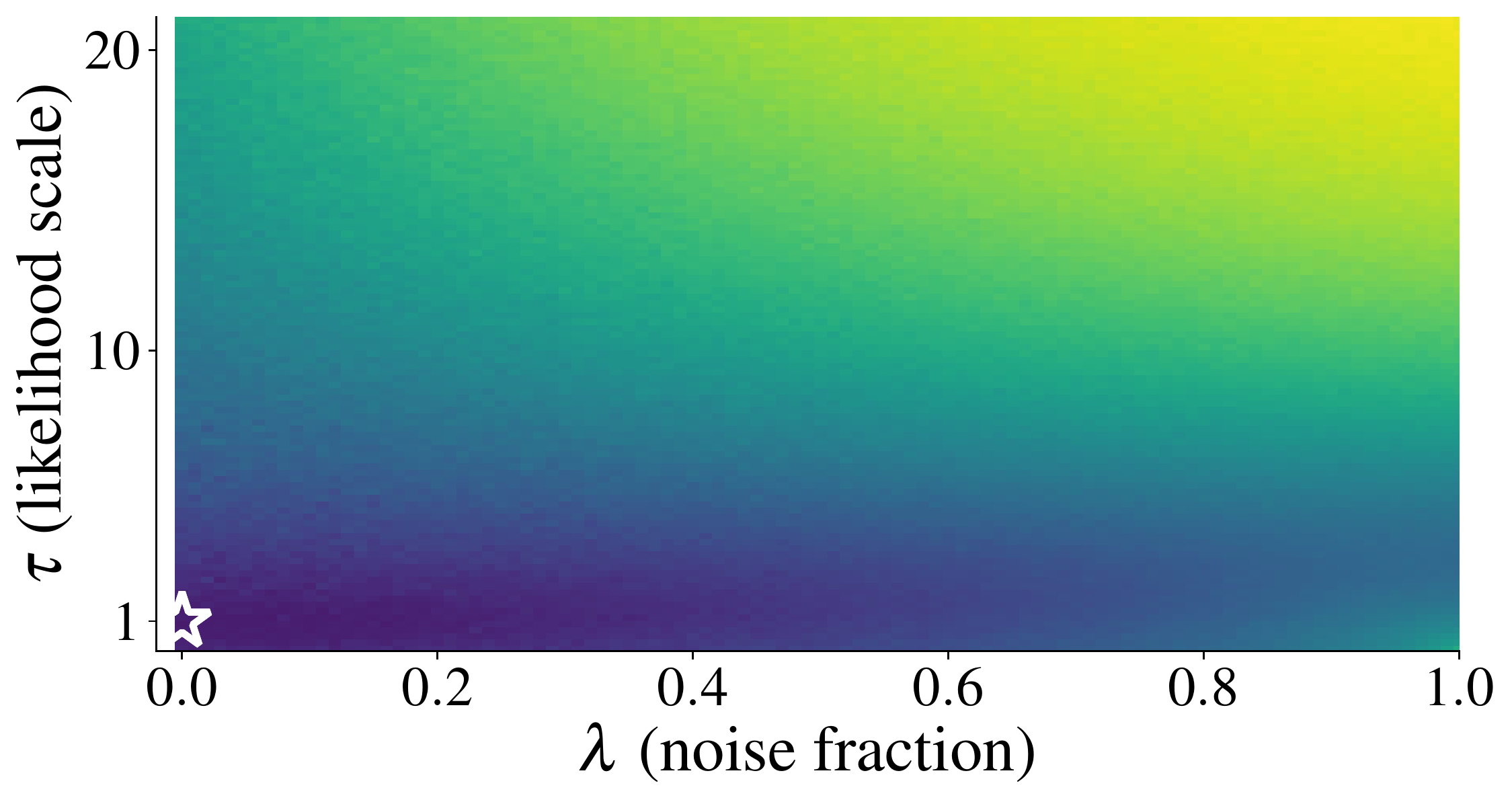}}
    \end{tabular}%
    \end{subfigure}%
    \begin{subfigure}[c]{0.06\linewidth}
    \includegraphics[width=\linewidth, clip, trim=9.8cm 0cm 0.2cm 0cm]{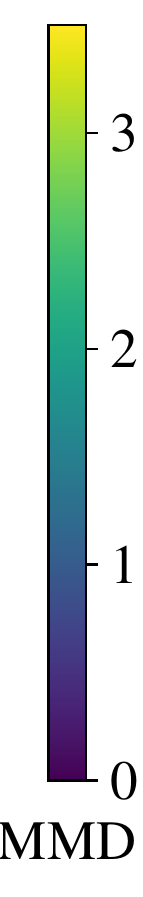}
    \end{subfigure}
    \caption{\textbf{Experiment \numberGaussianMeans.} Summary space MMD as a function of misspecification severity. White stars indicate the well-specified model configuration (i.e., equal to the training model $\M$).}
    \label{fig:mvn:mmd}
\end{figure}

\textit{Results.}
The BayesFlow network trained to minimize the augmented objective (Eq.~\ref{eq:bf_kl_mmd}) exhibits excellent recovery of the analytic posterior means when no misspecification is present (see \autoref{fig:mvn:well-specified-performance:recovery}). 
Furthermore, the posterior calibration \citep[SBC;][]{talts_validating_2020} remains excellent, as shown in \autoref{fig:mvn:well-specified-performance:calibration} via simultaneous confidence bands of rank ECDFs \cite{sailynoja_graphical_2021}.
All prior misspecifications manifest themselves in anomalies in the summary space which are directly detectable through visual inspection of the $2$-dimensional summary space in \autoref{fig:mvn:pairplot} (left). 
Note that the combined prior misspecification (location and scale) exhibits a summary space pattern that combines the location and scale of the respective location and scale misspecifications.
However, based on the $2$-dimensional summary space, misspecifications in the fixed parameters of the likelihood ($\tau$) and mixture noise are not detectable via an increased MMD (see \autoref{fig:mvn:mmd}, top right).

We further investigate the effect of an \emph{overcomplete} summary space with respect to the inference task, namely $S=4$ learned summary statistics with an otherwise equal architecture.
In addition to prior misspecifications, the overcomplete summary network also captures misspecifications in the noise and simulator via the MMD criterion (see \autoref{fig:mvn:mmd}, bottom row).
Furthermore, the induced misspecifications in the noise and simulator are visually detectable in the summary space samples (see \autoref{fig:app:mvn-means:overcomplete:pairplot} in the Appendix).
Recall that the $2$-dimensional summary space fails to capture these misspecifications (see \autoref{fig:mvn:mmd}, top right).
\begin{figure}[t]
\centering
    \begin{subfigure}[c]{0.9\linewidth}%
    \setlength\tabcolsep{2pt}%
    \begin{tabular}{ccccc}
    && \multicolumn{2}{c}{\textbf{Model Misspecification}} & \\
        & &
        \textbf{Prior} ($\M_P$) &
        \textbf{Simulator} ($\M_S$) \textbf{\& noise} ($\M_N$)
        \\
        \multirow{2}{*}{\hspace*{-0.1cm}\rotatebox[origin=c]{90}{\textbf{Summary Network}}} &
        \rotatebox[origin=c]{90}{\textbf{minimal}} &
        \raisebox{-0.48\height}{\includegraphics[width=0.45\linewidth]{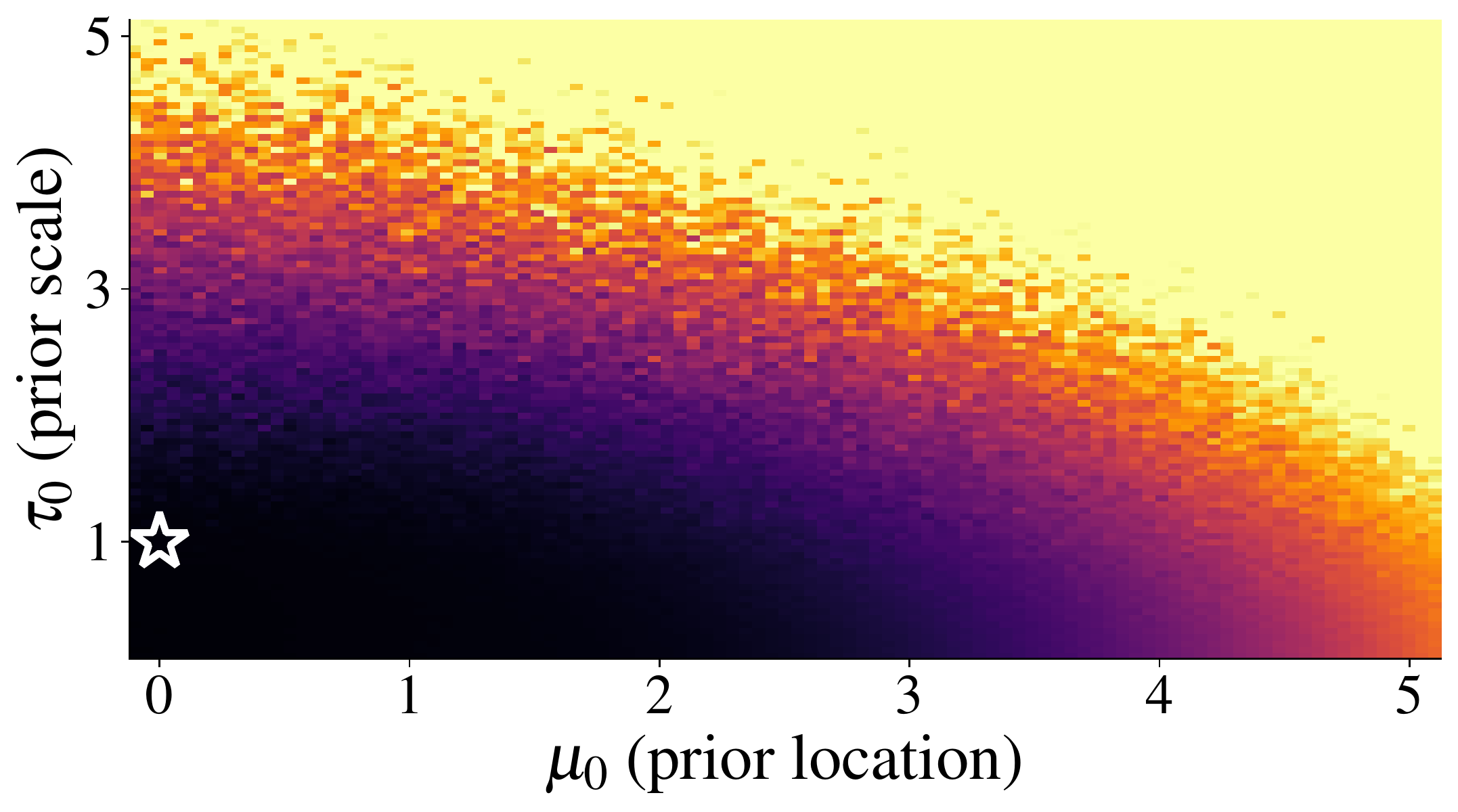}} &
        \raisebox{-0.48\height}{\includegraphics[width=0.46\linewidth]{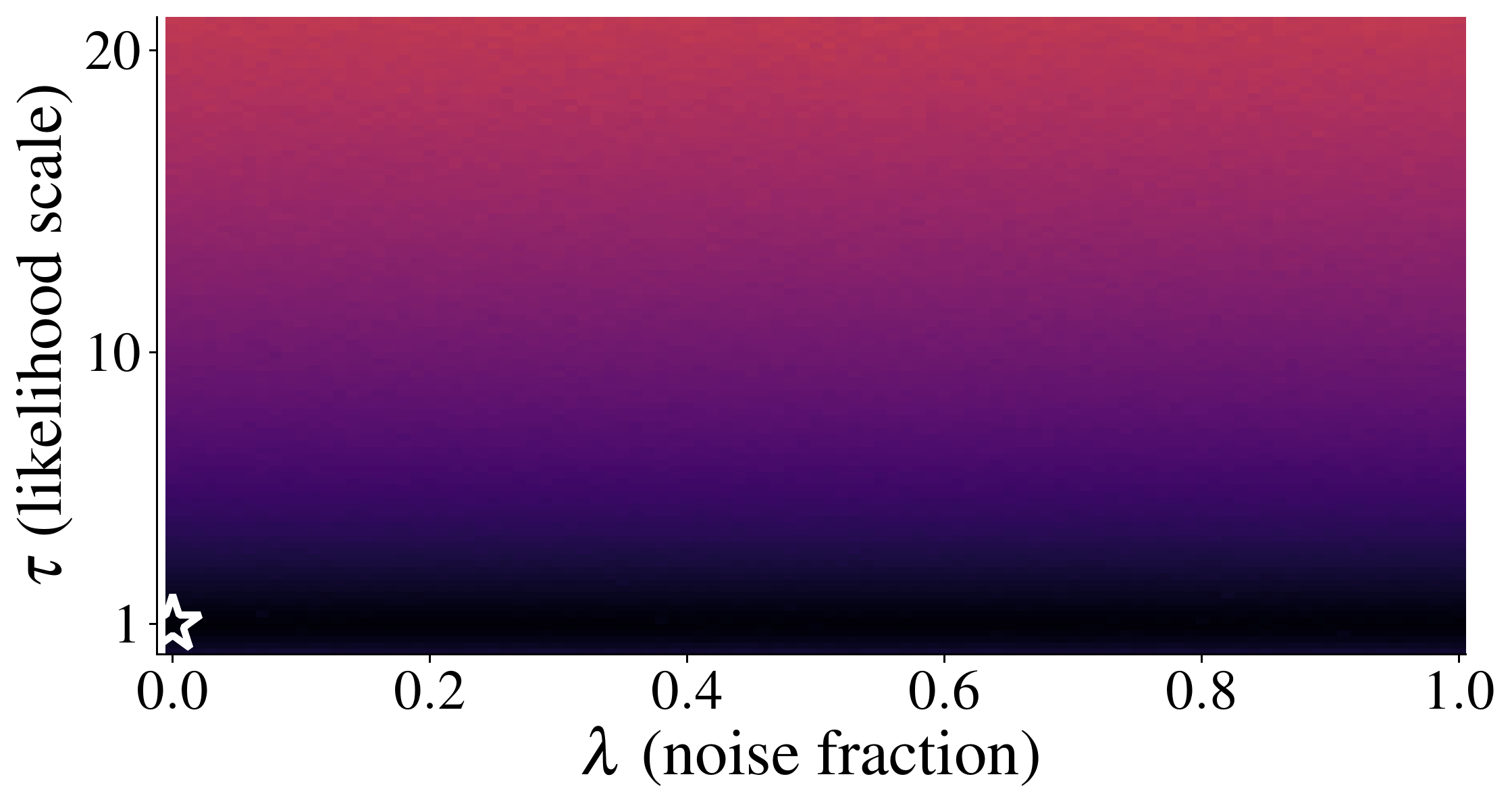}}
        \\
        &\rotatebox[origin=c]{90}{\textbf{overcomplete}} &
        \raisebox{-0.5\height}{\includegraphics[width=0.45\linewidth]{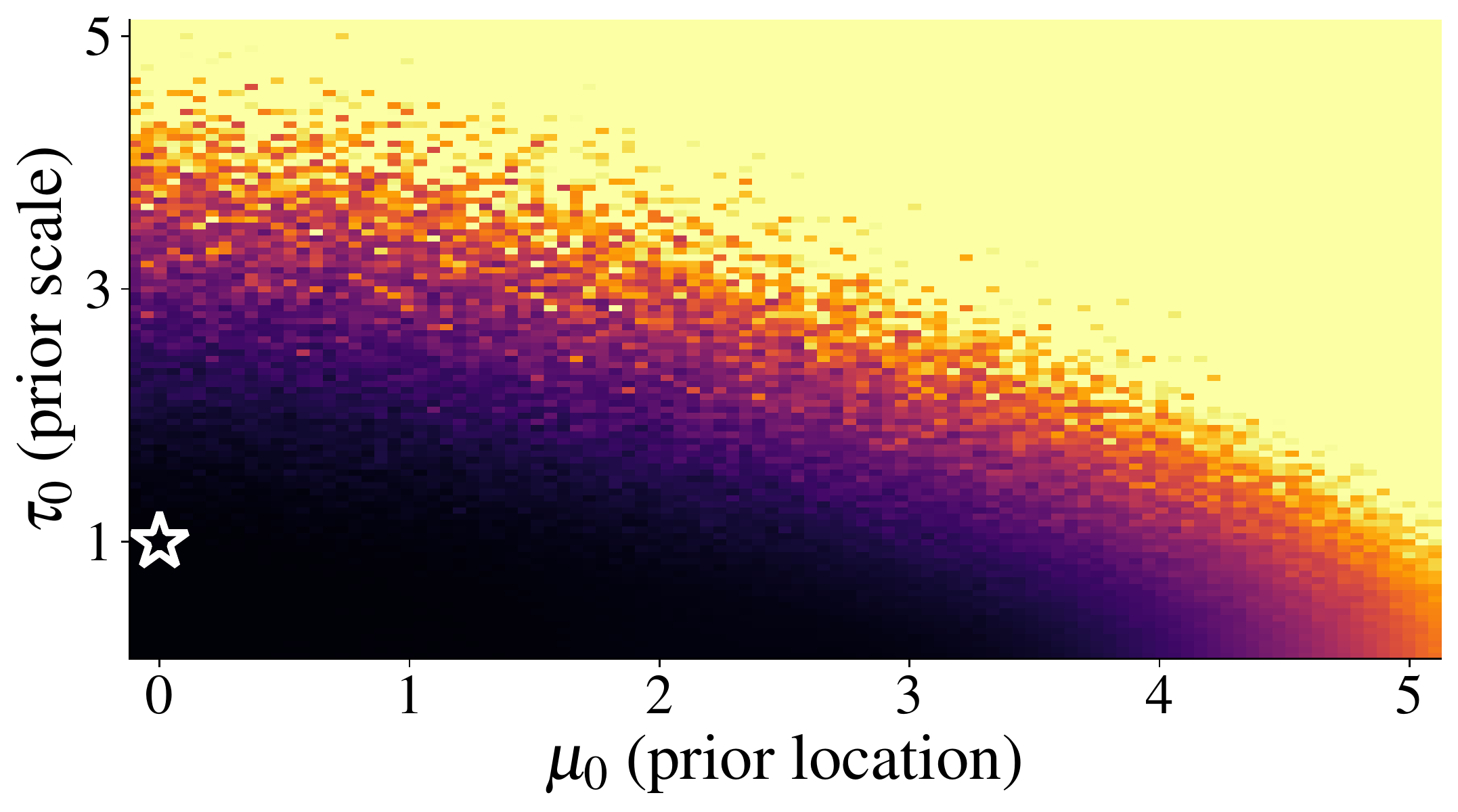}} &
        \raisebox{-0.5\height}{\includegraphics[width=0.46\linewidth]{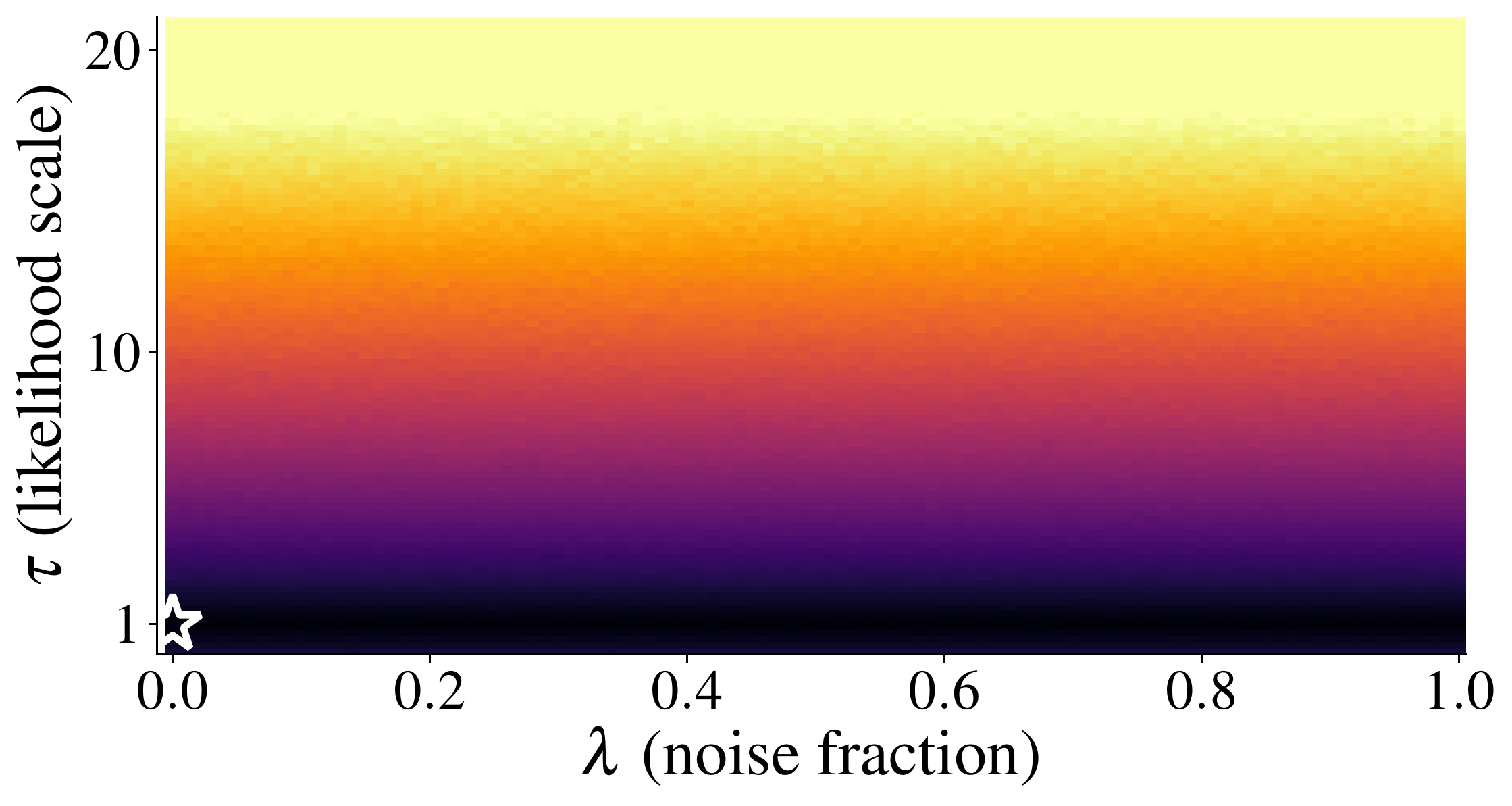}}
    \end{tabular}%
    \end{subfigure}%
    \begin{subfigure}[c]{0.08\linewidth}
    \includegraphics[width=\linewidth, clip, trim=9.5cm 0cm 0.2cm 0cm]{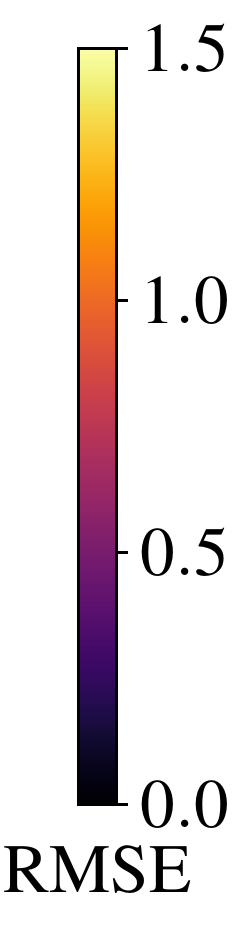}%
    \end{subfigure}
    \caption{\textbf{Experiment \numberGaussianMeans.} Posterior Error (RMSE of correct vs.\ analytic posterior means) as a function of misspecification severity. White stars indicate the well-specified model configuration (i.e., equal to the training model $\M$).}
    \label{fig:mvn:rmse}
\end{figure}
Finally, we compute the error in posterior recovery as a function of the misspecification severity.
To ease visualization, we use the RMSE of the approximated posterior mean from $L$ samples $\hat{\overline{\thetab}}^{(n)} = \frac{1}{L}\sum_{l=1}^L\thetab^{(l, n)}$ against the analytic posterior mean $\overline{\thetab}^{(n)}$ across a number of $N$ data sets.\footnote{Since the approximate posterior in the Gaussian model is likely to be symmetric---and the analytic posterior is symmetric by definition---we deem the posterior mean as an appropriate evaluation target for the RMSE across data sets.
In fact, error metrics over several posterior quantiles (\ie, $Q_{25}$, $Q_{50}$ and $Q_{75}$) in the place of posterior means yield identical results.
Other common metrics (\ie, MSE and MAE) yield identical results.}

\autoref{fig:mvn:rmse} illustrates that more severe model misspecifications generally coincide with a larger error in posterior estimation across all model misspecifications for both $S=2$ and $S=4$ learned summary statistics, albeit with fundamental differences, as explained in the following.
When processing data emerging from models with misspecified noise and simulator (see \autoref{fig:mvn:rmse}, right column), minimal and overcomplete summary networks exhibit a drastically different behavior:
While the minimal summary network cannot detect noise or simulator simulation gaps, its posterior estimation performance is not heavily impaired either (see \autoref{fig:mvn:rmse}, top right).
On the other hand, the overcomplete summary network is able to capture noise and simulator misspecifications, but also incurs larger posterior inference error (see \autoref{fig:mvn:rmse}, bottom right).
This might suggest a trade-off between model misspecification detection and posterior inference error, depending on the number of learnable summary statistics.

From a practical modeling perspective, researchers might wonder how to choose the number of learnable summary statistics.
While an intuitive heuristic might suggest ``the more, the merrier'', the observed results in this experiment beg to differ depending on the modeling goals.
If the focus in a critical application lies in detecting potential simulation gaps, it might be advantageous to utilize a large (overcomplete) summary vector.
However, modelers might also desire a network which is as robust as possible during test time, opting for a smaller number of summary statistics.
\textbf{Experiments \numberCovid} addresses this question for a complex non-linear time series model where the number of sufficient summary statistics is unknown.

\textit{SNPE-C (APT).}
Our method successfully detects model misspecification using SNPE-C \cite{apt} with the proposed MMD criterion and a structured summary space (see \autoref{app:snpe}).
The results are largely equivalent to those obtained with BayesFlow \cite{bayesflow}.
The nuanced differences are not soundly interpretable due to the architectural differences of the two frameworks.

\subsection{Experiment \numberGaussianMeansCov: 5D Gaussian Means and Covariance}
\begin{figure}[t]
    \begin{minipage}{.48\linewidth}
        \begin{subfigure}[t]{\linewidth}
            \includegraphics[width=\linewidth]{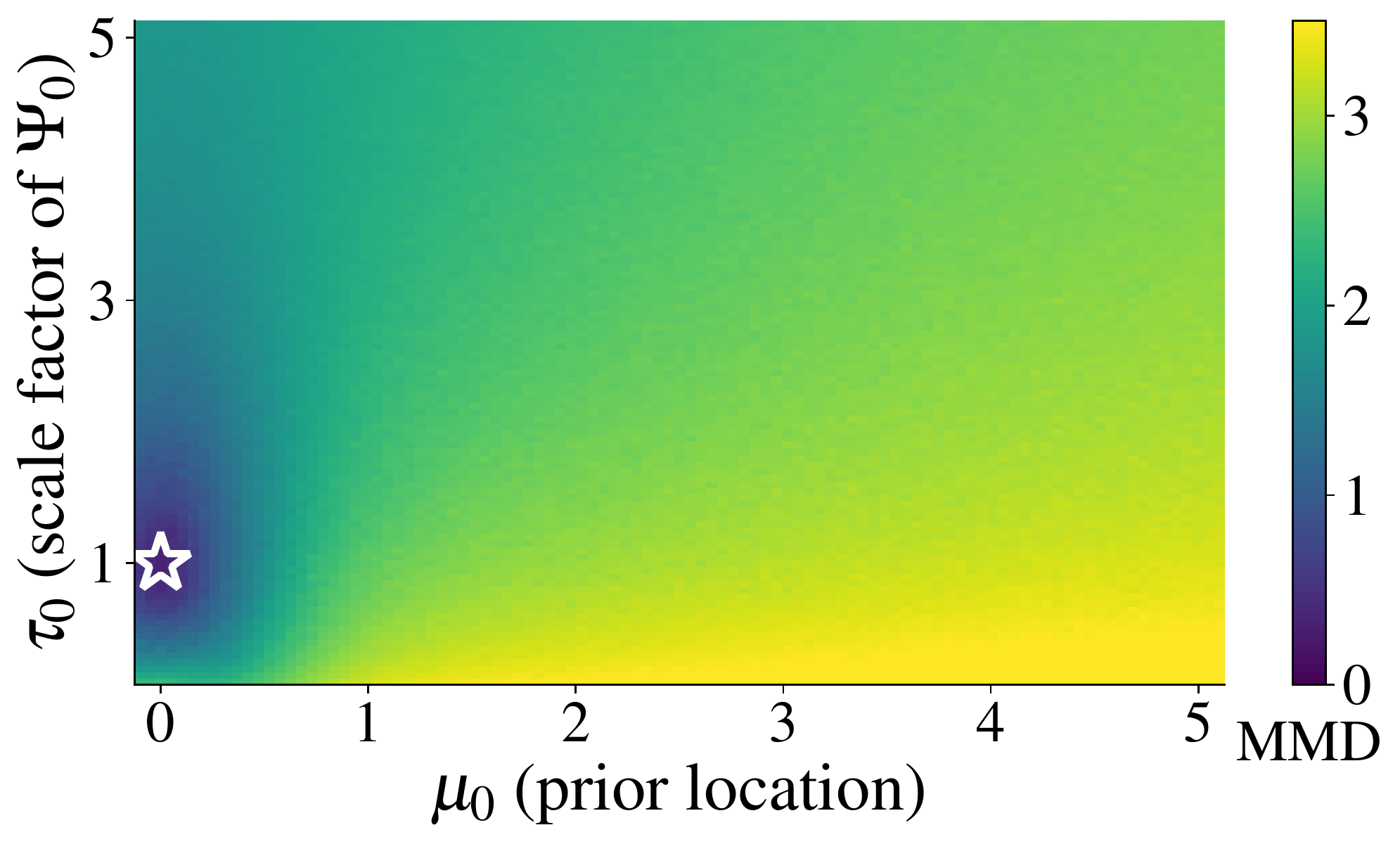}
            \caption{Prior misspecification.}
            \label{fig:app:mvn-full:MMD:prior}
        \end{subfigure}
    \end{minipage}
    \hfill
    \begin{minipage}{.48\linewidth}
        \begin{subfigure}[t]{\linewidth}
            \includegraphics[width=\linewidth]{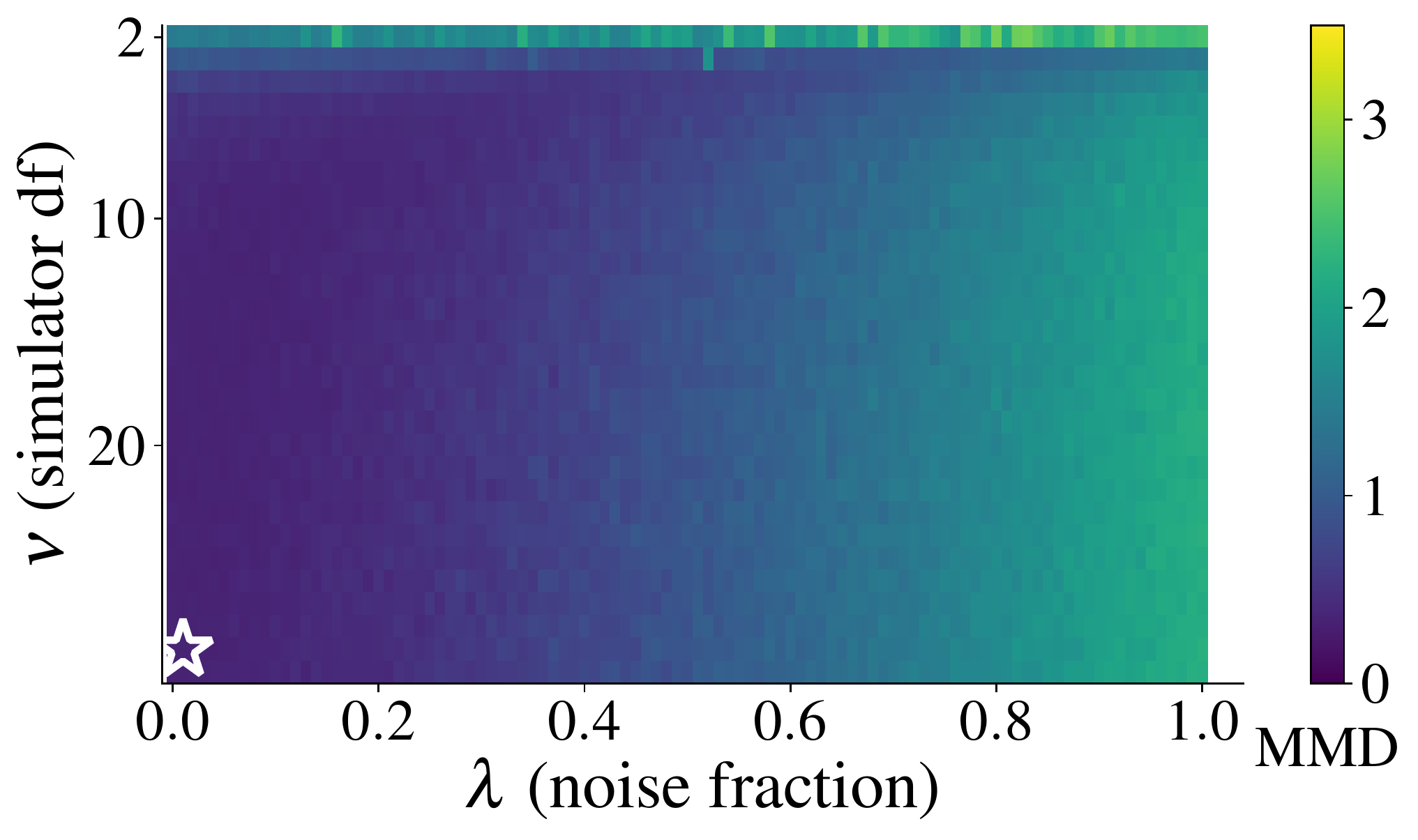}
            \caption{Simulator and noise misspecification.}
            \label{fig:app:mvn-full:MMD:likelihood-noise}
        \end{subfigure}
    \end{minipage}
    \hspace*{1cm}
\caption{\textbf{Experiment \numberGaussianMeansCov.} MMD as a function of model misspecification severity in the prior (left) as well as simulator and noise (right). 
All induced model misspecifications are detectable through an increased MMD.
White stars represent the configuration for the well-specified model (i.e., training model $\M$).
The deviating pattern in the top-most row of (\subref{fig:app:mvn-full:MMD:likelihood-noise}) is caused by the infinite variance of the Student $t$ likelihood with $\nu=2$ degrees of freedom: This is an extreme simulation gap with respect to the unit Gaussian model~$\M$, and consequently detected as such.}
\label{fig:app:mvn-full:MMD}
\end{figure}
This experiments extends the Gaussian conjugate model to higher dimensions and a more difficult task, i.e., recovering the means and the full covariance matrix of a $5$-dimensional Gaussian.
There is a total of $20$ inference parameters---5 means and 15 (co-)variances---meaning that $20$ summary statistics would suffice to solve the inference task.
The mean vector $\mub$ and the covariance matrix $\Sigmab$ are drawn from a joint prior, namely a normal-inverse-Wishart distribution \citep[$\NIW$;][]{Barnard2000}.
The normal-inverse-Wishart prior $\NIW(\mub, \Sigmab\given\mub_0, \lambda_0, \Psib_0, \nu_0)$ implies a hierarchical process, and the implementation details are described in Section~\ref{sec:app:mvn-full} in the Appendix.
We set the number of summary statistics to $S=40$ to balance the trade-off between posterior error and misspecification detection (see \textbf{Experiment \numberGaussianMeans}).
The model $\mathcal{M}$ used for training the networks as well as the induced model misspecifications (prior, simulator, and noise) are detailed in \autoref{tab:app:mvn-full-MMS} in the Appendix.

\textit{Results.} The converged posterior approximator can successfully recover the analytic posterior for all inference parameters when no model misspecification is present.
Thus, our method does not impede posterior inference when the models are well-specified.
Since the summary space comprises $S=40$ dimensions, visual inspection is no longer feasible, and we resort to the proposed MMD criterion.
Both induced prior misspecifications---i.e., location and variance---are detectable through an increased MMD (see \autoref{fig:app:mvn-full:MMD:prior}).
Model misspecifications via a heavy-tailed simulator---\ie, Student-$t$ with $\nu=2$ degrees of freedom---, as well as Beta noise, are also detectable with our MMD criterion (see \autoref{fig:app:mvn-full:MMD:likelihood-noise}).

\subsection{Experiment \numberCS: Cancer and Stromal Cell Model}
\begin{figure}[t]
    \centering%
    \begin{minipage}[b]{0.5\linewidth}%
        \begin{subfigure}[t]{\linewidth}%
            \includegraphics[width=\linewidth]{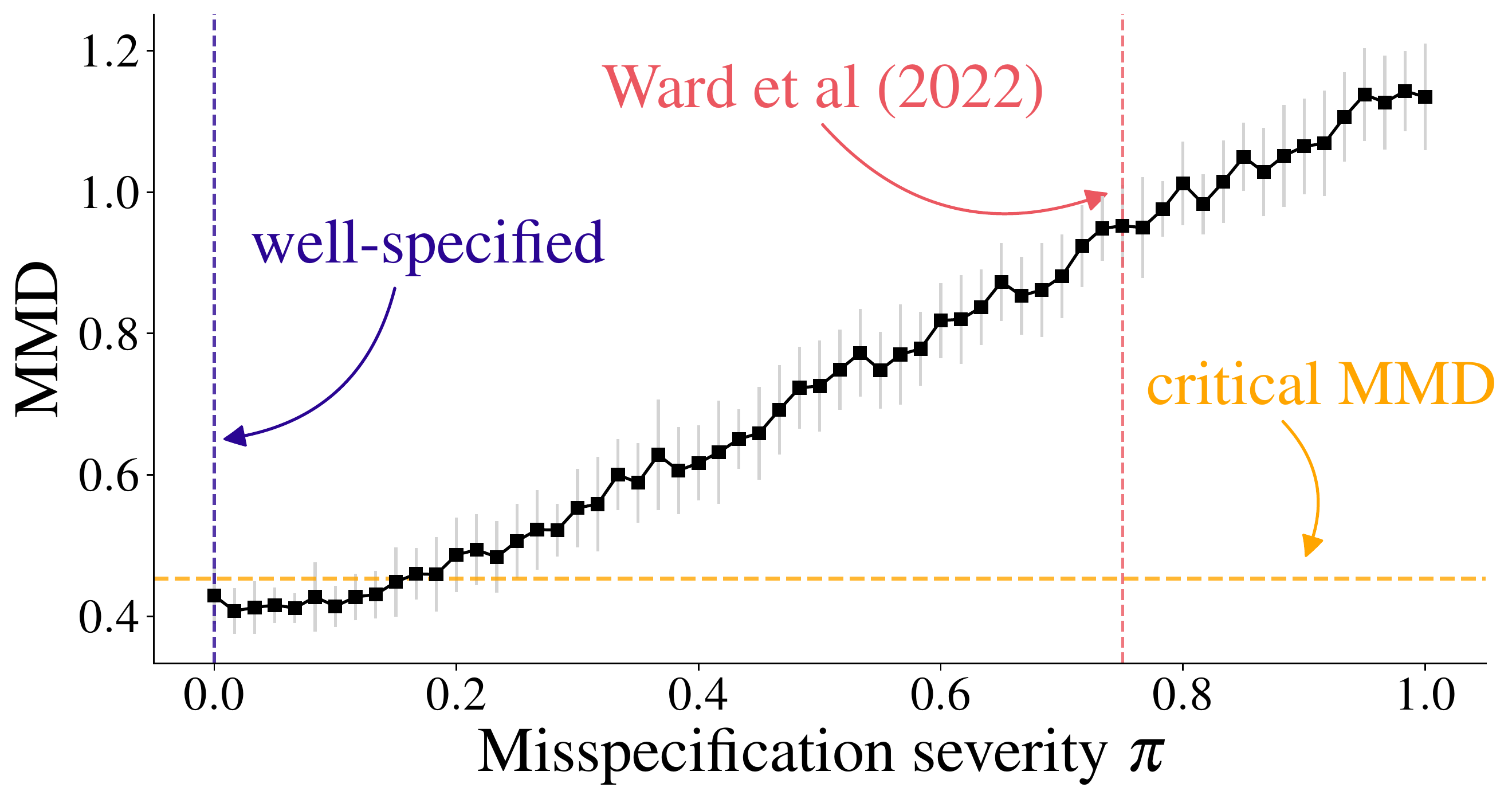}%
            \caption{MMD as a function of misspecification severity.}%
            \label{fig:cs:mms-mmd}%
        \end{subfigure}\\
        \begin{subfigure}[t]{\linewidth}%
                \includegraphics[width=\linewidth]{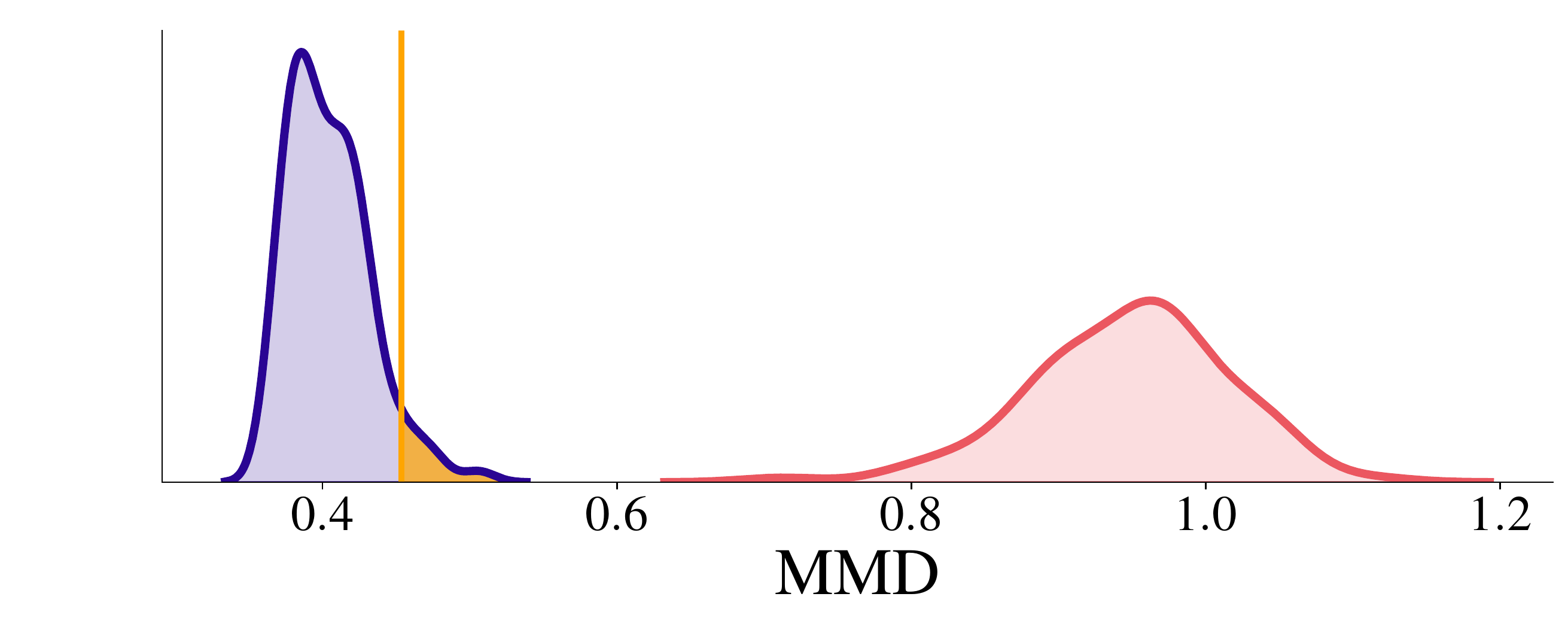}%
                \caption{Power for $\pi=0.75$.}%
            \label{fig:cs:power}%
        \end{subfigure}%
    \end{minipage}%
    \begin{subfigure}[t]{0.49\linewidth}%
        \includegraphics[width=\linewidth]{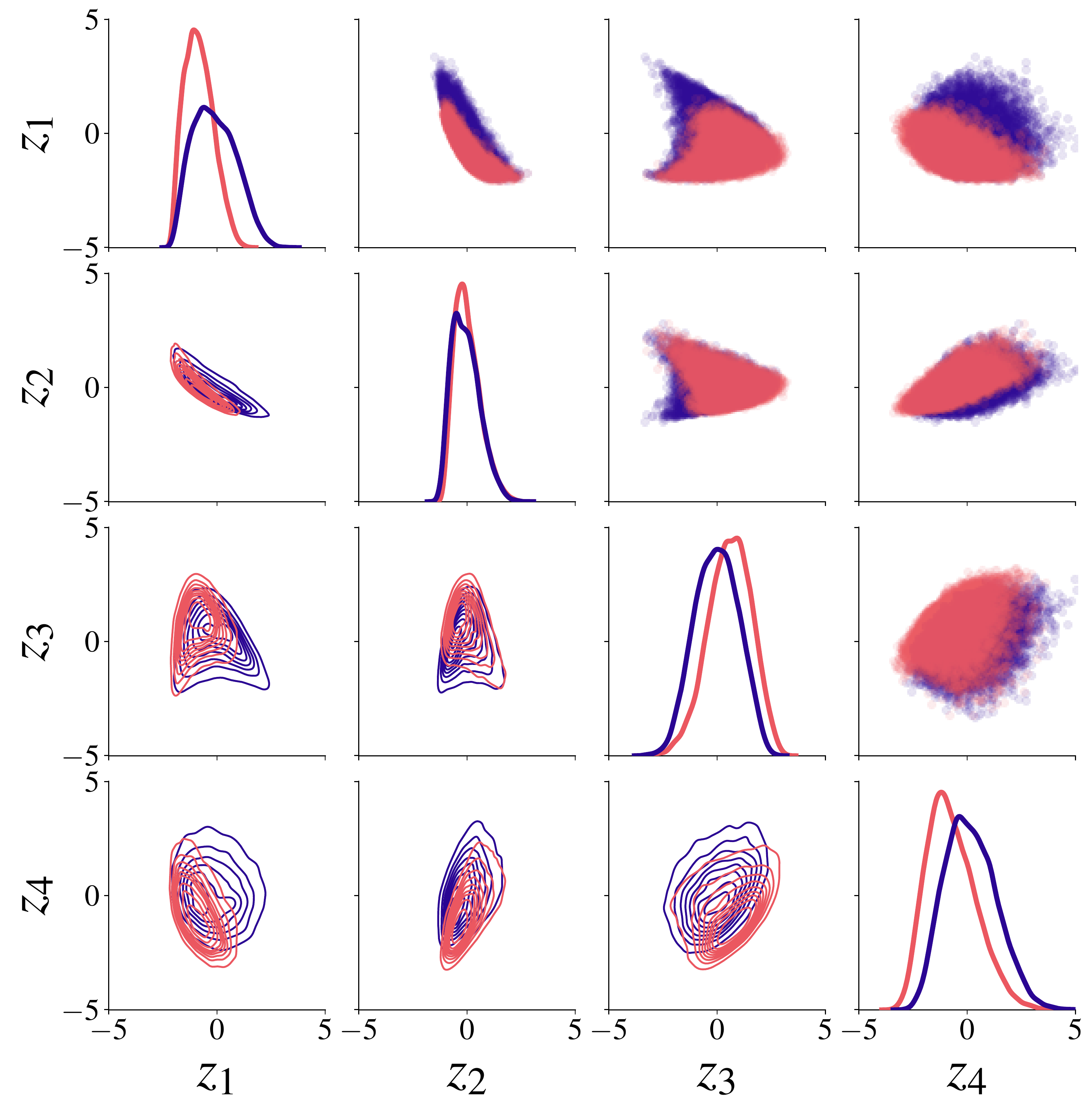}%
        \caption{Summary space for $\pi=0.75$.}%
        \label{fig:cs-experiment:pairplot}%
    \end{subfigure}\\
    \includegraphics[width=0.8\linewidth]{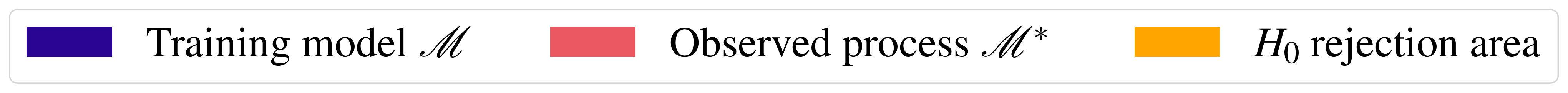}
    \caption{\textbf{Experiment \numberCS.} MMD increases for rising misspecification severity (\subref{fig:cs:mms-mmd}; mean and SD of 20 repetitions). The misspecification studied by \citet{ward_robust_2022} is detectable by our hypothesis test (\subref{fig:cs:power}) and via visual inspection (\subref{fig:cs-experiment:pairplot}).
    The critical MMD value (orange) in (\subref{fig:cs:mms-mmd}) corresponds to the critical MMD in (\subref{fig:cs:power}).
    }
\end{figure}
This experiment illustrates model misspecification detection in a marked point process model of cancer and stromal cells \cite{jones-todd_identifying_2019}.
We use the original implementation of \citet{ward_robust_2022} with hand-crafted summary statistics and showcase the applicability of our method in scenarios where good summary statistics are known.
The inference parameters are three Poisson rates $\lambda_c, \lambda_p, \lambda_d$, and the setup in \citet{ward_robust_2022} extracts four hand-crafted summary statistics from the 2D plane data: (1--2) number of cancer and stromal cells; (3--4) mean and maximum distance from stromal cells to the nearest cancer cell.
All implementation details are described in Section~\ref{app:cs} in the Appendix.

We achieve misspecification during inference by mimicking necrosis, which often occurs in core regions of tumors.
A Bernoulli distribution with parameter $\pi$ controls whether a cell is affected by necrosis or not.
Consequently, $\pi=0$ implies no necrosis (and thus no simulation gap), and $\pi=1$ entails that all cells are affected.
The experiments by \citet{ward_robust_2022} study a single misspecification, namely the case $\pi=0.75$ in our implementation.
In order to employ our proposed method for model misspecification detection, we add a small summary network $h_{\psib}:\mathbb{R}^4\to\mathbb{R}^4$ consisting of three hidden fully connected layers with $64$ units each.
This network $h_{\psib}$ merely transforms the hand-crafted summary statistics into a $4$-D unit Gaussian (cf.\ \autoref{alg:mms}).

\textit{Results.}
Our MMD misspecification score increases with increasingly severe model misspecification (i.e., increasing necrosis rate $\pi$), see \autoref{fig:cs:mms-mmd}.
What is more, for the single misspecification $\pi=0.75$ studied by \citet{ward_robust_2022}, we illustrate (i) the power of our proposed hypothesis test; and (ii) the summary space distribution for misspecified data.
The power ($1-\beta$) essentially equals $1$, as shown in \autoref{fig:cs:power}: The MMD sampling distributions under the training model ($H_0$) and under the observed data generating process ($\M^*$) are completely separated.
In fact, \autoref{fig:cs-experiment:pairplot} unveils that the induced model misspecification is directly visible in the outputs of the summary network $h_{\psib}$.

\subsection{Experiment \numberDDM: Drift Diffusion Model}
\label{sec:ddm-experiment}
\begin{figure}[t]
    \begin{subfigure}[t]{0.33\linewidth}
        \includegraphics[width=\linewidth]{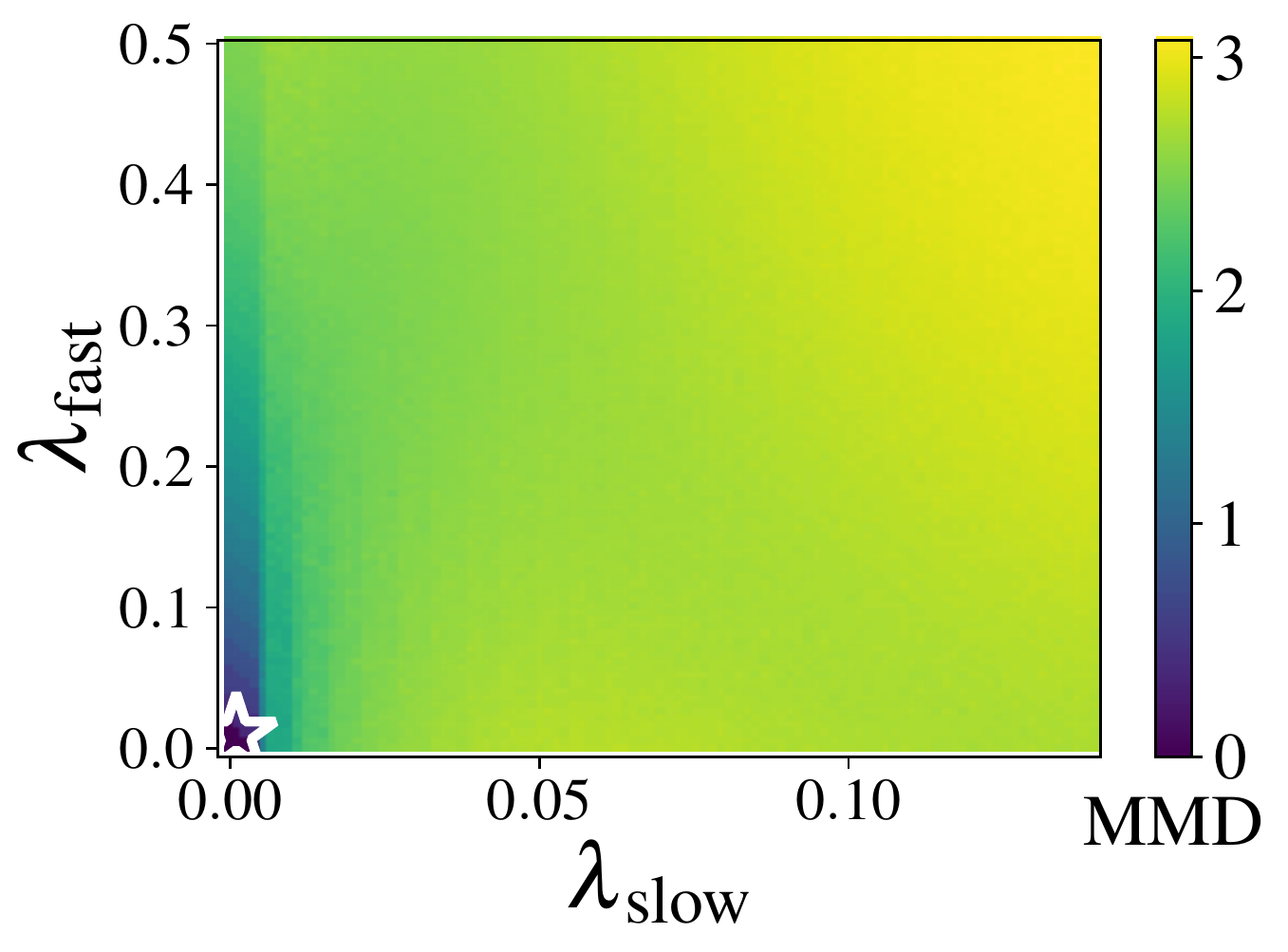}
        \caption{MMD as a function of misspecification severity.}
        \label{fig:exp:ddm-mmd-contamination}
    \end{subfigure}
    \hfill
    \begin{subfigure}[t]{0.62\linewidth}
        \includegraphics[width=\linewidth]{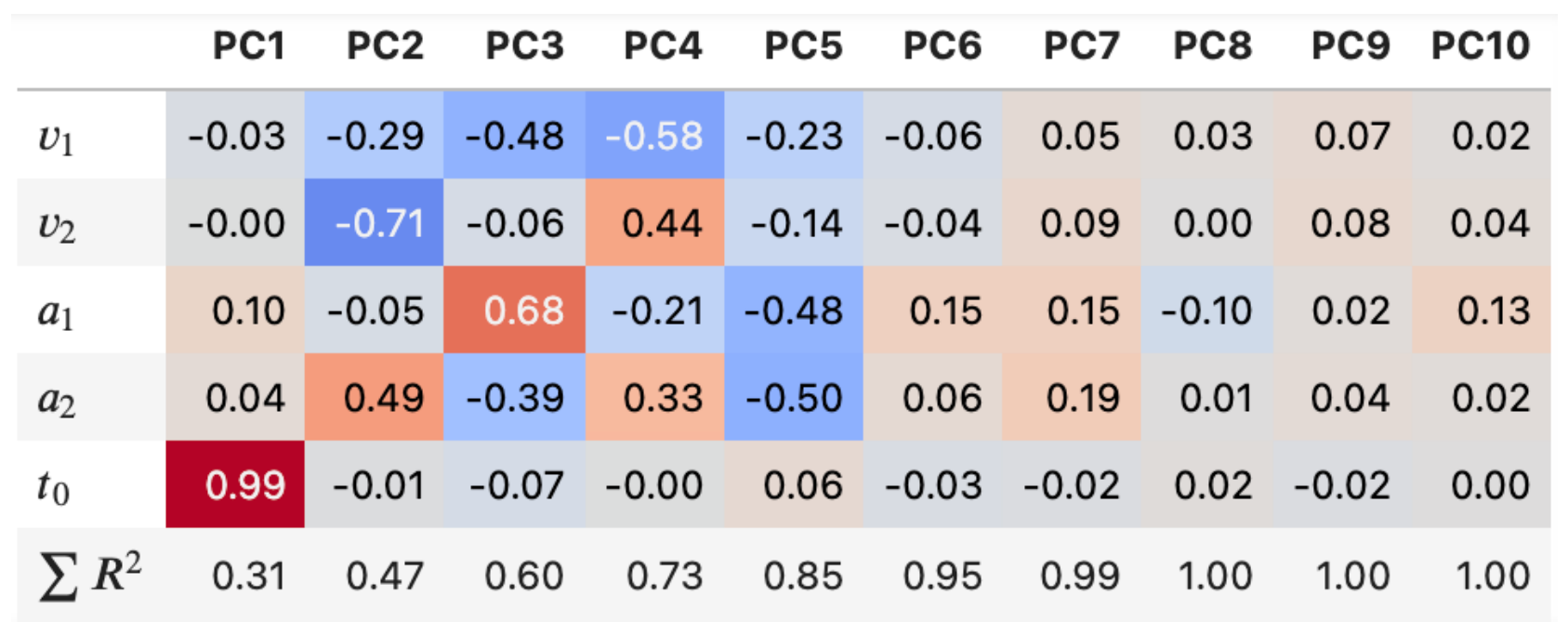}
        \caption{Correlation between parameters $\thetab$ and principal components (PCs) of learned summary statistics.}
        \label{fig:exp:ddm-pca}
    \end{subfigure}
    \caption{\textbf{Experiment \numberDDM.} Contamination of reaction times is detectable with our method (\subref{fig:exp:ddm-mmd-contamination}), and the principal components from the learned summary statistics coincide with the true parameters $\thetab$ (\subref{fig:exp:ddm-pca}).
    White star indicates the configuration for the well-specified model (i.e., training model $\M$ without contamination).}
    \label{fig:exp:ddm-figure}
\end{figure}
This experiment aims to (i) apply the new optimization objective to a complex model of decision making; (ii) illustrate the effect of dimensionality reduction (principal component analysis); (iii) tackle strategies to determine the required number of learned summary statistics in more complex applications; and (iv) compare the posterior estimation of BayesFlow under a misspecified model with the estimation provided by the Stan implementation of HMC-MCMC \cite{Carpenter2017, Stan2022} as a current gold-standard for Bayesian inference.

We focus on the drift diffusion model (DDM)---a cognitive model describing reaction times (RTs) in binary decision tasks \cite{Ratcliff2008} which is well amenable to amortized inference \cite{Radev2020bayesflow-cognition}.
The DDM assumes that perceptual information for a choice alternative accumulates continuously according to a Wiener diffusion process. 
Thus, the change in information $\diff x_j$ in experimental condition $j$ follows a random walk with drift and Gaussian noise: $\mathrm{d}x_j = v\mathrm{d}t + \xi \sqrt{\mathrm{d}t}$ with $\xi\sim\mathcal{N}(0, 1)$.
Our model implementation assumes five free parameters $\thetab = (v_1, v_2, a_1, a_2, t_0)$ which produce $2$-dimensional data stemming from two simulated conditions.
The summary network is a permutation-invariant network which reduces $i.\,i.\,d.\ $RT data sets to $S=10$ summary statistics each.
We realize a simulation gap by simulating typically observed contaminants: fast guesses (\eg, due to inattention), very slow responses (\eg, due to mind wandering), or a combination of the two.
The parameter $\lambda$ controls the fraction of the observed data which is contaminated  (see Section~\ref{sec:app:ddm} in the Appendix for implementation details).
For the comparison with Stan, we simulate $100$ uncontaminated DDM data sets and three scenarios (fast guesses, slow responses, fast and slow combined) with a fraction of $\lambda = 10\%$ contaminants.

\textit{Results.} During inference, our criterion reliably detects the induced misspecifications: Increasing fractions $\lambda$ of contaminants (fast, slow, and combined) manifest themselves in increasing MMD values (see \autoref{fig:exp:ddm-mmd-contamination}).
The results of applying PCA to the summary network outputs $\{\observed{\z}^{(n)}\}$ for the well-specified model (no contamination) are illustrated in \autoref{fig:exp:ddm-pca}.
We observe that the first five principal components exhibit a large overlap with the true model parameters $\thetab$ and jointly account for 85\% of the variance in the summary output.
Furthermore, the drift rates and decision thresholds within conditions are entangled (\ie, $v_1, a_1$ and $v_2, a_2$).
This entanglement mimics the strong posterior correlations observed between these two parameters.
In practical applications, dimensionality reduction might act as a guideline for determining the number of minimally sufficient summary statistics or parameter redundancies for a given problem.

\begin{table}[b]
    \centering
    \renewcommand{\arraystretch}{1.2}
    \begin{tabular}{l|c|c}
        \textbf{Model (Contamination)} & \textbf{Posterior error MMD} & \textbf{Summary space MMD}\\
        \hline
        $\M_{\ }$: Uncontaminated                  & $0.25\,[0.13, 0.56]$    & $0.45\,[0.42, 0.52]$ \\
        $\M_{1}$: Fast contaminants               & $2.66\,[1.44, 3.40]$    & $2.68\,[2.61, 2.74]$ \\
        $\M_{2}$: Slow contaminants               & $0.55\,[0.23, 1.01]$    & $1.18\,[1.13, 1.26]$ \\
        $\M_{3}$: Fast \& slow contaminants      & $1.90\,[0.83, 3.18]$    & $2.33\,[2.19, 2.43]$
    \end{tabular}
    \caption{
    \textbf{Experiment \numberDDM.} Posterior error of the approximate neural posterior (MMD to Stan samples; median and 95\% CI).
    The bootstrapped MMD values for the summary statistics of the $100$ investigated data sets and $1\,000$ samples from the uncontaminated model $\M$
    illustrate that posterior errors are mirrored by anomalies in the summary space and thus detectable.
    }
    \label{tab:ddm:stan-bf}
\end{table}

For the comparison with Stan, we juxtapose $4\,000$ samples from the approximate neural posterior with $4\,000$ samples obtained from the Stan sampler after ensuring MCMC convergence and sufficient sampling efficiency for each data set (see \autoref{fig:exp:ddm:stan-bf} for an illustration). 
Because Stan is currently considered state-of-the-art for likelihood-based Bayesian inference, we assume the Stan samples are representative of the \emph{correct posterior under the potentially misspecified model} (see Section~\ref{sec:posterior-inference-errors}).
In order to quantify the difference between the posterior samples from BayesFlow and Stan, we use the MMD criterion as well.
When no model misspecification is present, the posterior samples from BayesFlow and Stan match almost perfectly (see \autoref{fig:exp:ddm:stan-bf:clean}).
This means that our augmented optimization objective still enables correct posterior approximation under well-specified models.
In contrast, the results in \autoref{fig:exp:ddm:stan-bf:slow} and \autoref{tab:ddm:stan-bf} clearly indicate that the amortized posteriors deteriorate as a result of the induced misspecification.
Moreover, these results closely mirror the overall detectability of misspecification obtained by matching the summary representations of $1000$ data sets from the uncontaminated process with the representations of the $100$ data sets for each of the above scenarios via MMD (see \autoref{tab:ddm:stan-bf}).

\subsection{Experiment \numberCovid: Epidemiological Model for COVID-19}\label{sec:experiment-covid}
\begin{figure}[t]
\centering
    \begin{subfigure}[t]{.53\linewidth}
        \includegraphics[width=\linewidth]{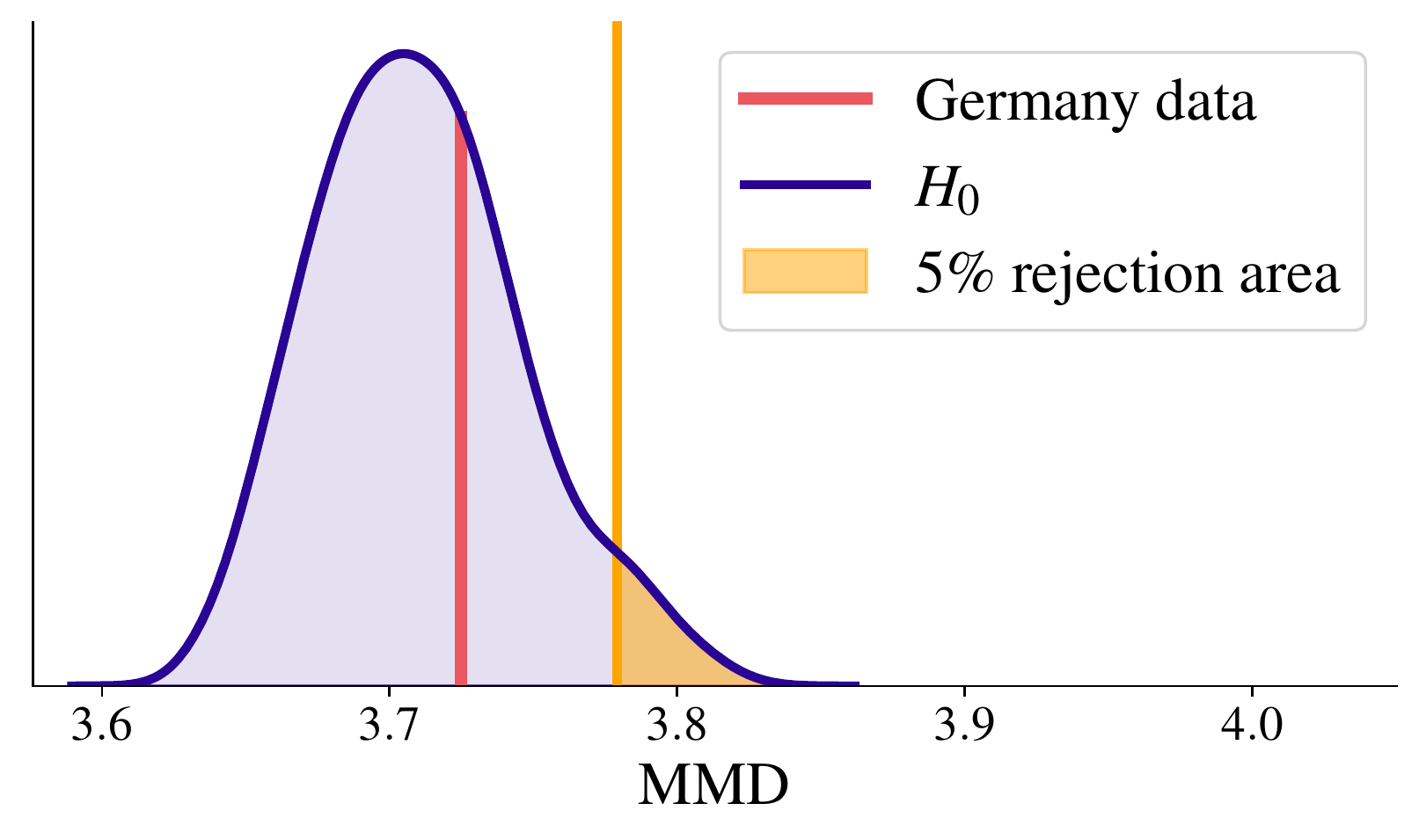}
        \caption{Representation of Germany's COVID-19 time series with respect to the MMD distribution under the null hypothesis $H_0: p^*(\x)=p(\x\given\mathcal{M})$.}
        \label{fig:exp:covid:mmd-real-data}
    \end{subfigure}
    \hfill
    \begin{subfigure}[t]{.44\linewidth}
        \includegraphics[width=\linewidth]{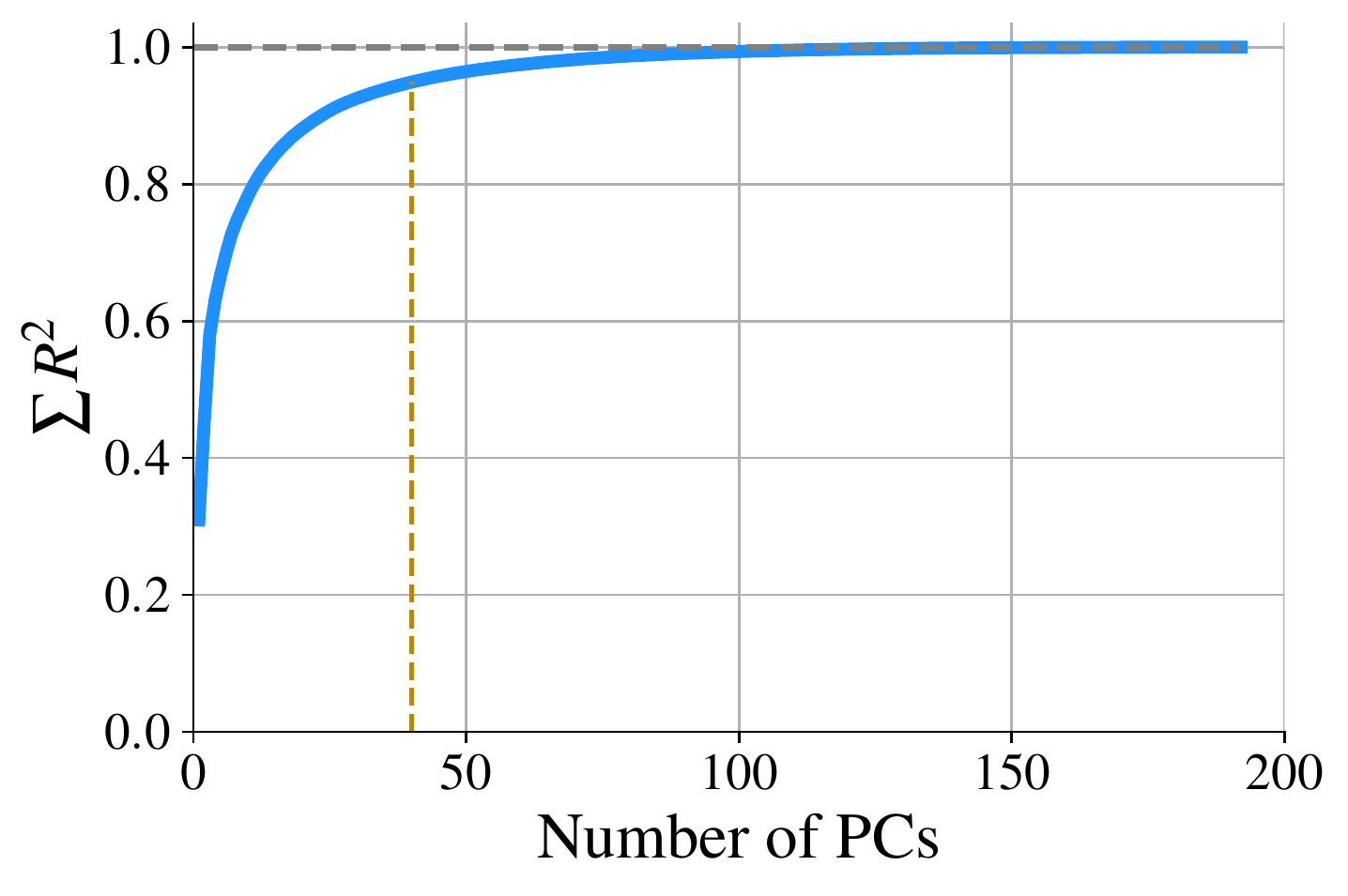}
        \caption{Cumulative explained variance ratio for the summary network output as a function of the number of principal components (PC).}
        \label{fig:exp:covid:PCA-exp-var}
    \end{subfigure}
    \caption{\textbf{Experiment \numberCovid}. The observed German COVID-19 time series is well-specified with respect to the training model $\M$ (\subref{fig:exp:covid:mmd-real-data}), and the first $40$ principal components jointly account for $95\%$ of the variance in the summary network outputs (\subref{fig:exp:covid:PCA-exp-var}; golden dashed line).}
    \label{fig:exp:COVID:real-data-and-PCA}
\end{figure}

Compartmental models in epidemiology are very popular for inferring relevant disease parameters, simulating possible outbreak scenarios, and projecting future outcomes \cite{covid_germany}.
Given the abundance of such models and their increasing complexity, the importance of detecting simulation gaps for trustworthy inference is two-fold. 
First, since substantial conclusions are based on the posterior distributions of model parameters, it is important that these distributions are formally correct even when models do not capture all relevant real-world factors. 
Second, given the dynamic aspect of these models, it is important to detect if an initially well-specified model becomes misspecified at a later time, so the model and its predictions can be amended.

\begin{figure}[t]
    \centering
    \begin{tabular}{cccc}
    & {\Large$N=1$} & {\Large$N=2$} & {\Large$N=5$}\\
    \rotatebox[origin=c]{90}{\raisebox{0.1cm}{\colsquare{observedcolor}}\Large$\M_1$} & 
    \raisebox{-0.48\height}{\includegraphics[width=.28\linewidth]{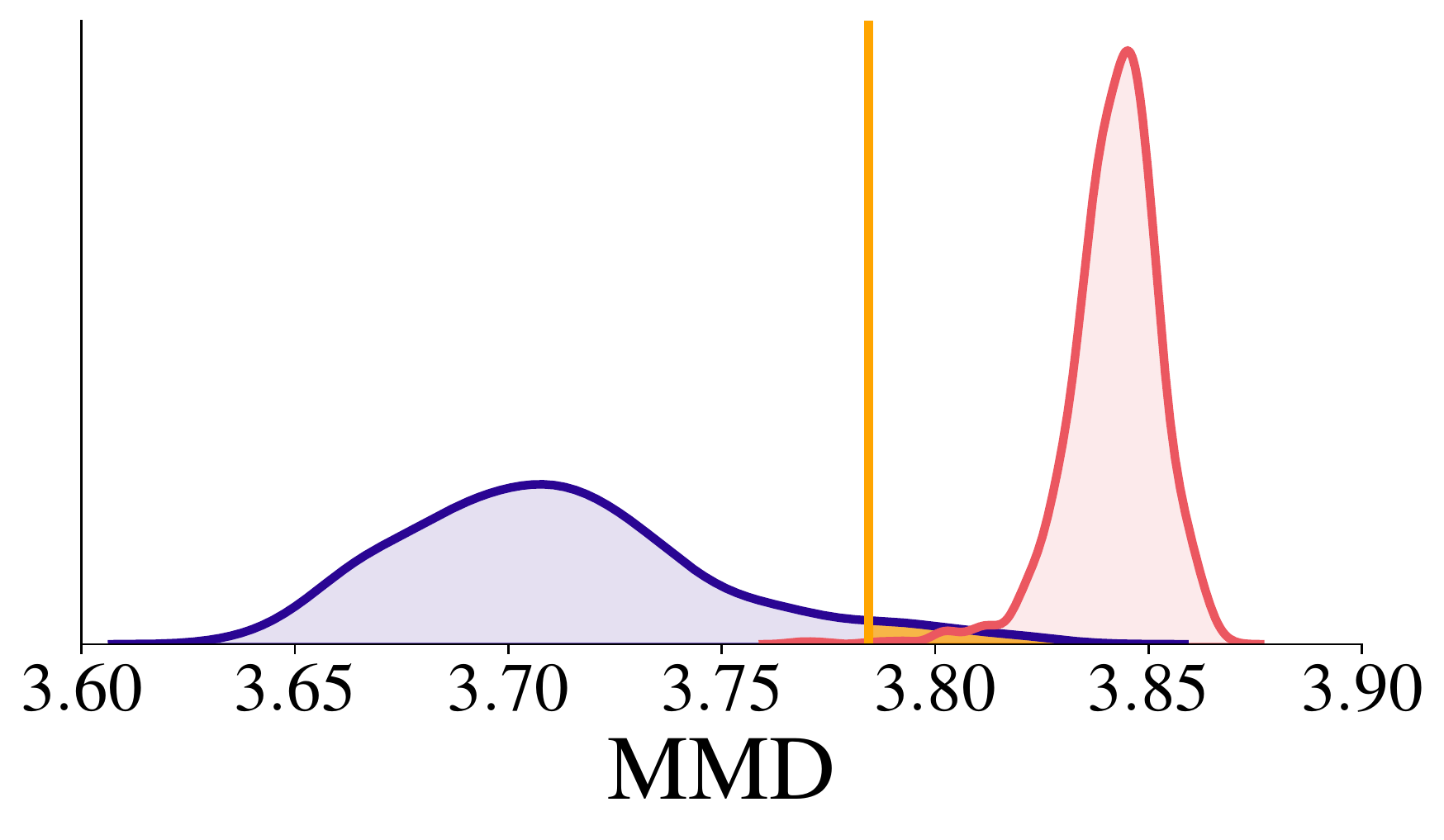}} &
    \raisebox{-0.48\height}{\includegraphics[width=.28\linewidth]{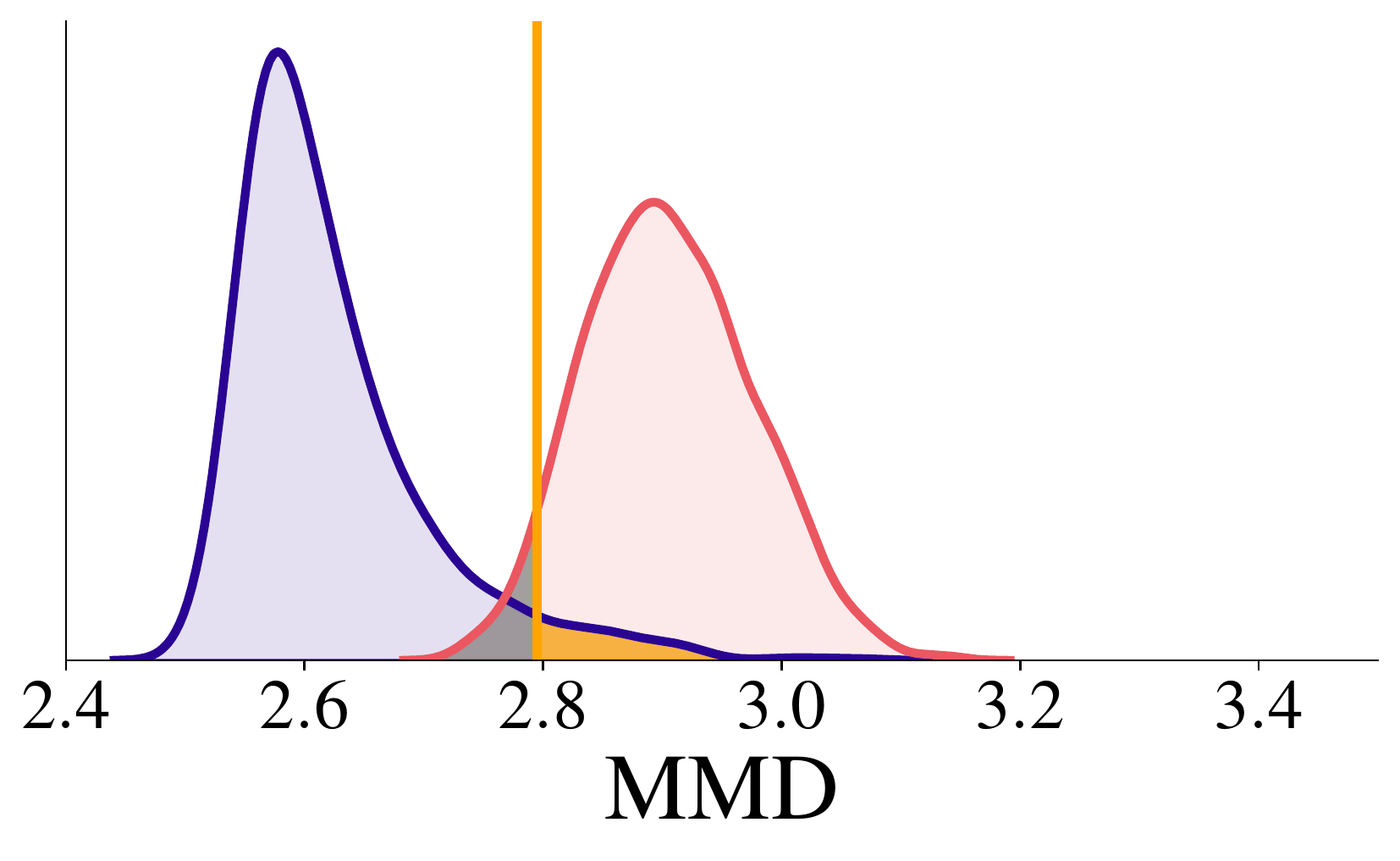}} &
    \raisebox{-0.48\height}{\includegraphics[width=.28\linewidth]{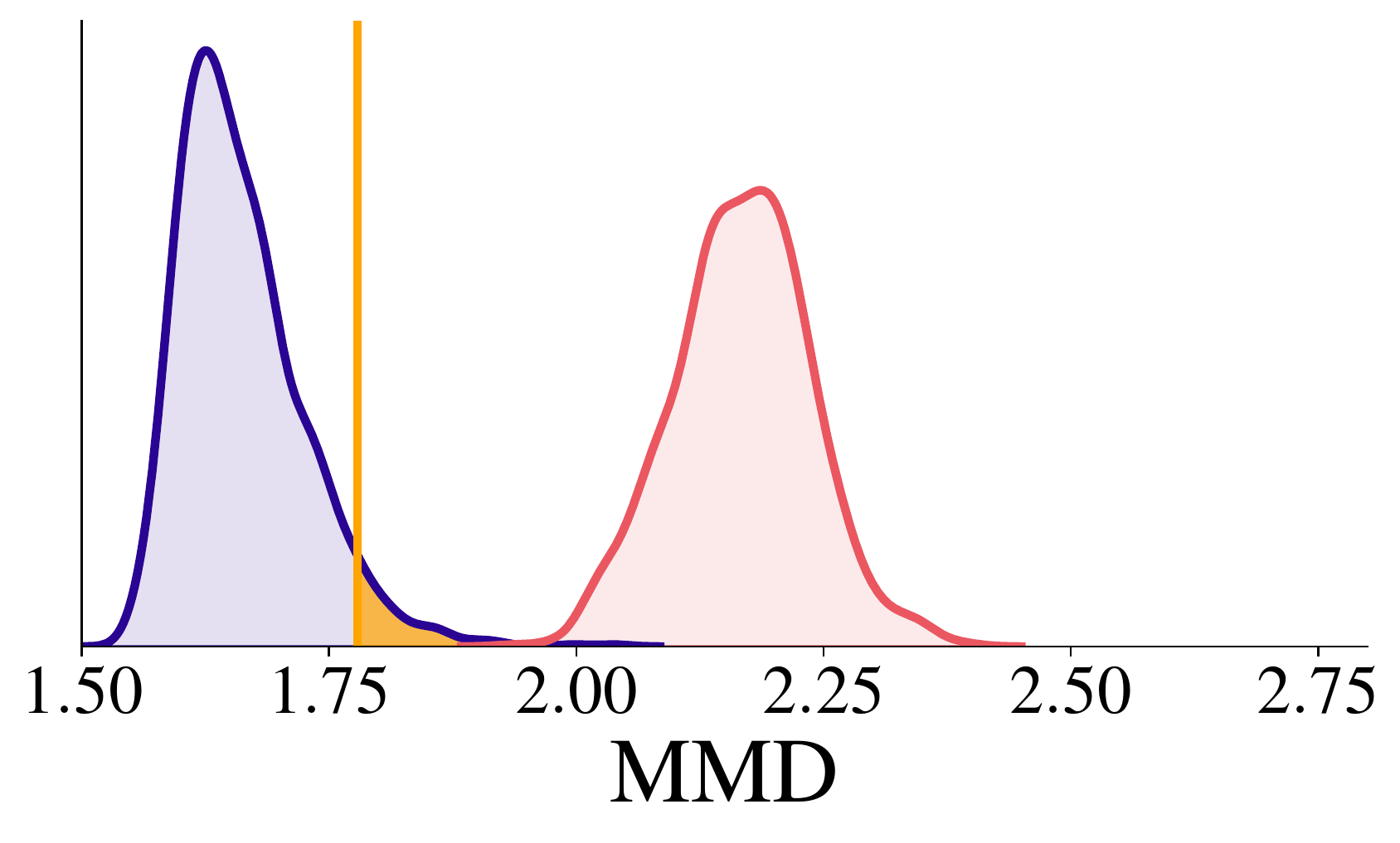}}\\
    \rotatebox[origin=c]{90}{\raisebox{0.1cm}{\colsquare{observedcolor}}\Large$\M_2$} & 
    \raisebox{-0.48\height}{\includegraphics[width=.28\linewidth]{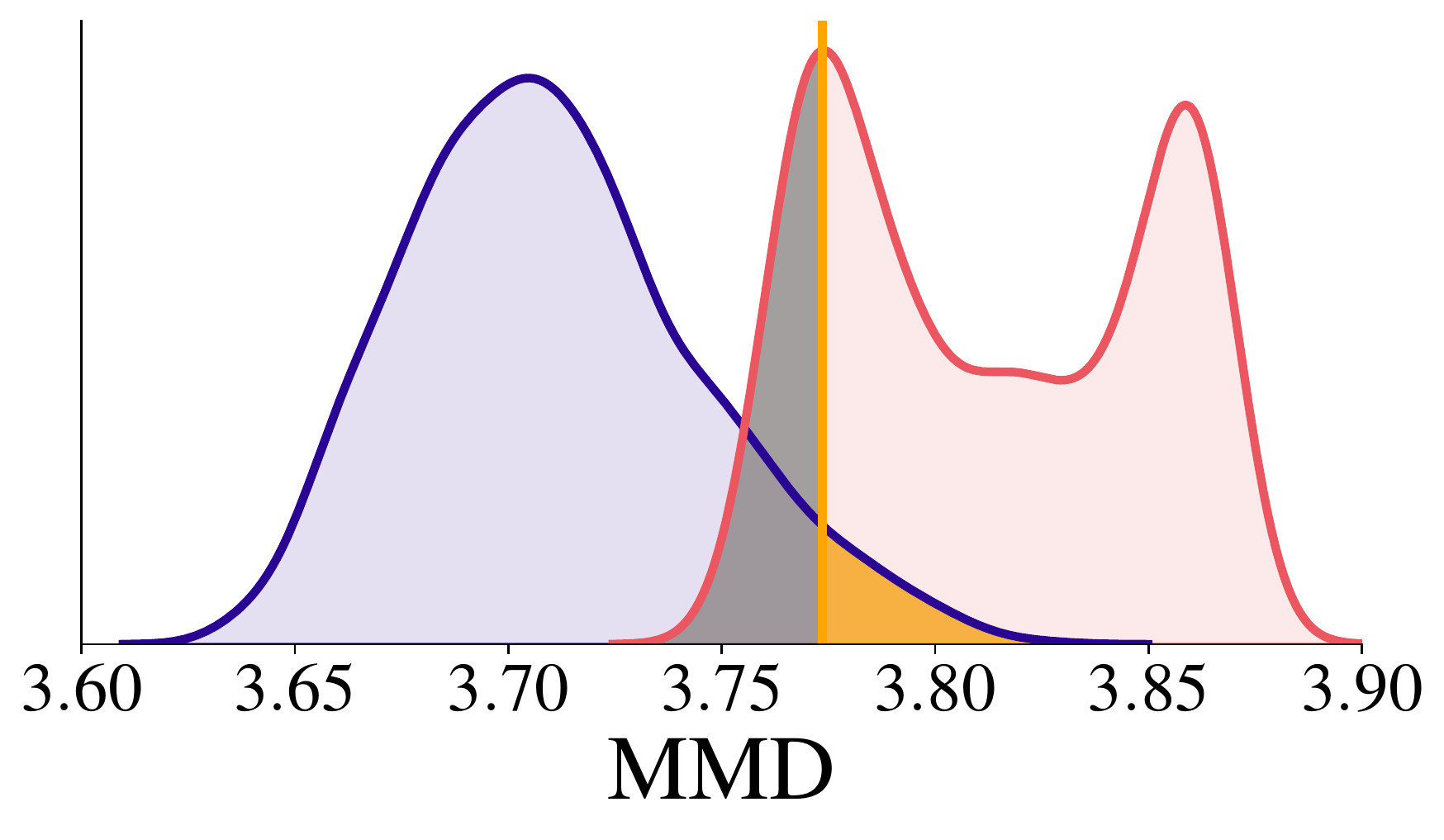}} &
    \raisebox{-0.48\height}{\includegraphics[width=.28\linewidth]{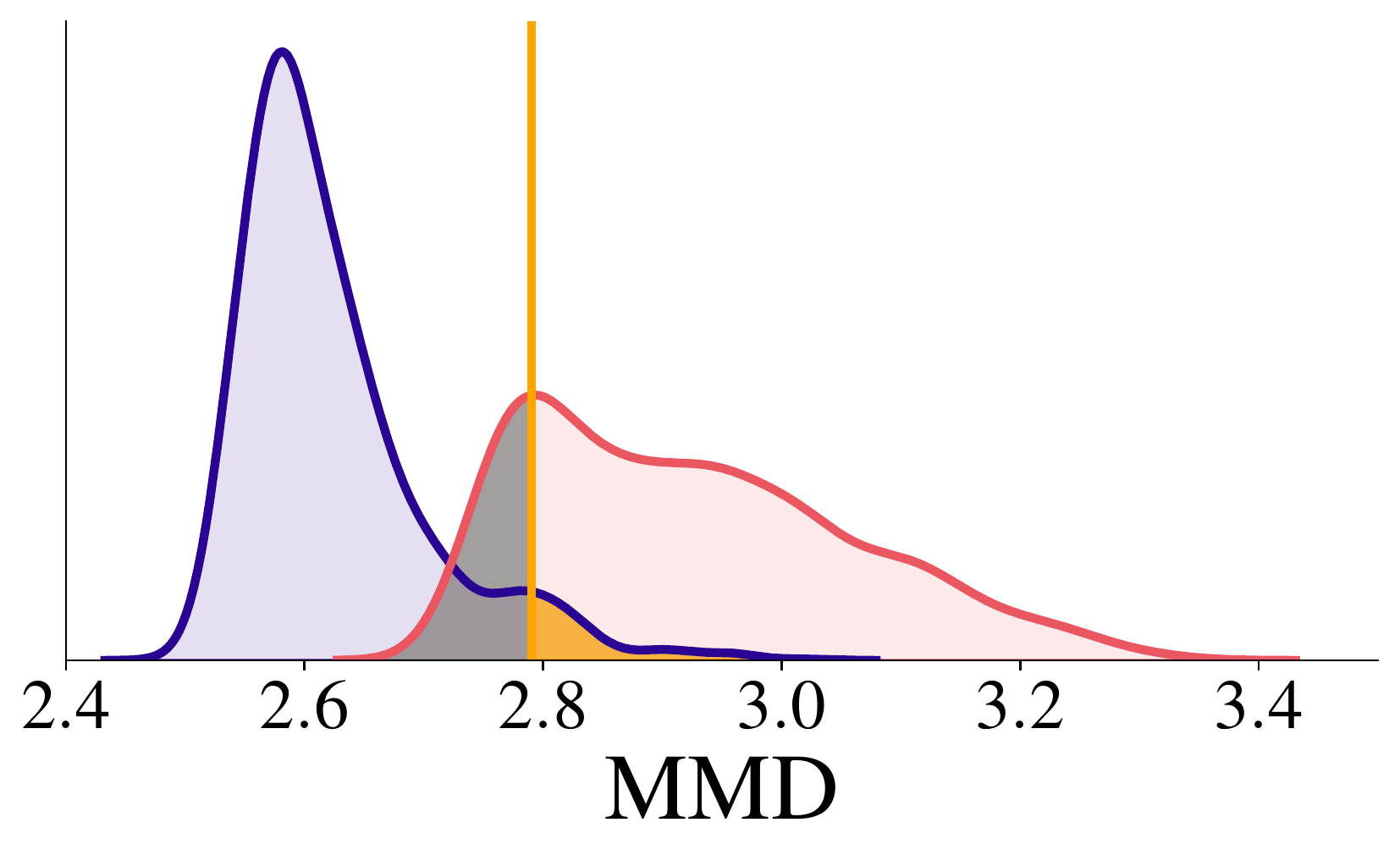}} &
    \raisebox{-0.48\height}{\includegraphics[width=.28\linewidth]{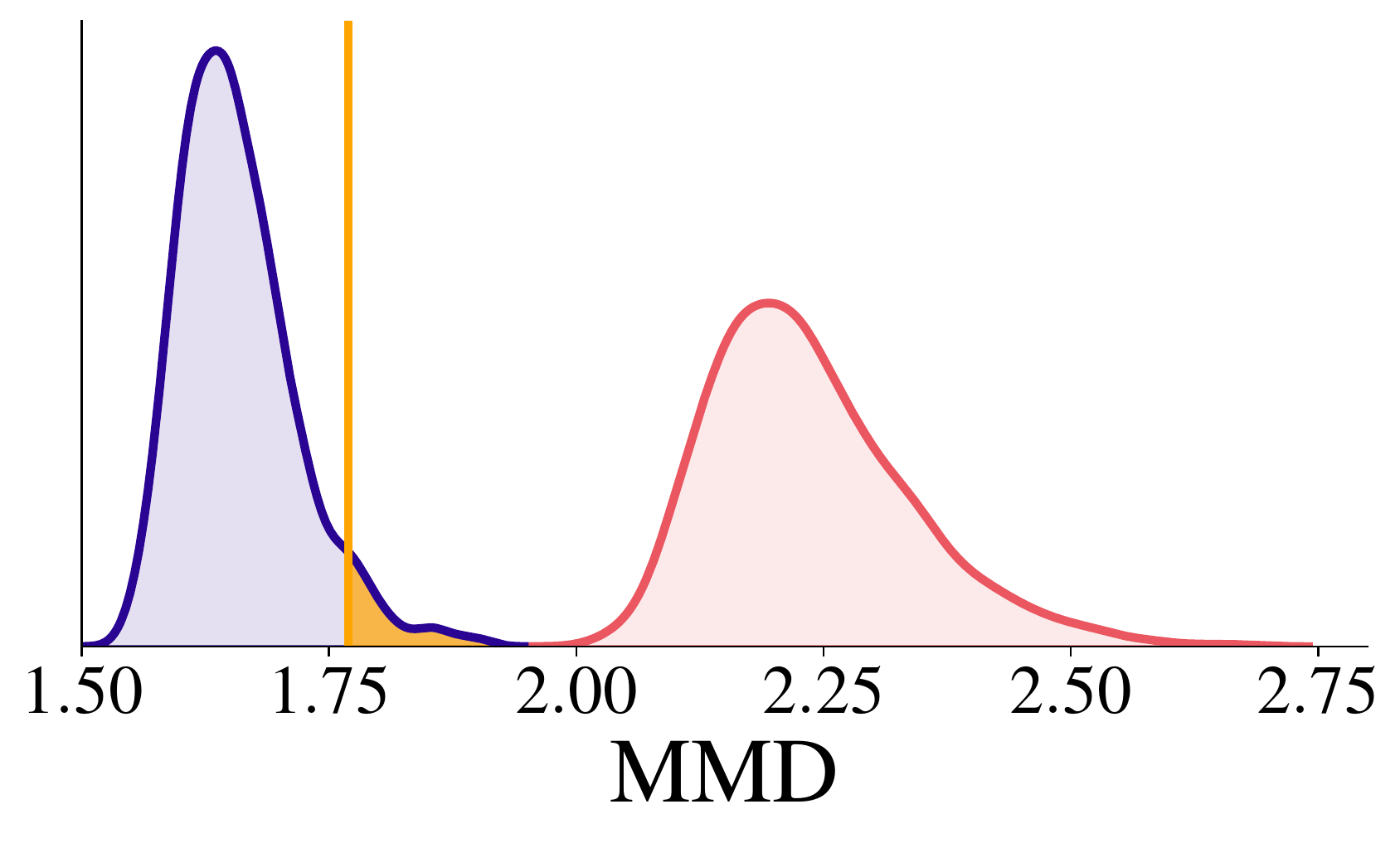}}\\
    \rotatebox[origin=c]{90}{\raisebox{0.1cm}{\colsquare{observedcolor}}\Large$\M_3$} & 
    \raisebox{-0.48\height}{\includegraphics[width=.28\linewidth]{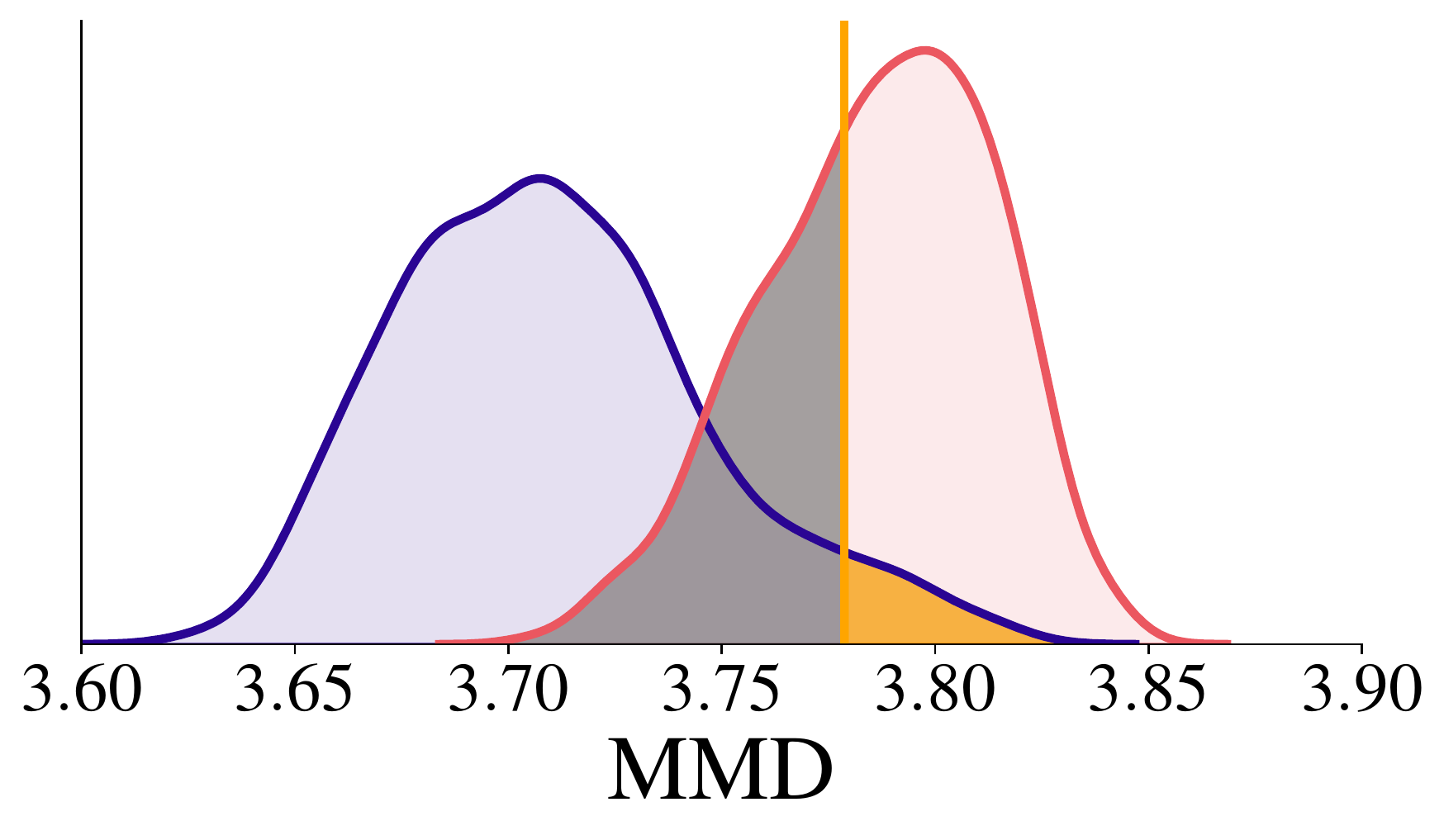}} &
    \raisebox{-0.48\height}{\includegraphics[width=.28\linewidth]{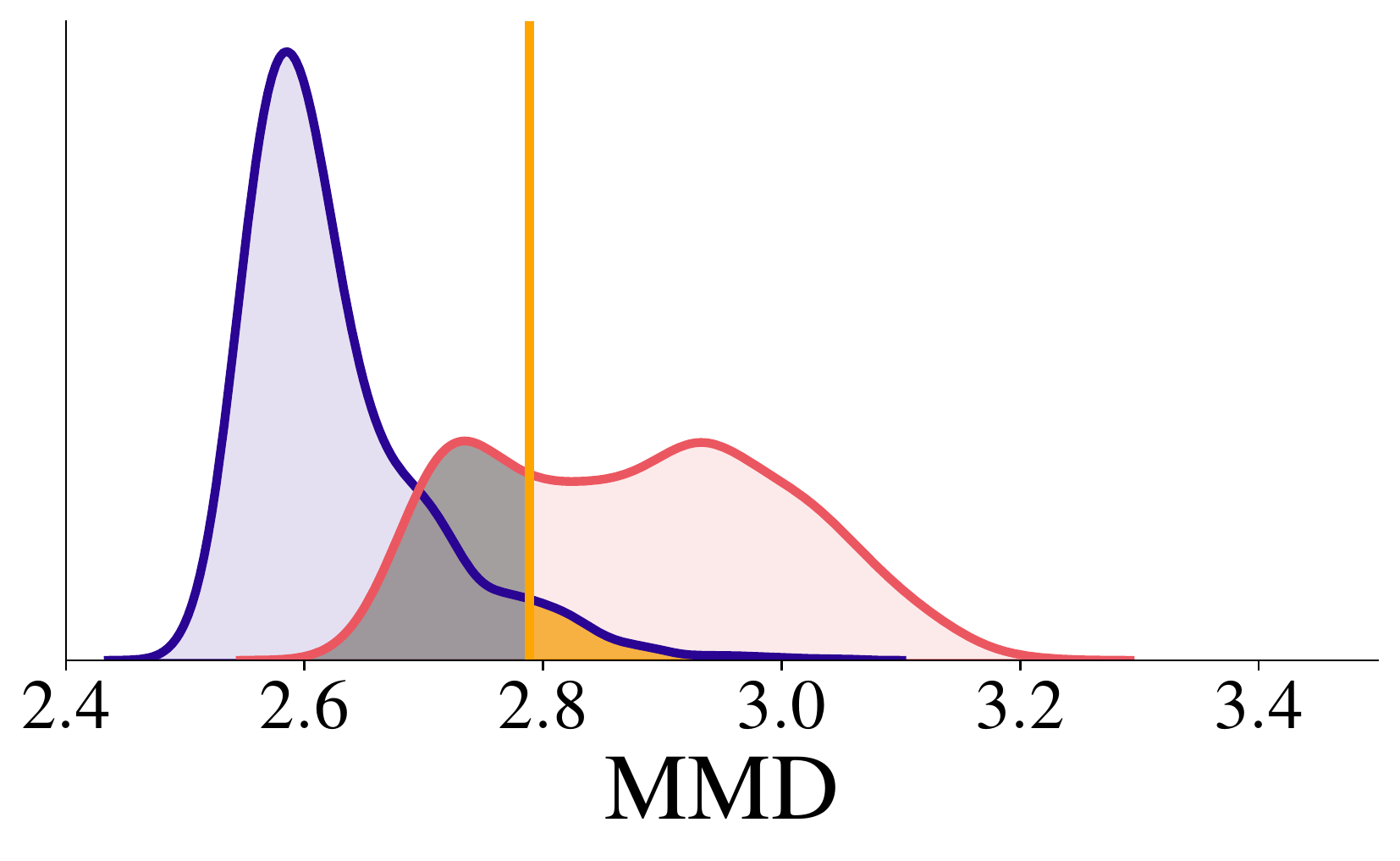}} &
    \raisebox{-0.48\height}{\includegraphics[width=.28\linewidth]{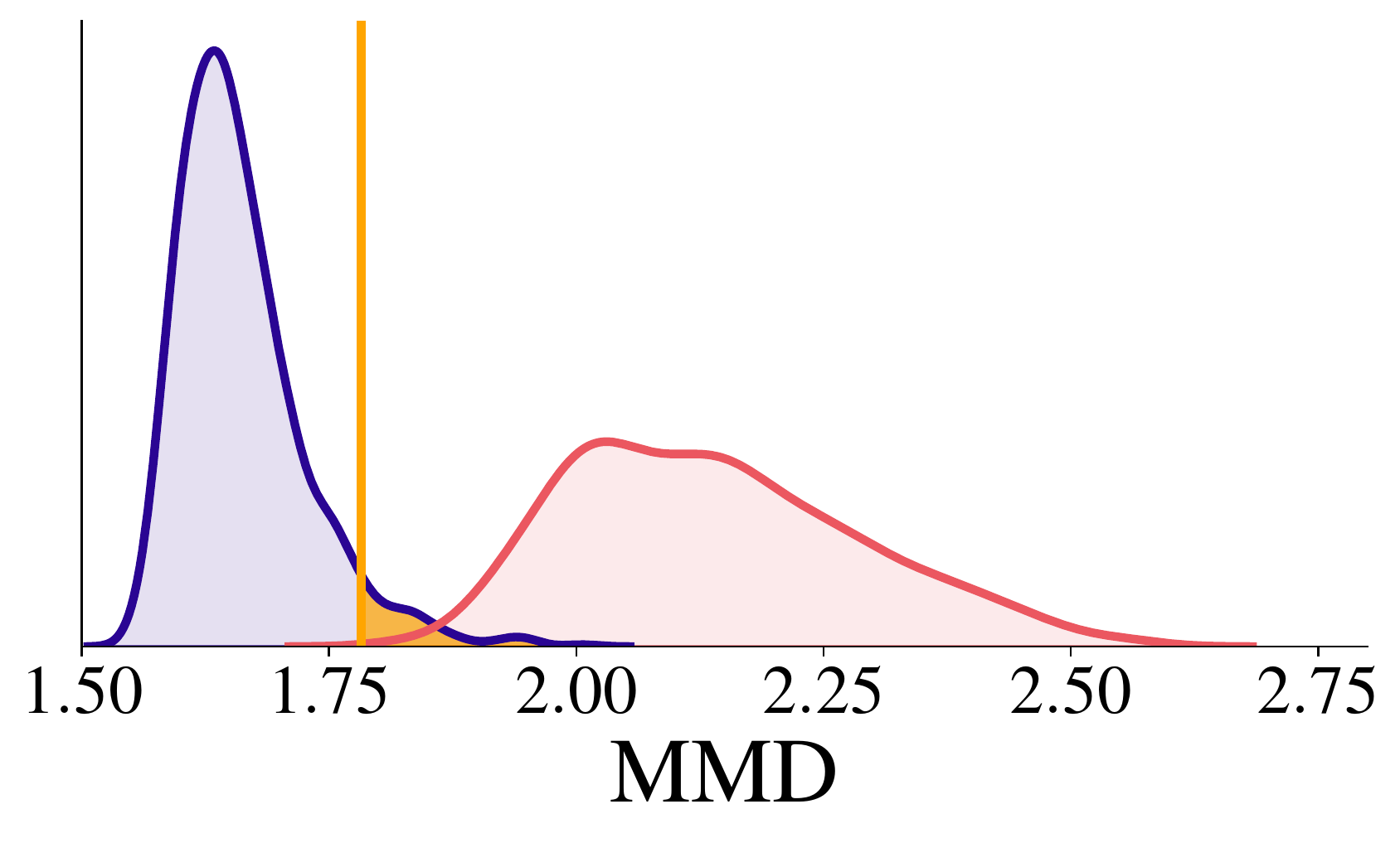}}\\
    \end{tabular}\\
    \includegraphics[width=1.0\linewidth]{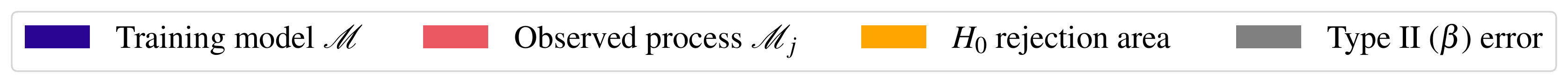}
    \caption{\textbf{Experiment \numberCovid.} Detailed illustration of the power analysis. \autoref{tab:covid19-models-mmd} contains the test power ($1-\beta$) for each scenario, and as few as $N=5$ observed data sets suffice to achieve a negligible type II error $\beta\approx 0$.}
    \label{fig:app:covid:power}
\end{figure}

As a final real-world example, we thus focus on a high-dimensional compartmental model representing the early months of the COVID-19 pandemic in Germany \cite{outbreak}.
Here, we investigate the utility of our distribution matching method to detect simulation gaps in a much more realistic and non-trivial extension of the SIR settings in \citet{lueckmann_benchmarking_2021} and  \citet{ward_robust_2022} with substantially increased complexity.\footnote{The SIR experiment from \citet{ward_robust_2022} induces misspecification through delayed reporting. 
This is modeled via $\M_2$ in our experiment.
The additional models $\M_1, \M_3$ in our experiment are extensions to the existing literature on model misspecification.}
Moreover, we perform inference on real COVID-19 data from Germany and use our new method to test whether the model used in \citet{outbreak} is misspecified, possibly undermining the trustworthiness of political conclusions that are based on the inferred posteriors.
To achieve this, we train a BayesFlow setup identical to \citet{outbreak} but using our new optimization objective (Eq.~\ref{eq:bf_kl_mmd}) to encourage a structured summary space.
We then simulate $1000$ time series from the training model $\mathcal{M}$ and $1000$ time series from three misspecified models: (i) a model $\mathcal{M}_1$ without an intervention sub-model; (ii) a model $\mathcal{M}_2$ without an observation sub-model; (iii) a model $\mathcal{M}_3$ without a latent ``carrier'' compartment \cite{covid_germany}.
\begin{table*}[b]
    \centering
    \renewcommand{\arraystretch}{1.2}
    \begin{tabular}{c|ccc|ccc}
        \multirow{2}{*}{\bfseries Model}& 
        \multicolumn{3}{c|}{\bfseries Bootstrap MMD} & 
        \multicolumn{3}{c}{ \bfseries Power ($1-\beta$)} \\
        & $N=1$  & $N=2$ & $N=5$ & $N=1$ & $N=2$ & $N=5$\\
        \hline
        $\mathcal{M}_{\ }$ & $3.70\,[3.65, 3.79]$ & $2.61\,[2.54, 2.91]$ & $1.66\,[1.59, 1.84]$ & --- & --- & ---\\
        $\mathcal{M}_1$ & $3.76\,[3.72, 3.80]$ & $2.86\,[2.62, 3.16]$ & $2.11\,[1.82, 2.50]$ & $.998$ & $.958$ & $\approx1.0$ \\
        $\mathcal{M}_2$ & $3.80\,[3.73, 3.83]$ & $2.81\,[2.65, 3.00]$ & $2.01\,[1.82, 2.19]$ & $.789$ & $.804$ & $\approx1.0$ \\
        $\mathcal{M}_3$ & $3.78\,[3.74, 3.83]$ & $2.81\,[2.68, 3.11]$ & $2.07\,[1.92, 2.41]$ & $.631$ & $.690$ & $\approx1.0$ \\
    \end{tabular}
    \caption{
    \textbf{Experiment \numberCovid.} Results for different variations of the COVID-19 compartmental model.
    We report the median and 95\% CI of 100 bootstrap samples. 
    }
    \label{tab:covid19-models-mmd}
\end{table*}
\textit{Results.}
\autoref{tab:covid19-models-mmd} shows the MMD between the summary representation of $N=1,2,5$ bootstrapped time series from each model and the summary representation of the $1000$ time series from model $\mathcal{M}$ (see Section~\ref{sec:app:bootstrap} for bootstrapping details).
We also calculate the power ($1-\beta$) of our hypothesis test for each misspecified model under the sampling distribution estimated from $1\,000$ samples of the $1\,000$ time series from $\mathcal{M}$ at a type I error probability of $\alpha=.05$.
We observe that the power of the test rapidly increases with more data sets and the Type II error probability ($\beta$) is essentially zero for as few as $N=5$ time series (see \autoref{fig:app:covid:power}).

As a next step, we pass the reported COVID-19 data between 1 March and 21 April 2020 \citep[data from][under CC BY 4.0 license]{dong_interactive_2020} through the summary network and compute the critical MMD value for a sampling-based hypothesis test with an $\alpha$ level of $.05$ (see \autoref{fig:exp:covid:mmd-real-data}). 
The MMD of the Germany data is well below the critical MMD value (it essentially lies in the bulk of the distribution), leading to the conclusion that the assumed training model $\M$ is well-specified for this time period.
Finally, we perform linear dimensionality reduction (PCA) on the summary space and find that the first 40 principal components jointly explain $95\%$ of the variance in the $192$-dimensional summary space outputs (see \autoref{fig:exp:covid:PCA-exp-var}). 
Thus, a $40$-dimensional learned summary vector might provide a good approximation of the true (unknown) minimally sufficient summary statistics and render inference less fragile in the face of potential misspecifications (cf.\ \textbf{Experiment \numberGaussianMeans}).

\section{Conclusions}\label{sec:discussion}

With this work, we approached a fundamental problem in amortized simulation-based Bayesian inference, namely, capturing posterior errors due to model misspecification.
We argued that misspecified models might cause so-called \emph{simulation gaps}, resulting in deviations between simulations during training time and actual observed data at test time.
We further showed that simulation gaps can be detrimental for the performance and faithfulness of simulation-based inference relying on neural networks.
We proposed to increase the networks' awareness of posterior errors by compressing simulations into a structured latent summary space induced by a modified optimization objective in an unsupervised fashion. 
We then applied the maximum mean discrepancy (MMD) estimator, equipped with a sampling-based hypothesis test, as a criterion to spotlight discrepancies between model-implied and actually observed distributions in summary space.
While we focused on the application to SNPE-C and BayesFlow, the proposed method can be easily integrated into other frameworks with learned or hand-crafted summary statistics as well.

The proposed method can be extended and modified in multiple ways.
While we optimized the summary space to follow a spherical Gaussian distribution, our method is likewise applicable to other latent distributions.
For example, a heavy-tailed summary distribution (\eg, $\alpha$-stable distributions with tunable tail parameters) would reduce the impact of outliers in latent space.
Furthermore, the sampling-based hypothesis test in summary space can be readily replaced with a more sophisticated statistical testing regime for out-of-distribution detection \cite{bergamin_model-agnostic_2022}.
The idea of \citet{frazier_model_2020} to detect simulation gaps by posterior discrepancies in an ensemble of differently configured approximator instances is interesting, and future work might incorporate this into simulation-based neural inference.
In addition, considerations on information geometry and non-Euclidean spaces might guide future research into building more flexible latent spaces and distance metrics \cite{info_geo}.
Our software implementations are openly available at \href{https://www.github.com/marvinschmitt/ModelMisspecificationBF}{\textrm{github.com/marvinschmitt/ModelMisspecificationBF}} and can be seamlessly integrated into an end-to-end workflow for amortized simulation-based inference.

\acks{This research was supported by the Cyber Valley Research Fund (grant number: CyVy-RF-2021-16), the Deutsche Forschungsgemeinschaft (DFG, German Research Foundation) under Germany’s Excellence Strategy -- EXC-2075 - 390740016 (the Stuttgart Cluster of Excellence SimTech) and EXC-2181 - 390900948 (the Heidelberg Cluster of Excellence STRUCTURES), the Deutsche Forschungsgemeinschaft (DFG, German Research Foundation; grant number GRK 2277 ``Statistical Modeling in Psychology''), and the Informatics for Life initiative funded by the Klaus Tschira Foundation.
The authors gratefully acknowledge the support and funding.
}


\appendix
\clearpage
\begin{center}
    \Large\bfseries Appendix
\end{center}

\counterwithin{figure}{section}

\section{Implementation Details}

\subsection{5D Gaussian with Covariance (Experiment \numberGaussianMeansCov)}
\label{sec:app:mvn-full}
The normal-inverse-Wishart prior $\NIW(\mub, \Sigmab\given\mub_0, \lambda_0, \Psib_0, \nu_0)$ implies a hierarchical prior. 
Suppose the covariance matrix $\Sigmab$ has an inverse Wishart distribution $\mathcal{W}^{-1}(\Sigmab\given\Psib_0, \nu_0)$  and the mean vector $\mub$ has a multivariate normal distribution $\mathcal{N}(\mub\given\mub_0, \frac{1}{\lambda_0}\Sigmab)$, then the tuple $(\mub, \Sigmab)$ has a normal-inverse-Wishart distribution $\NIW(\mub, \Sigmab\given\mub_0, \lambda_0, \Psib_0, \nu_0)$.
Finally, the likelihood is Gaussian:$\x_k\sim\mathcal{N}(\mub, \Sigmab)\;\text{for}\; k=1,\ldots, K$.

For a multivariate Gaussian with unknown mean and unknown covariance matrix, the analytic joint posterior $p(\mub_p, \Sigmab_p\given\x)$ follows a normal-inverse Wishart distribution again:
\begin{equation}
\begin{aligned}
    (\mub_p, \Sigmab_p \given \x) & \sim\text{N-}\mathcal{W}^{-1}(\mub_p, \Sigmab_p\given\mub_K, \lambda_K, \Psib_K, \nu_K)\quad\text{with}\\
    \mub_K & = \dfrac{\lambda_0\mub_0 + K\bar{\x}}{\lambda_0+K},\quad 
    \lambda_K =\lambda_0+K,\quad
    \nu_K  = \nu_0+K,\\
    \Psib_K & = \Psib_0+\sum\limits_{k=1}^K(\x_k-\bar{\x})(\x_k-\bar{\x})^T+\dfrac{\lambda_0 K}{\lambda_0 + K}(\bar{\x}-\mub_0)(\bar{\x}-\mub_0)^T
\end{aligned}
\end{equation}
The marginal posteriors for $\mub_p$ and $\Sigmab_p$ then follow as \cite{Murphy2007}:
\begin{equation}\label{eq:mvn-full-cov:analytic-posterior:marginal}
\begin{aligned}
    \mub_p & \sim t_{\nu_K-D-1}\Big(\mub_p\,\Big|\,\mub_K, \frac{\Psib_K^{-1}}{\lambda_K(\nu_K-D+1)}\Big)\\
    \Sigmab_p & \sim \mathcal{W}^{-1}(\Sigmab_p\given\Psib_K, \nu_K)
\end{aligned}
\end{equation}

The model $\mathcal{M}$ used for training the networks as well as the types of induced model misspecifications are outlined in \autoref{tab:app:mvn-full-MMS}.
\begin{table*}[b]
    \centering
    \scriptsize
    \begin{tabular}{l|ll}
        \bfseries Model &\bfseries Prior &\bfseries Likelihood\\
        \hline
        $\mathcal{M}$ (No MMS) &
        $\big(\mub, \Sigmab\big)\sim\NIW(\mub_0=\0, \lambda_0=5, \Psib=\mathbb{I},\nu=10)$ &
        $\x_k\sim\mathcal{N}(\mub, \Sigmab)$
        \\
        $\mathcal{M}_P$ (Prior) &
        $\big(\mub, \Sigmab\big)\sim\NIW(\mub_0\neq\0, \lambda_0=5, \Psib=\tau_0\mathbb{I},\nu=10)$&
        $\x_k\sim\mathcal{N}(\mub, \Sigmab)$\\
        $\mathcal{M}_S$ (Simulator) &
        $\big(\mub, \Sigmab\big)\sim\NIW(\mub_0=\0, \lambda_0=5, \Psib=\mathbb{I},\nu=10)$&
        $\x_k\sim t_{\text{df}}(\mub, \Sigmab),\quad\text{df}\in\mathbb{N}_{>0}$
        \\
        $\mathcal{M}_N$ (Noise)&
        $\big(\mub, \Sigmab\big)\sim\NIW(\mub_0=\0, \lambda_0=5, \Psib=\mathbb{I},\nu=10)$& 
        $\x_k\sim\lambda\cdot\mathrm{Beta}(2, 5)+(1-\lambda)\cdot\mathcal{N}(\mub, \Sigmab)
        $
    \end{tabular}
        \caption{Investigated model misspecifications (MMS) for the $5$-dimensional Gaussian with fully estimated covariance matrix. 
    }
    \label{tab:app:mvn-full-MMS}
\end{table*}
In the evaluation, we compare the means of the approximate posterior samples with the first moment of the respective marginal analytic posterior from Eq.~\ref{eq:mvn-full-cov:analytic-posterior:marginal}. 
We evaluate correlation matrices with standard deviations on the diagonal. 
For the $t$ distributed posterior mean and inverse-Wishart distributed posterior covariance, we obtain \cite{Mardia1979}:
\begin{equation}
        \mathbb{E}(\mub_p) =\mub_K,\quad
        \mathbb{E}(\Sigmab_p) =\frac{\Psib_K}{\nu_K-D-1}
\end{equation}

\subsection{Cancer and Stromal Cells (Experiment \numberCS)}\label{app:cs}

As implemented by \citet{ward_robust_2022}, the CS model simulates the development of cancer and stromal cells.
The respective cell counts (total $N_c$, unobserved parents $N_p$, daughters for each parent $N_d^{(i)}$) are stochastically defined through Poisson distributions
\begin{equation}
        N_c \sim \mathrm{Poisson}(\lambda_c),\quad N_p \sim \mathrm{Poisson}(\lambda_c),\quad N_d^{(i)} \sim \mathrm{Poisson}(\lambda_c),\quad i=1,\ldots, N_p.
\end{equation}

The distance based metrics for the hand-crafted summary statistics 3--4 are empirically approximated from 50 stromal cells to avoid computing the full distance matrix \cite{ward_robust_2022}.
The prior distributions are specified as 
\begin{equation}
        \lambda_c \sim \mathrm{Gamma}(25, 0.03),\quad
        \lambda_d \sim \mathrm{Gamma}(5, 0.5),\quad
        \lambda_p \sim \mathrm{Gamma}(45, 3)
\end{equation}
where $\mathrm{Gamma}(a, b)$ denotes the Gamma distribution with location $a$ and rate $b$.
To induce model misspecification through necrosis, a Bernoulli variable $w_i\sim\mathrm{Bernoulli}(\pi)$ is sampled, removing cancer cells within a specified radius around the parent cell \cite{ward_robust_2022}.
Thus, the Bernoulli parameter $\pi$ controls the degree of misspecification, ranging from no misspecification ($\pi=0$; no necrosis) to maximal misspecification with respect to that parameter ($\pi=1$; necrosis of all susceptible cells).

\subsection{DDM (Experiment \numberDDM)}\label{sec:app:ddm}
The starting point of the evidence accumulation process is unbiased, $x_{t=0}=\frac{a}{2}$.
During training, all parameters are drawn from Gamma prior distributions:
\begin{equation}
    v_1, v_2, a_1, a_2, t_0 \sim\Gamma(5, 0.5).
\end{equation}
We first generate uncontaminated data $\x^*$ from the well-specified generative model $\M$. 
Second, we randomly choose a fraction $\lambda\in[0,1]$ of the data $\x^*$.
Third, we replace this fraction with data-dependent contaminants $\xi$,
\begin{equation}
    \begin{aligned}
    \text{Fast guesses:}\quad\xi & \sim\mathcal{U}\big(0.1, Q_{10}(\x^*)\big)\\
    \text{Slow responses:}\quad\xi & \sim\mathcal{U}\big(Q_{75}(\x^*), 10\big),
    \end{aligned}
\end{equation}
where $Q_k(\x^*)$ denotes the $k^{\text{th}}$ percentile of $\x^*$.
The asymmetry in percentiles between fast and slow responses arises from the inherent positive skewness of reaction time distributions. 
The fixed upper limit of slow response contamination is motivated by the maximum number of iterations of the utilized diffusion model simulator. 
The contamination procedure is executed separately for each condition and response type. 
If an experiment features both fast and slow contamination, the fraction $\lambda$ is equally split between fast and slow contamination. 
The uncontaminated data set is generated once and acts as a baseline for all analyses of an experiment, resulting in a baseline MMD of 0 since $\x^*$ is unaltered if $\lambda=0$.

\section{Bootstrapping Procedure}
\label{sec:app:bootstrap}
In \textbf{Experiment \numberCovid}, we estimate a sampling distribution of the MMD between samples from the specified training model $\M$ with $\x\sim p(\x\given\M^*)$ and samples from the (opaque) observed model $\M^*$ with $\observed{x}\sim p^*(\x)$.
Since simulating time series from the compartmental models is time-consuming, we opt for bootstrapping \cite{Stine1989} on $M=1\,000$ pre-simulated time series $\{\x^{(m)}\}$ from $\mathcal{M}$ and $N=1\,000$ pre-simulated time series $\{\observed{\x}^{(n)}\}$ from $\M^*$.
In each bootstrapping iteration, we draw $M=1\,000$ samples with replacement from $\{\x^{(m)}\}$ as well as $N_B\in\{1,2,5\}$ samples (with replacement) from $\{\observed{\x}^{(n)}\}$ and calculate the MMD between the sets of bootstrap samples.

\section{2D Gaussian Means: Overcomplete Summary Statistics}

\begin{figure}[t]
    \centering
    \includegraphics[width=.46\linewidth]{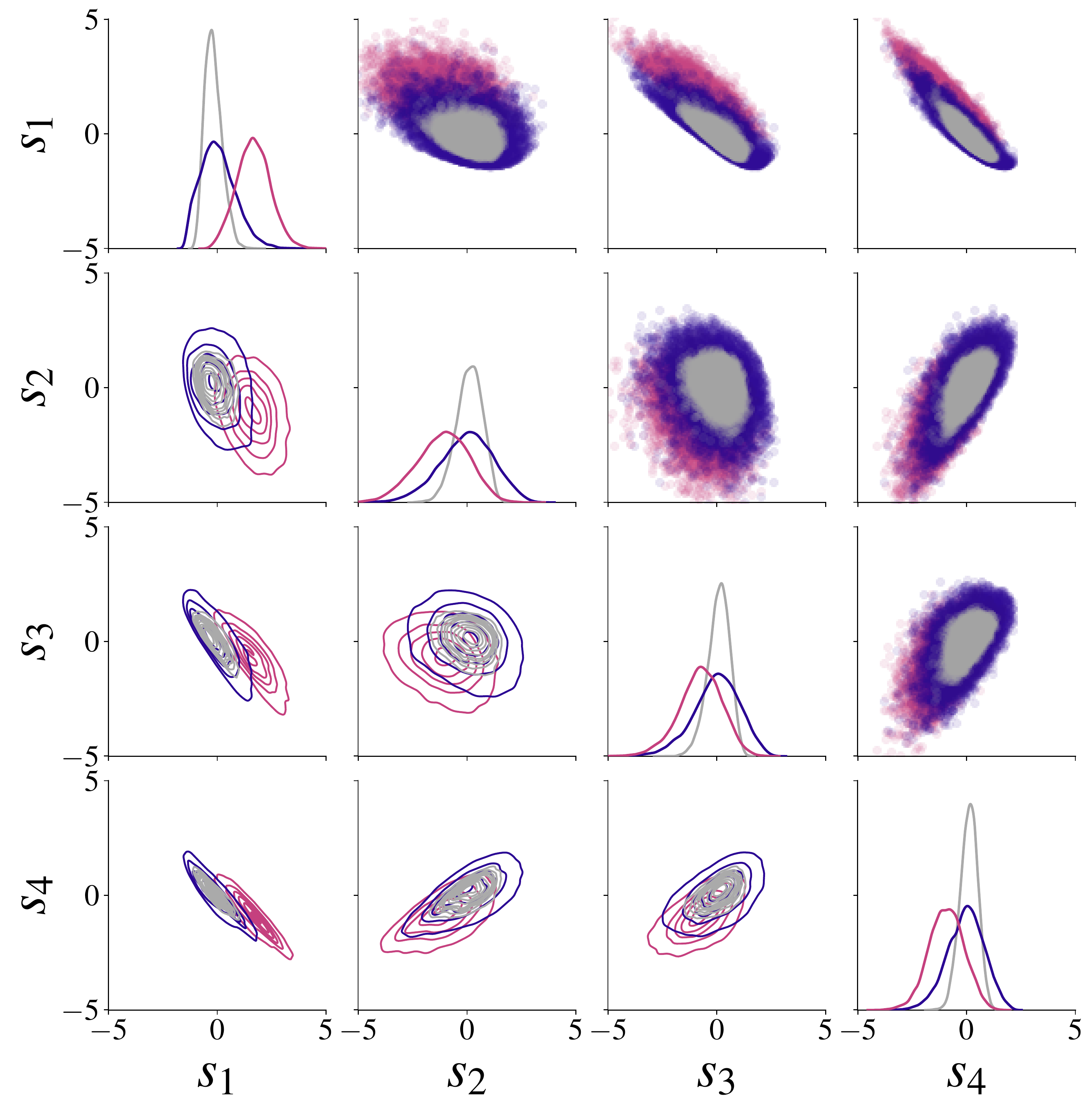}\\
    \includegraphics[width=0.7\linewidth]{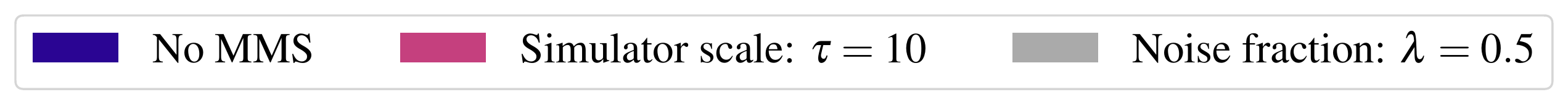}
    \caption{Pairplot of $10\,000$ latent summary space samples from the overcomplete summary network. Both noise (orange) and simulator (pink) misspecifications are distinguishable from the typical latent generative space (blue).}
    \label{fig:app:mvn-means:overcomplete:pairplot}
\end{figure}

\autoref{fig:app:mvn-means:overcomplete:pairplot} shows the latent summary space when overcomplete sufficient summary statistics ($S=4$) are used in \textbf{Experiment \numberGaussianMeans} to recover the means of a $2$-dimensional Gaussian.
Model misspecification with respect to both simulator and noise is detectable through anomalies in the latent summary space.
A network with $S=2$ summary statistics and otherwise equivalent architecture could not capture these types of model misspecification. 

\section{Replication of Experiment \numberGaussianMeans~with SNPE-C}
\label{app:snpe}

In the following, we show the results of repeating \textbf{Experiment \numberGaussianMeans} with SNPE-C \citep[APT; ][]{apt} instead of BayesFlow for posterior inference.
The results are largely equivalent to those obtained with the BayesFlow framework.

\begin{figure}[t]
    \centering
    \begin{subfigure}[t]{0.45\linewidth}
        \includegraphics[width=\linewidth]{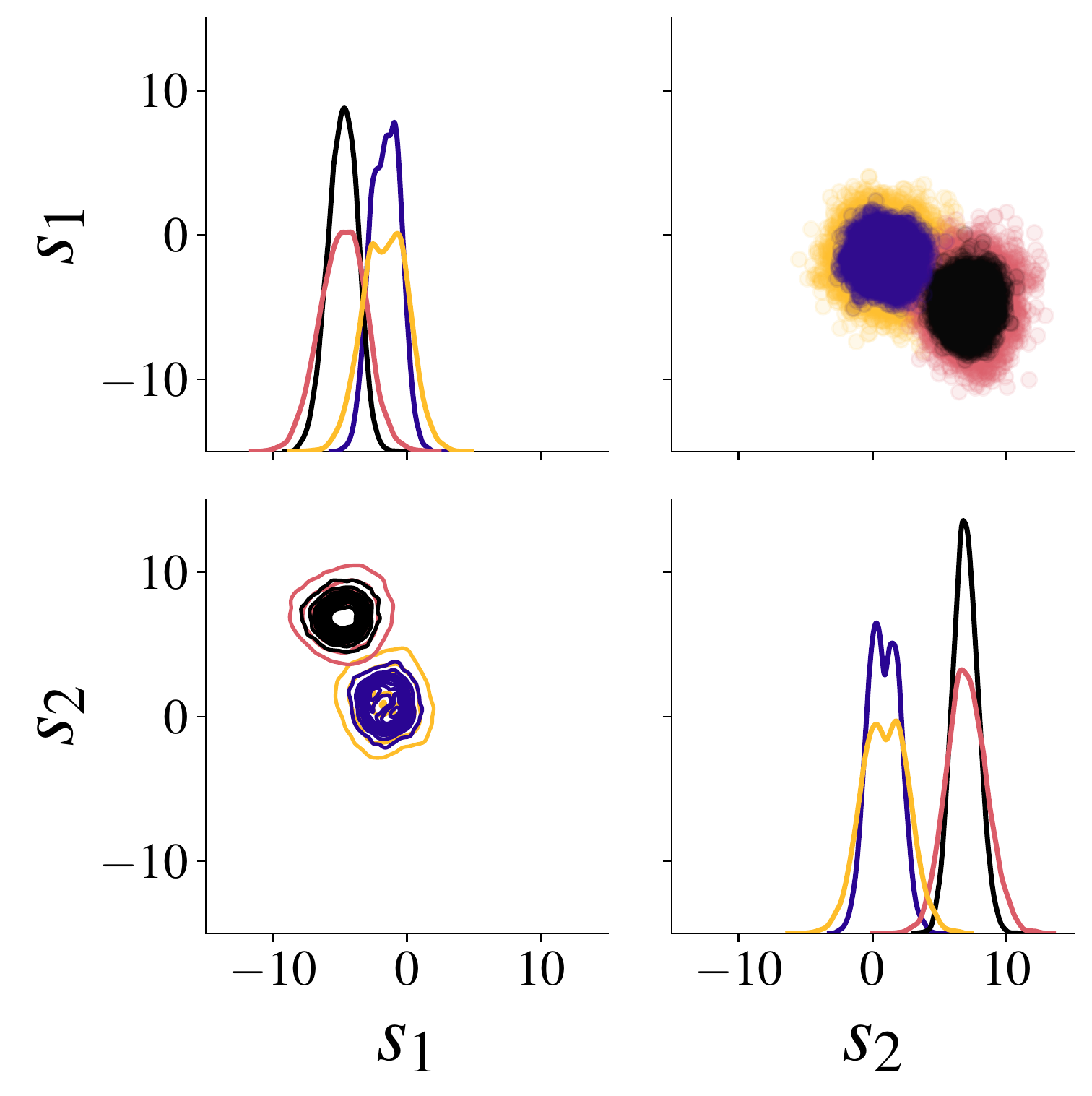}
    \end{subfigure}\hfill
    \begin{subfigure}[t]{0.45\linewidth}
        \includegraphics[width=\linewidth]{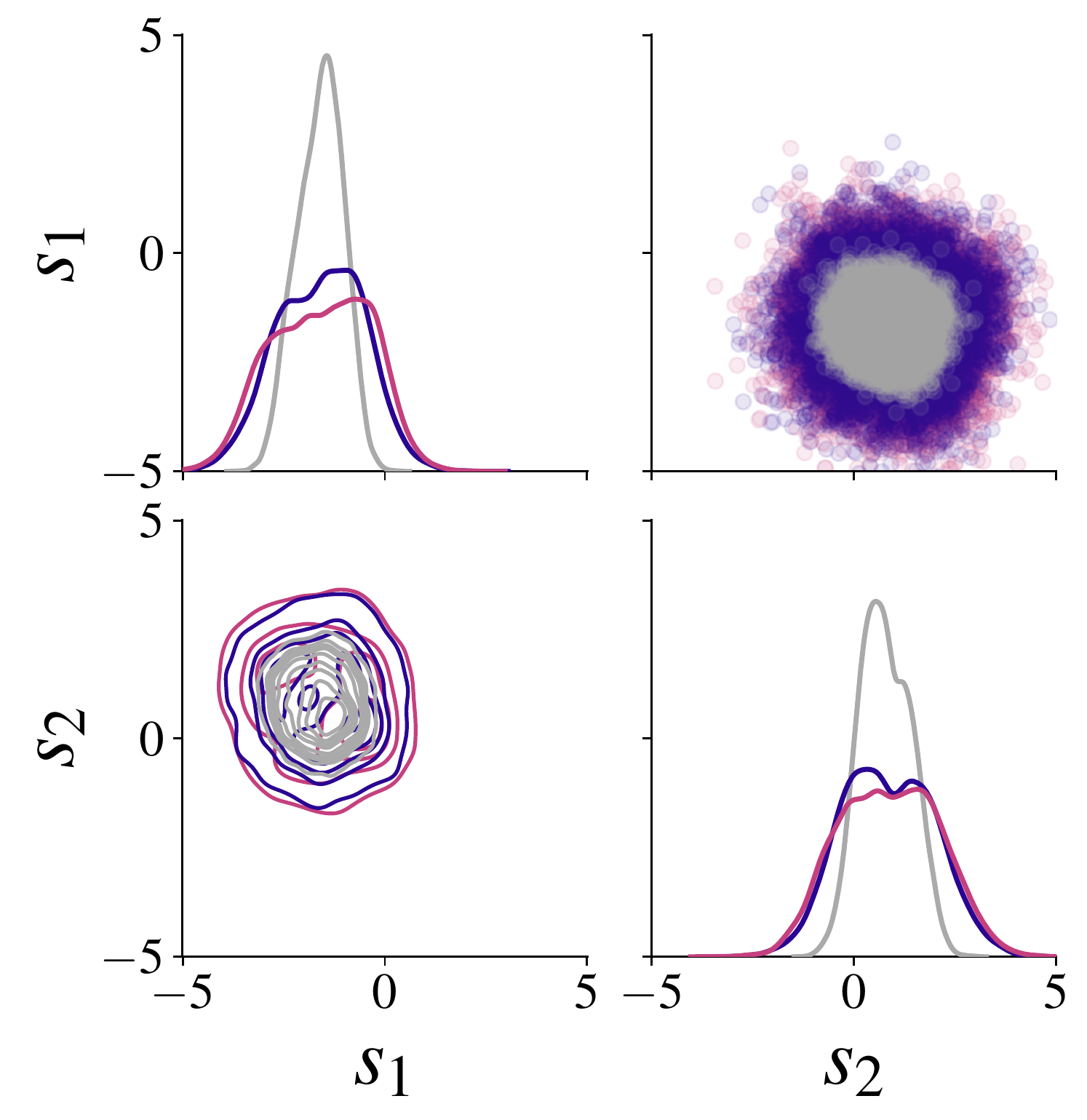}
    \end{subfigure}\\
    \includegraphics[width=0.95\linewidth]{plots/abf_mvn_means_sufficient_pairplot_MMD_legend_new.pdf}
    \caption{\textbf{Experiment \numberGaussianMeans, SNPE-C.} Summary space samples for the minimal sufficient summary network ($S=2$) from a well-specified model $\M$ (blue) and misspecified configurations. \textbf{Left:} Prior misspecification can be detected. \textbf{Right:} Simulator scale misspecification is indistinguishable from the validation summary statistics.}
\end{figure}

\begin{figure}[t]
    \centering
    \begin{subfigure}[c]{0.9\linewidth}%
    \setlength\tabcolsep{2pt}%
    \begin{tabular}{ccccc}
    && \multicolumn{2}{c}{\textbf{Model Misspecification}} & \\
        & &
        \textbf{Prior} ($\M_P$) &
        \textbf{Simulator} ($\M_S$) \textbf{\& noise} ($\M_N$)
        \\
        \multirow{2}{*}{\hspace*{-0.1cm}\rotatebox[origin=c]{90}{\textbf{Summary Network}}} &
        \rotatebox[origin=c]{90}{\textbf{minimal}} &
        \raisebox{-0.48\height}{\includegraphics[width=0.40\linewidth]{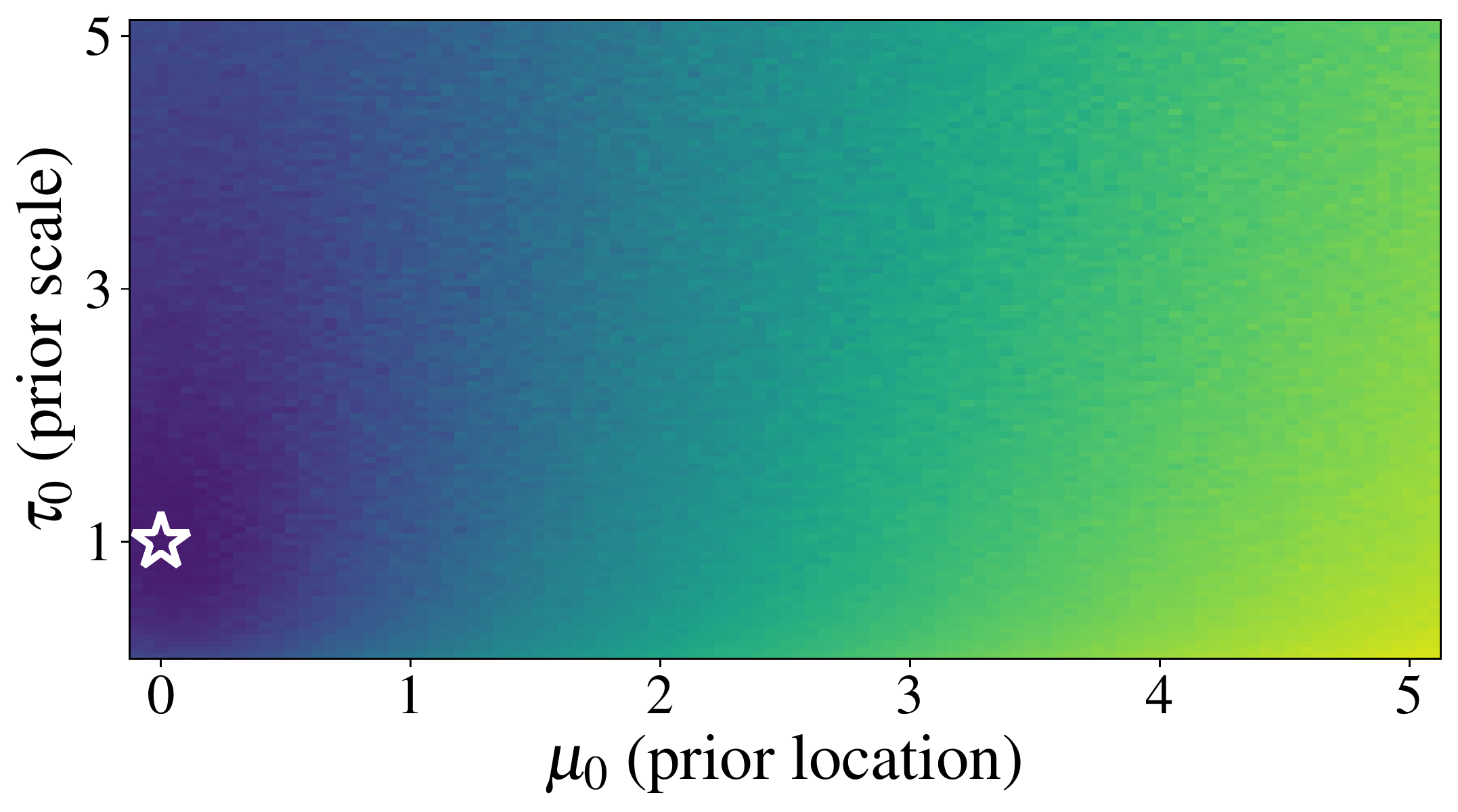}} &
        \raisebox{-0.48\height}{\includegraphics[width=0.40\linewidth]{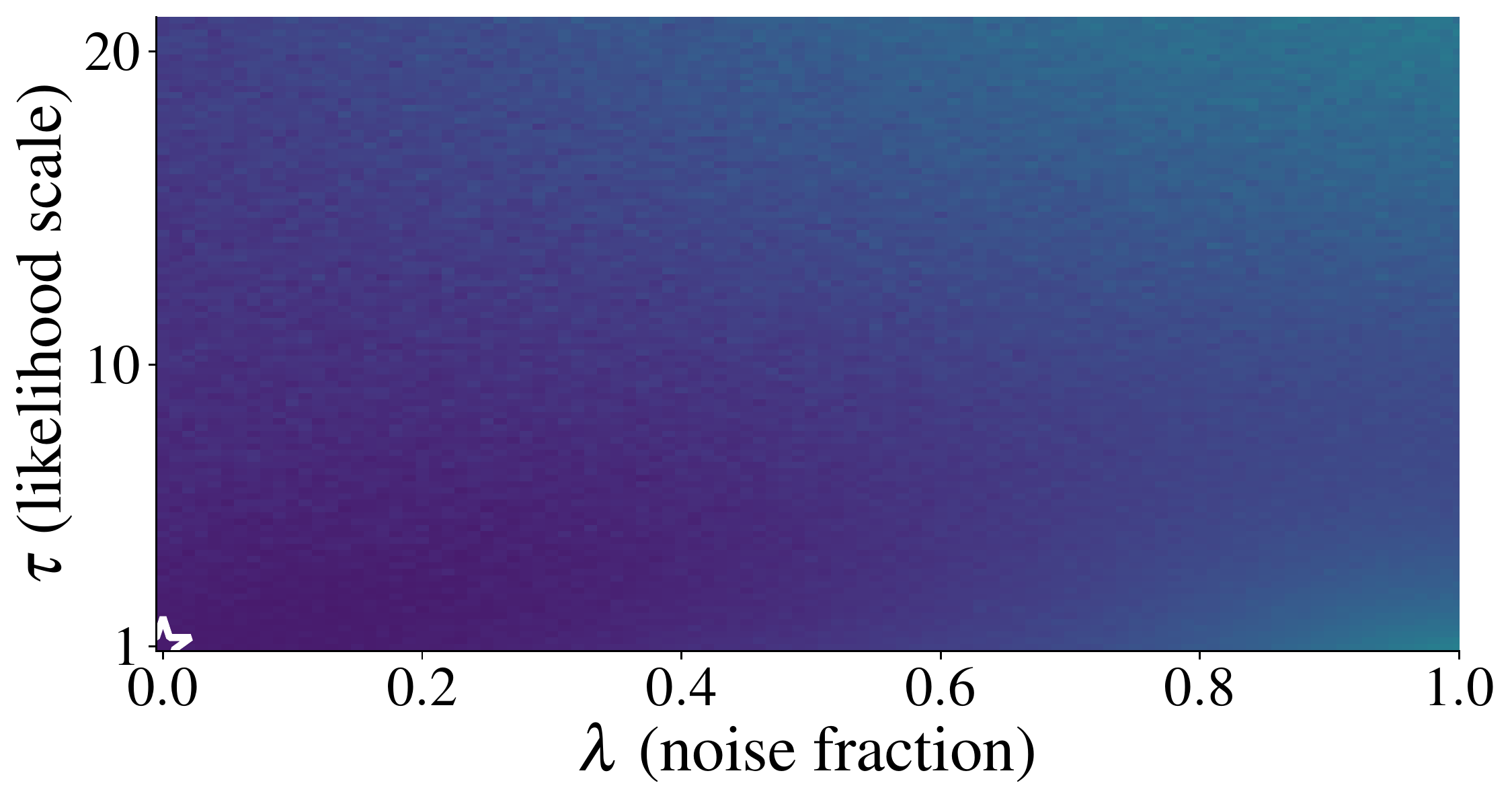}}
        \\
        &\rotatebox[origin=c]{90}{\textbf{overcomplete}} &
        \raisebox{-0.5\height}{\includegraphics[width=0.40\linewidth]{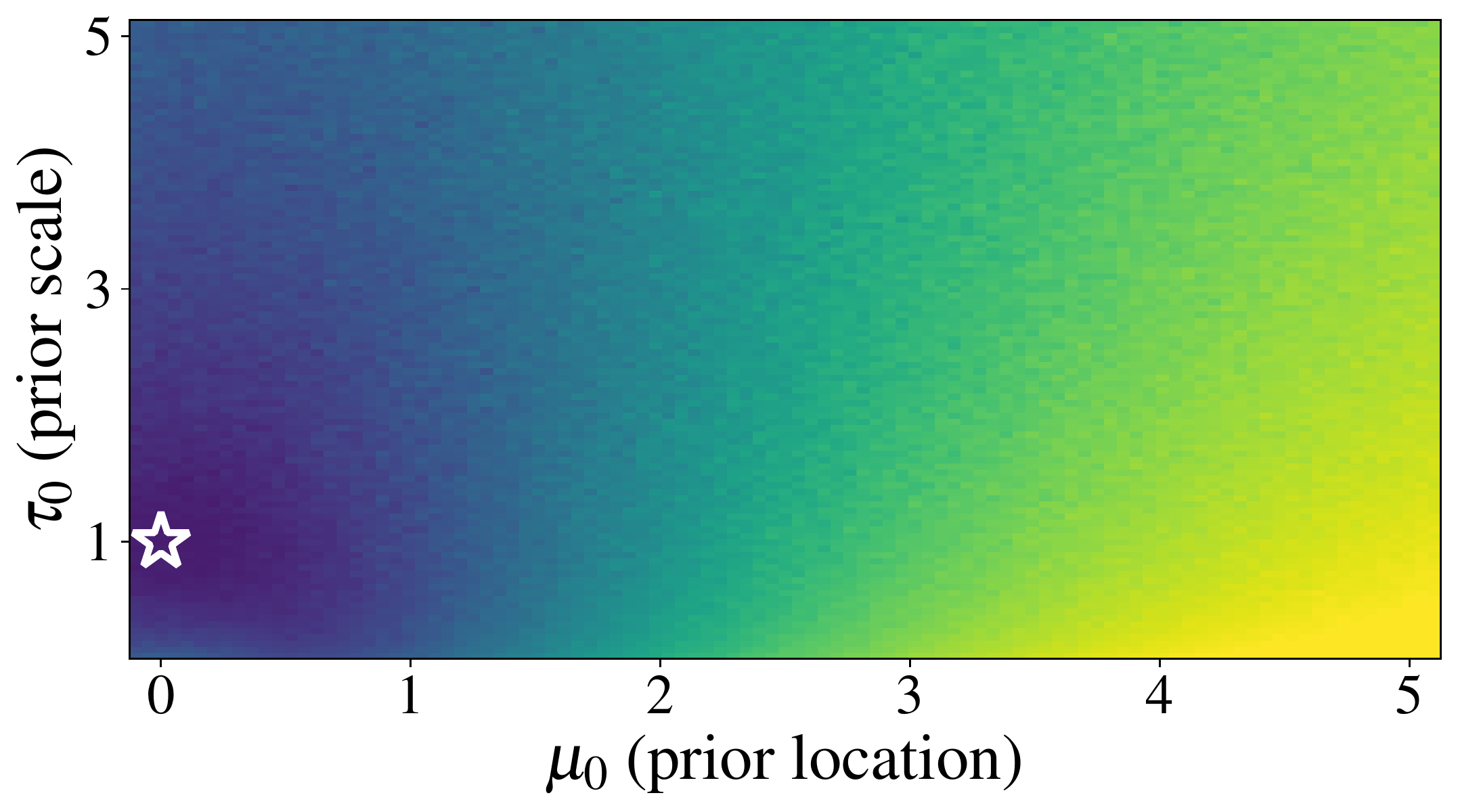}} &
        \raisebox{-0.5\height}{\includegraphics[width=0.40\linewidth]{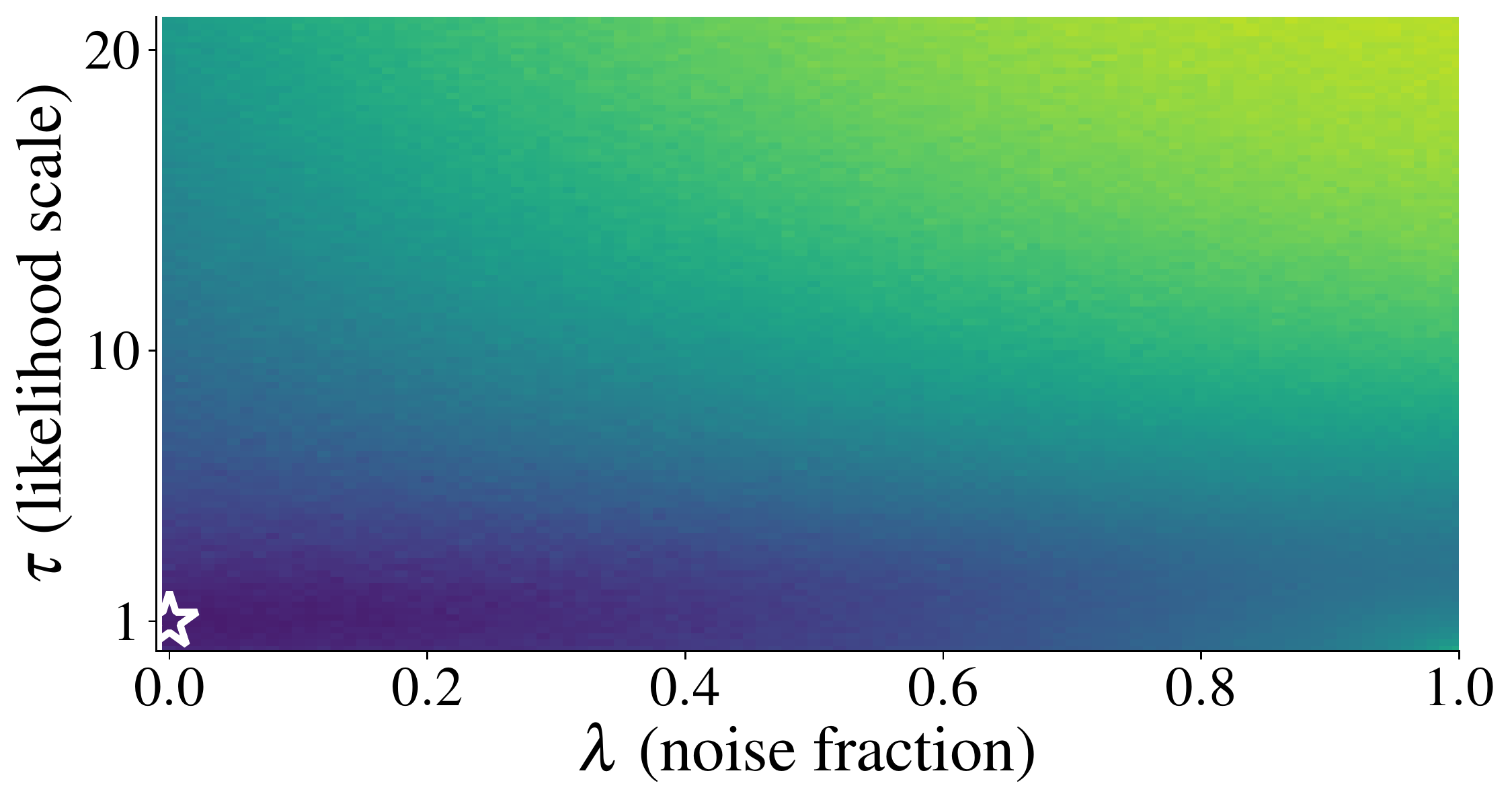}}
    \end{tabular}%
    \end{subfigure}%
    \begin{subfigure}[c]{0.06\linewidth}
    \includegraphics[width=\linewidth, clip, trim=9.8cm 0cm 0.2cm 0cm]{plots/abf_mvn_means_mmd_heatmaps_colorbar.pdf}%
    \end{subfigure}
    \caption{\textbf{Experiment \numberGaussianMeans, SNPE-C.} Summary space MMD as a function of misspecification severity. White stars indicate the well-specified model configuration (i.e., equal to the training model $\M$).}
\end{figure}

\FloatBarrier
\bibliography{references}

\end{document}

%% file: plots/simulation_gap.tex
\begin{tikzpicture}[
every text node part/.style={align=center, font={\Large}},
dot/.style={draw, circle, minimum width=0.5cm},
network-box/.style={draw, rectangle, fill=networkcolor!30, minimum height=5cm, minimum width = 2.5cm, inner sep=0.3cm},
posterior-box/.style={draw, rectangle, rounded corners = .10cm, minimum width = 3cm, inner sep=0.5cm},
arrow/.style = {->, very thick}
]

    \node[draw=none, label={Training Set}] (typical-generative-set) {
    \begin{tikzpicture}[scale=0.6]
    \filldraw[draw=black,fill=viridisgreen!30]  plot[smooth, tension=.8, fill=viridisgreen!30] coordinates {(-3.5,0.5) (-3,2.5) (-1,3.5) (1.5,3) (4,3.5) (5,2.5) (5,-2) (2.5,-3) (0.2,-1.5) (-3,-1.5) (-3.5,0.5)};
    \end{tikzpicture}
    };
    \node[dot, fill=darkgreen!80, left=1.3, label=above right:{$\x\sim p(\x\given\mathcal{M})$}] (x-star) at (typical-generative-set) {};
    
    \node[draw, color=viridisgreen,
    fit={(typical-generative-set)}, 
    label={[anchor=south west]south west:\color{forestlikegreen}Generative Model $\mathcal{M}$}, 
    rounded corners=.30cm, inner sep=0.8cm
    ] (generative-model) {};

    \node[dot, fill=errorcolor!80, below right = 0.4cm and -2cm of generative-model, label=below right:{$\observed{\x}\sim p^*(\x)$}] (x-obs) {};

    \node[network-box, right = of generative-model] (summary-network) {$\huge{h_{\psib}}$\\Summary\\Network};
    
    \node[draw=none, right = of summary-network, label={Summary Space}] (kde) {
        \begin{tikzpicture}
            \draw[thick,->] (-3.5, -3.5) -- (-3.5, 3);
            \draw[thick,->] (-3.5, -3.5) -- (3, -3.5);
            \foreach \x/\alpha in {2/10, 1.6/20, 1.2/30, 0.8/45, 0.4/60}
                \fill[viridisgreen!\alpha] (0, 0) circle (\x);
        \end{tikzpicture}
    };
    
    \node[dot, fill=darkgreen!80,
    above right = 0.25cm and 0.3cm of kde,
    label=above:{$h_{\psib}(\x)$}
    ] (s-star) at (kde) {};
    \node[dot, 
    fill=errorcolor!80, 
    below left = -1.5cm and -1.4cm of kde,
    label=below:{$h_{\psib}(\observed{\x})$}
    ] (s-obs) {};
    
    \node[network-box, right = of kde] (inference-network) {$\huge{f_{\phib}}$\\Inference\\Network};

    \matrix[right = of inference-network, row sep = 0.2cm] {
    \node[posterior-box, fill=viridisgreen!20]  (correct-posterior) {Correct\\Posterior}; \\ 
    \node[posterior-box, fill=errorcolor!20] (incorrect-posterior) {Incorrect\\Posterior}; \\
    };

    
    \draw [dashed, thick] (typical-generative-set) -- (x-obs) node [sloped,midway](M){};;
    
    \node[below left = of M] (simulation-gap) {Misspecification};
    \draw [arrow] (simulation-gap)  -- (M);
    
    \node[below left=-2.6cm and -2.6cm of kde] (kde-outer) {};
    \draw [dashed, thick] (s-obs) -- (kde-outer) node [sloped,midway](M2){};;
    
    \node[below right = 2cm and 0.3cm of M2] (simulation-gap-detected) {Misspecification detected};
    \draw [arrow] (simulation-gap-detected)  -- (M2);

    \draw [arrow, dashed] (x-star) -- (summary-network);
    \draw [arrow, dashed, color=errorcolor] (x-obs) -- (summary-network);
    
    \draw [arrow, dashed] (summary-network) -- (s-star);
    \draw [arrow, color=errorcolor, dashed] (summary-network) -- (s-obs);
    
    
    \draw [arrow, dashed] (s-star) -- (inference-network);
    \draw [arrow, dashed, color=errorcolor] (s-obs) -- (inference-network);
    
    \draw [arrow, dashed] (inference-network) -- (correct-posterior.west);
    \draw [arrow, color=errorcolor, dashed] (inference-network) -- (incorrect-posterior.west);
\end{tikzpicture}

%% file: main.bbl
\begin{thebibliography}{79}
\providecommand{\natexlab}[1]{#1}
\providecommand{\url}[1]{\texttt{#1}}
\expandafter\ifx\csname urlstyle\endcsname\relax
  \providecommand{\doi}[1]{doi: #1}\else
  \providecommand{\doi}{doi: \begingroup \urlstyle{rm}\Url}\fi

\bibitem[Alquier and Ridgway(2019)]{alquier_concentration_2019}
Pierre Alquier and James Ridgway.
\newblock Concentration of tempered posteriors and of their variational
  approximations.
\newblock \emph{arXiv:1706.09293 [cs, math, stat]}, April 2019.
\newblock arXiv: 1706.09293.

\bibitem[Ardizzone et~al.(2018)Ardizzone, Kruse, Wirkert, Rahner, Pellegrini,
  Klessen, Maier{-}Hein, Rother, and K{\"{o}}the]{Ardizzone2018}
Lynton Ardizzone, Jakob Kruse, Sebastian~J. Wirkert, Daniel Rahner, Eric~W.
  Pellegrini, Ralf~S. Klessen, Lena Maier{-}Hein, Carsten Rother, and Ullrich
  K{\"{o}}the.
\newblock {Analyzing Inverse Problems with Invertible Neural Networks}.
\newblock \emph{CoRR}, abs/1808.04730, 2018.

\bibitem[Ardizzone et~al.(2019)Ardizzone, Lüth, Kruse, Rother, and
  Köthe]{Ardizzone2019}
Lynton Ardizzone, Carsten Lüth, Jakob Kruse, Carsten Rother, and Ullrich
  Köthe.
\newblock {Guided Image Generation with Conditional Invertible Neural
  Networks}, 2019.

\bibitem[Arvanitidis et~al.(2021)Arvanitidis, Gonz{\'a}lez-Duque, Pouplin,
  Kalatzis, and Hauberg]{info_geo}
Georgios Arvanitidis, Miguel Gonz{\'a}lez-Duque, Alison Pouplin, Dimitris
  Kalatzis, and S{\o}ren Hauberg.
\newblock {Pulling back information geometry}.
\newblock \emph{arXiv preprint arXiv:2106.05367}, 2021.

\bibitem[Barnard et~al.(2000)Barnard, McCulloch, and Meng]{Barnard2000}
John Barnard, Robert McCulloch, and Xiao-Li Meng.
\newblock {Modelling Covariance Matrices in Terms of Standard Deviations and
  Correlations, with Application To Shrinkage}.
\newblock \emph{Statistica Sinica}, 10:\penalty0 1281--1311, 10 2000.

\bibitem[Bergamin et~al.(2022)Bergamin, Mattei, Havtorn, Senetaire, Schmutz,
  Maaløe, Hauberg, and Frellsen]{bergamin_model-agnostic_2022}
Federico Bergamin, Pierre-Alexandre Mattei, Jakob~D. Havtorn, Hugo Senetaire,
  Hugo Schmutz, Lars Maaløe, Søren Hauberg, and Jes Frellsen.
\newblock Model-agnostic out-of-distribution detection using combined
  statistical tests, March 2022.

\bibitem[Berger and Wolpert(1988)]{berger_likelihood_1988}
James~O. Berger and Robert~L. Wolpert.
\newblock \emph{The likelihood principle}.
\newblock Number v. 6 in Lecture notes-monograph series. Institute of
  Mathematical Statistics, Hayward, Calif, 2nd ed edition, 1988.
\newblock ISBN 978-0-940600-13-3.

\bibitem[Betancourt(2017)]{betancourt2017}
Michael Betancourt.
\newblock {A conceptual introduction to Hamiltonian Monte Carlo}.
\newblock \emph{arXiv preprint}, 2017.

\bibitem[Bieringer et~al.(2021)Bieringer, Butter, Heimel, H{\"o}che, K{\"o}the,
  Plehn, and Radev]{bayesflow_qcd}
Sebastian Bieringer, Anja Butter, Theo Heimel, Stefan H{\"o}che, Ullrich
  K{\"o}the, Tilman Plehn, and Stefan~T Radev.
\newblock {Measuring QCD Splittings with Invertible Networks}.
\newblock \emph{SciPost Physics Proceedings}, 10\penalty0 (6), 2021.

\bibitem[Bloem-Reddy and Teh(2020)]{invariant}
Benjamin Bloem-Reddy and Yee~Whye Teh.
\newblock {Probabilistic Symmetries and Invariant Neural Networks.}
\newblock \emph{J. Mach. Learn. Res.}, 21:\penalty0 90--1, 2020.

\bibitem[B{\"u}rkner et~al.(2022)B{\"u}rkner, Scholz, and
  Radev]{burkner2022some}
Paul-Christian B{\"u}rkner, Maximilian Scholz, and Stefan Radev.
\newblock Some models are useful, but how do we know which ones? towards a
  unified bayesian model taxonomy.
\newblock \emph{arXiv preprint arXiv:2209.02439}, 2022.

\bibitem[Butter et~al.(2022)Butter, Plehn, Schumann, Badger, Caron, Cranmer,
  Di~Bello, Dreyer, Forte, Ganguly, et~al.]{butter2022machine}
Anja Butter, Tilman Plehn, Steffen Schumann, Simon Badger, Sascha Caron, Kyle
  Cranmer, Francesco~Armando Di~Bello, Etienne Dreyer, Stefano Forte, Sanmay
  Ganguly, et~al.
\newblock Machine learning and lhc event generation.
\newblock \emph{arXiv preprint arXiv:2203.07460}, 2022.

\bibitem[Bürkner et~al.(2020)Bürkner, Gabry, and
  Vehtari]{burkner_approximate_2020}
Paul-Christian Bürkner, Jonah Gabry, and Aki Vehtari.
\newblock Approximate leave-future-out cross-validation for {Bayesian} time
  series models.
\newblock \emph{Journal of Statistical Computation and Simulation}, 90\penalty0
  (14):\penalty0 2499--2523, September 2020.
\newblock \doi{10.1080/00949655.2020.1783262}.
\newblock arXiv:1902.06281 [stat].

\bibitem[Cannon et~al.(2022)Cannon, Ward, and
  Schmon]{cannon_investigating_2022}
Patrick Cannon, Daniel Ward, and Sebastian~M. Schmon.
\newblock Investigating the {Impact} of {Model} {Misspecification} in {Neural}
  {Simulation}-based {Inference}, September 2022.
\newblock arXiv:2209.01845 [cs, stat].

\bibitem[Carpenter et~al.(2017)Carpenter, Gelman, Hoffman, Lee, Goodrich,
  Betancourt, Brubaker, Guo, Li, and Riddell]{Carpenter2017}
Bob Carpenter, Andrew Gelman, Matthew~D Hoffman, Daniel Lee, Ben Goodrich,
  Michael Betancourt, Marcus Brubaker, Jiqiang Guo, Peter Li, and Allen
  Riddell.
\newblock {Stan: A probabilistic programming language}.
\newblock \emph{Journal of statistical software}, 76\penalty0 (1), 2017.

\bibitem[Cranmer et~al.(2020)Cranmer, Brehmer, and Louppe]{frontier}
Kyle Cranmer, Johann Brehmer, and Gilles Louppe.
\newblock {The frontier of simulation-based inference}.
\newblock \emph{Proceedings of the National Academy of Sciences}, 117\penalty0
  (48):\penalty0 30055--30062, 2020.

\bibitem[Dehning et~al.(2020)Dehning, Zierenberg, Spitzner, Wibral, Neto,
  Wilczek, and Priesemann]{covid_germany}
Jonas Dehning, Johannes Zierenberg, F~Paul Spitzner, Michael Wibral,
  Joao~Pinheiro Neto, Michael Wilczek, and Viola Priesemann.
\newblock {Inferring change points in the spread of COVID-19 reveals the
  effectiveness of interventions}.
\newblock \emph{Science}, 369\penalty0 (6500), 2020.

\bibitem[Delaunoy et~al.(2022)Delaunoy, Hermans, Rozet, Wehenkel, and
  Louppe]{delaunoy_towards_2022}
Arnaud Delaunoy, Joeri Hermans, François Rozet, Antoine Wehenkel, and Gilles
  Louppe.
\newblock Towards {Reliable} {Simulation}-{Based} {Inference} with {Balanced}
  {Neural} {Ratio} {Estimation}, August 2022.
\newblock arXiv:2208.13624 [cs, stat].

\bibitem[Dellaporta et~al.(2022)Dellaporta, Knoblauch, Damoulas, and
  Briol]{dellaporta_robust_2022}
Charita Dellaporta, Jeremias Knoblauch, Theodoros Damoulas, and
  François-Xavier Briol.
\newblock Robust {Bayesian} {Inference} for {Simulator}-based {Models} via the
  {MMD} {Posterior} {Bootstrap}.
\newblock 2022.
\newblock \doi{10.48550/ARXIV.2202.04744}.

\bibitem[Dong et~al.(2020)Dong, Du, and Gardner]{dong_interactive_2020}
Ensheng Dong, Hongru Du, and Lauren Gardner.
\newblock An interactive web-based dashboard to track {COVID}-19 in real time.
\newblock \emph{The Lancet Infectious Diseases}, 20\penalty0 (5):\penalty0
  533--534, May 2020.
\newblock \doi{10.1016/S1473-3099(20)30120-1}.

\bibitem[Durkan et~al.(2020)Durkan, Murray, and Papamakarios]{contrastive}
Conor Durkan, Iain Murray, and George Papamakarios.
\newblock {On contrastive learning for likelihood-free inference}.
\newblock In \emph{International Conference on Machine Learning}, pages
  2771--2781. PMLR, 2020.

\bibitem[Frazier and Drovandi(2021)]{frazier_robust_2021}
David~T. Frazier and Christopher Drovandi.
\newblock Robust {Approximate} {Bayesian} {Inference} {With} {Synthetic}
  {Likelihood}.
\newblock \emph{Journal of Computational and Graphical Statistics}, 30\penalty0
  (4):\penalty0 958--976, October 2021.
\newblock \doi{10.1080/10618600.2021.1875839}.

\bibitem[Frazier et~al.(2020)Frazier, Robert, and Rousseau]{frazier_model_2020}
David~T. Frazier, Christian~P. Robert, and Judith Rousseau.
\newblock Model misspecification in approximate {Bayesian} computation:
  consequences and diagnostics.
\newblock \emph{Journal of the Royal Statistical Society: Series B (Statistical
  Methodology)}, 82\penalty0 (2):\penalty0 421--444, April 2020.
\newblock \doi{10.1111/rssb.12356}.

\bibitem[Gabry et~al.(2019{\natexlab{a}})Gabry, Simpson, Vehtari, Betancourt,
  and Gelman]{bayes_ppc}
Jonah Gabry, Daniel Simpson, Aki Vehtari, Michael Betancourt, and Andrew
  Gelman.
\newblock {Visualization in Bayesian workflow}.
\newblock \emph{Journal of the Royal Statistical Society: Series A (Statistics
  in Society)}, 182\penalty0 (2):\penalty0 389--402, 2019{\natexlab{a}}.

\bibitem[Gabry et~al.(2019{\natexlab{b}})Gabry, Simpson, Vehtari, Betancourt,
  and Gelman]{gabry_visualization_2019}
Jonah Gabry, Daniel Simpson, Aki Vehtari, Michael Betancourt, and Andrew
  Gelman.
\newblock Visualization in {Bayesian} workflow.
\newblock \emph{Journal of the Royal Statistical Society: Series A (Statistics
  in Society)}, 182\penalty0 (2):\penalty0 389--402, 2019{\natexlab{b}}.
\newblock Publisher: Wiley Online Library.

\bibitem[Ghaderi-Kangavari et~al.(2022)Ghaderi-Kangavari, Rad, and
  Nunez]{ghaderi-kangavari_general_2022}
Amin Ghaderi-Kangavari, Jamal~Amani Rad, and Michael~D. Nunez.
\newblock A general integrative neurocognitive modeling framework to jointly
  describe {EEG} and decision-making on single trials.
\newblock preprint, PsyArXiv, August 2022.

\bibitem[Giummol{\`e} et~al.(2019)Giummol{\`e}, Mameli, Ruli, and
  Ventura]{giummole2019objective}
Federica Giummol{\`e}, Valentina Mameli, Erlis Ruli, and Laura Ventura.
\newblock Objective bayesian inference with proper scoring rules.
\newblock \emph{Test}, 28\penalty0 (3):\penalty0 728--755, 2019.

\bibitem[Gon{\c{c}}alves et~al.(2020)Gon{\c{c}}alves, Lueckmann, Deistler,
  Nonnenmacher, {\"O}cal, Bassetto, Chintaluri, Podlaski, Haddad, Vogels,
  et~al.]{gonccalves2020training}
Pedro~J Gon{\c{c}}alves, Jan-Matthis Lueckmann, Michael Deistler, Marcel
  Nonnenmacher, Kaan {\"O}cal, Giacomo Bassetto, Chaitanya Chintaluri,
  William~F Podlaski, Sara~A Haddad, Tim~P Vogels, et~al.
\newblock {Training deep neural density estimators to identify mechanistic
  models of neural dynamics}.
\newblock \emph{Elife}, 9:\penalty0 e56261, 2020.

\bibitem[Greenberg et~al.(2019)Greenberg, Nonnenmacher, and Macke]{apt}
David Greenberg, Marcel Nonnenmacher, and Jakob Macke.
\newblock {Automatic posterior transformation for likelihood-free inference}.
\newblock In \emph{International Conference on Machine Learning}, pages
  2404--2414. PMLR, 2019.

\bibitem[Gretton et~al.(2012)Gretton, Borgwardt, Rasch, Schölkopf, and
  Smola]{Gretton2012}
A~Gretton, K.~Borgwardt, Malte Rasch, Bernhard Schölkopf, and AJ~Smola.
\newblock {A Kernel Two-Sample Test}.
\newblock \emph{The Journal of Machine Learning Research}, 13:\penalty0
  723--773, 03 2012.

\bibitem[Gr{\"u}nwald et~al.(2017)Gr{\"u}nwald, Van~Ommen,
  et~al.]{bayesian_miss}
Peter Gr{\"u}nwald, Thijs Van~Ommen, et~al.
\newblock {Inconsistency of Bayesian inference for misspecified linear models,
  and a proposal for repairing it}.
\newblock \emph{Bayesian Analysis}, 12\penalty0 (4):\penalty0 1069--1103, 2017.

\bibitem[Hermans et~al.(2020)Hermans, Begy, and Louppe]{ratios}
Joeri Hermans, Volodimir Begy, and Gilles Louppe.
\newblock {Likelihood-free mcmc with amortized approximate ratio estimators}.
\newblock In \emph{International Conference on Machine Learning}, pages
  4239--4248. PMLR, 2020.

\bibitem[Hermans et~al.(2021)Hermans, Delaunoy, Rozet, Wehenkel, and
  Louppe]{hermans2021averting}
Joeri Hermans, Arnaud Delaunoy, Fran{\c{c}}ois Rozet, Antoine Wehenkel, and
  Gilles Louppe.
\newblock Averting a crisis in simulation-based inference.
\newblock \emph{arXiv preprint arXiv:2110.06581}, 2021.

\bibitem[Holmes and Walker(2017)]{holmes2017assigning}
Chris~C Holmes and Stephen~G Walker.
\newblock Assigning a value to a power likelihood in a general bayesian model.
\newblock \emph{Biometrika}, 104\penalty0 (2):\penalty0 497--503, 2017.

\bibitem[Jones-Todd et~al.(2019)Jones-Todd, Caie, Illian, Stevenson, Savage,
  Harrison, and Bown]{jones-todd_identifying_2019}
Charlotte~M. Jones-Todd, Peter Caie, Janine~B. Illian, Ben~C. Stevenson, Anne
  Savage, David~J. Harrison, and James~L. Bown.
\newblock Identifying prognostic structural features in tissue sections of
  colon cancer patients using point pattern analysis: {Point} pattern analysis
  of colon cancer tissue sections.
\newblock \emph{Statistics in Medicine}, 38\penalty0 (8):\penalty0 1421--1441,
  April 2019.
\newblock \doi{10.1002/sim.8046}.

\bibitem[Knoblauch et~al.(2019)Knoblauch, Jewson, and
  Damoulas]{knoblauch2019generalized}
Jeremias Knoblauch, Jack Jewson, and Theodoros Damoulas.
\newblock Generalized variational inference: Three arguments for deriving new
  posteriors.
\newblock \emph{arXiv preprint arXiv:1904.02063}, 2019.

\bibitem[Leclercq(2022)]{leclercq_simulation-based_2022}
Florent Leclercq.
\newblock Simulation-based inference of {Bayesian} hierarchical models while
  checking for model misspecification, September 2022.
\newblock arXiv:2209.11057 [astro-ph, q-bio, stat].

\bibitem[Loaiza-Maya et~al.(2021)Loaiza-Maya, Martin, and
  Frazier]{loaiza2021focused}
Ruben Loaiza-Maya, Gael~M Martin, and David~T Frazier.
\newblock Focused bayesian prediction.
\newblock \emph{Journal of Applied Econometrics}, 36\penalty0 (5):\penalty0
  517--543, 2021.

\bibitem[Lotfi et~al.(2022)Lotfi, Izmailov, Benton, Goldblum, and
  Wilson]{lotfi2022bayesian}
Sanae Lotfi, Pavel Izmailov, Gregory Benton, Micah Goldblum, and Andrew~Gordon
  Wilson.
\newblock Bayesian model selection, the marginal likelihood, and
  generalization.
\newblock \emph{arXiv preprint arXiv:2202.11678}, 2022.

\bibitem[Lueckmann et~al.(2017)Lueckmann, Goncalves, Bassetto, {\"O}cal,
  Nonnenmacher, and Macke]{bayes_lstm}
Jan-Matthis Lueckmann, Pedro~J Goncalves, Giacomo Bassetto, Kaan {\"O}cal,
  Marcel Nonnenmacher, and Jakob~H Macke.
\newblock {Flexible statistical inference for mechanistic models of neural
  dynamics}.
\newblock \emph{Advances in Neural Information Processing Systems}, 30, 2017.

\bibitem[Lueckmann et~al.(2021{\natexlab{a}})Lueckmann, Boelts, Greenberg,
  Goncalves, and Macke]{lueckmann2021benchmarking}
Jan-Matthis Lueckmann, Jan Boelts, David Greenberg, Pedro Goncalves, and Jakob
  Macke.
\newblock Benchmarking simulation-based inference.
\newblock In \emph{International Conference on Artificial Intelligence and
  Statistics}, pages 343--351. PMLR, 2021{\natexlab{a}}.

\bibitem[Lueckmann et~al.(2021{\natexlab{b}})Lueckmann, Boelts, Greenberg,
  Goncalves, and Macke]{lueckmann_benchmarking_2021}
Jan-Matthis Lueckmann, Jan Boelts, David Greenberg, Pedro Goncalves, and Jakob
  Macke.
\newblock Benchmarking {Simulation}-{Based} {Inference}.
\newblock In Arindam Banerjee and Kenji Fukumizu, editors, \emph{Proceedings of
  {The} 24th {International} {Conference} on {Artificial} {Intelligence} and
  {Statistics}}, volume 130 of \emph{Proceedings of {Machine} {Learning}
  {Research}}, pages 343--351. PMLR, April 2021{\natexlab{b}}.

\bibitem[Mardia et~al.(1979)Mardia, Kent, and Bibby]{Mardia1979}
{Kantilal Varichand} Mardia, {John T.} Kent, and {John M.} Bibby.
\newblock \emph{{Multivariate analysis}}.
\newblock Probability and mathematical statistics. Acad. Press, London, 1979.

\bibitem[Masegosa(2020)]{masegosa2020learning}
Andres Masegosa.
\newblock Learning under model misspecification: Applications to variational
  and ensemble methods.
\newblock \emph{Advances in Neural Information Processing Systems},
  33:\penalty0 5479--5491, 2020.

\bibitem[Matsubara et~al.(2022)Matsubara, Knoblauch, Briol, and
  Oates]{matsubara_robust_2022}
Takuo Matsubara, Jeremias Knoblauch, François-Xavier Briol, and Chris~J.
  Oates.
\newblock Robust {Generalised} {Bayesian} {Inference} for {Intractable}
  {Likelihoods}, January 2022.
\newblock arXiv:2104.07359 [math, stat].

\bibitem[Muandet et~al.(2017)Muandet, Fukumizu, Sriperumbudur, and
  Sch\"{o}lkopf]{Muandet2017}
Krikamol Muandet, Kenji Fukumizu, Bharath Sriperumbudur, and Bernhard
  Sch\"{o}lkopf.
\newblock {Kernel Mean Embedding of Distributions: A Review and Beyond}.
\newblock \emph{Foundations and Trends{\textregistered} in Machine Learning},
  10\penalty0 (1-2):\penalty0 1--141, 2017.
\newblock \doi{10.1561/2200000060}.

\bibitem[Murphy(2007)]{Murphy2007}
Kevin Murphy.
\newblock {Conjugate Bayesian analysis of the Gaussian distribution}, 11 2007.

\bibitem[Nguyen et~al.(2019)Nguyen, Cressie, and Hobbs]{nguyen2019sensitivity}
Hai Nguyen, Noel Cressie, and Jonathan Hobbs.
\newblock Sensitivity of optimal estimation satellite retrievals to
  misspecification of the prior mean and covariance, with application to oco-2
  retrievals.
\newblock \emph{Remote Sensing}, 11\penalty0 (23):\penalty0 2770, 2019.

\bibitem[Nowak and Guthke(2016)]{Nowak2016}
Wolfgang Nowak and Anneli Guthke.
\newblock {Entropy-based experimental design for optimal model discrimination
  in the geosciences}.
\newblock \emph{Entropy}, 18\penalty0 (11):\penalty0 409--434, 2016.

\bibitem[Pacchiardi and Dutta(2022{\natexlab{a}})]{pacchiardi2022likelihood}
Lorenzo Pacchiardi and Ritabrata Dutta.
\newblock Likelihood-free inference with generative neural networks via scoring
  rule minimization.
\newblock \emph{arXiv preprint arXiv:2205.15784}, 2022{\natexlab{a}}.

\bibitem[Pacchiardi and Dutta(2022{\natexlab{b}})]{pacchiardi_score_2022}
Lorenzo Pacchiardi and Ritabrata Dutta.
\newblock Score {Matched} {Neural} {Exponential} {Families} for
  {Likelihood}-{Free} {Inference}, January 2022{\natexlab{b}}.
\newblock arXiv:2012.10903 [stat].

\bibitem[Pang et~al.(2022)Pang, Shen, Cao, and Hengel]{pang_deep_2022}
Guansong Pang, Chunhua Shen, Longbing Cao, and Anton van~den Hengel.
\newblock Deep {Learning} for {Anomaly} {Detection}: {A} {Review}.
\newblock \emph{ACM Computing Surveys}, 54\penalty0 (2):\penalty0 1--38, March
  2022.
\newblock \doi{10.1145/3439950}.
\newblock arXiv:2007.02500 [cs, stat].

\bibitem[Papamakarios and Murray(2016)]{papamakarios2016fast}
George Papamakarios and Iain Murray.
\newblock Fast $\varepsilon$-free inference of simulation models with bayesian
  conditional density estimation.
\newblock \emph{Advances in neural information processing systems}, 29, 2016.

\bibitem[Papamakarios et~al.(2017)Papamakarios, Pavlakou, and
  Murray]{papamakarios2017masked}
George Papamakarios, Theo Pavlakou, and Iain Murray.
\newblock Masked autoregressive flow for density estimation.
\newblock \emph{Advances in neural information processing systems}, 30, 2017.

\bibitem[Papamakarios et~al.(2019)Papamakarios, Sterratt, and Murray]{snle}
George Papamakarios, David Sterratt, and Iain Murray.
\newblock {Sequential neural likelihood: Fast likelihood-free inference with
  autoregressive flows}.
\newblock In \emph{The 22nd International Conference on Artificial Intelligence
  and Statistics}, pages 837--848. PMLR, 2019.

\bibitem[Pudlo et~al.(2016)Pudlo, Marin, Estoup, Cornuet, Gautier, and
  Robert]{pudlo2016reliable}
Pierre Pudlo, Jean-Michel Marin, Arnaud Estoup, Jean-Marie Cornuet, Mathieu
  Gautier, and Christian~P Robert.
\newblock Reliable abc model choice via random forests.
\newblock \emph{Bioinformatics}, 32\penalty0 (6):\penalty0 859--866, 2016.

\bibitem[Radev et~al.(2020{\natexlab{a}})Radev, Mertens, Voss, Ardizzone, and
  K{\"o}the]{bayesflow}
Stefan~T Radev, Ulf~K Mertens, Andreas Voss, Lynton Ardizzone, and Ullrich
  K{\"o}the.
\newblock {BayesFlow: Learning complex stochastic models with invertible neural
  networks}.
\newblock \emph{IEEE Transactions on Neural Networks and Learning Systems},
  2020{\natexlab{a}}.

\bibitem[Radev et~al.(2020{\natexlab{b}})Radev, Voss, Wieschen, and
  Buerkner]{Radev2020bayesflow-cognition}
Stefan~T. Radev, Andreas Voss, Eva~Marie Wieschen, and Paul-Christian Buerkner.
\newblock {Amortized Bayesian Inference for Models of Cognition},
  2020{\natexlab{b}}.

\bibitem[Radev et~al.(2021{\natexlab{a}})Radev, D'Alessandro, Mertens, Voss,
  K{\"o}the, and B{\"u}rkner]{amortized_bmc}
Stefan~T Radev, Marco D'Alessandro, Ulf~K Mertens, Andreas Voss, Ullrich
  K{\"o}the, and Paul-Christian B{\"u}rkner.
\newblock {Amortized bayesian model comparison with evidential deep learning}.
\newblock \emph{IEEE Transactions on Neural Networks and Learning Systems},
  2021{\natexlab{a}}.

\bibitem[Radev et~al.(2021{\natexlab{b}})Radev, Graw, Chen, Mutters, Eichel,
  B{\"a}rnighausen, and K{\"o}the]{outbreak}
Stefan~T Radev, Frederik Graw, Simiao Chen, Nico~T Mutters, Vanessa~M Eichel,
  Till B{\"a}rnighausen, and Ullrich K{\"o}the.
\newblock {OutbreakFlow: Model-based Bayesian inference of disease outbreak
  dynamics with invertible neural networks and its application to the COVID-19
  pandemics in Germany}.
\newblock \emph{PLOS Computational Biology}, 17\penalty0 (10):\penalty0
  e1009472, 2021{\natexlab{b}}.

\bibitem[Ramesh et~al.(2022)Ramesh, Lueckmann, Boelts, Tejero-Cantero,
  Greenberg, Gon{\c{c}}alves, and Macke]{ramesh2022gatsbi}
Poornima Ramesh, Jan-Matthis Lueckmann, Jan Boelts, {\'A}lvaro Tejero-Cantero,
  David~S Greenberg, Pedro~J Gon{\c{c}}alves, and Jakob~H Macke.
\newblock Gatsbi: Generative adversarial training for simulation-based
  inference.
\newblock \emph{arXiv preprint arXiv:2203.06481}, 2022.

\bibitem[Ratcliff and McKoon(2008)]{Ratcliff2008}
Roger Ratcliff and Gail McKoon.
\newblock {The Diffusion Decision Model: Theory and Data for Two-Choice
  Decision Tasks}.
\newblock \emph{Neural Computation}, 20\penalty0 (4):\penalty0 873--922, April
  2008.
\newblock \doi{10.1162/neco.2008.12-06-420}.

\bibitem[Ruff et~al.(2018)Ruff, Vandermeulen, Goernitz, Deecke, Siddiqui,
  Binder, Müller, and Kloft]{ruff_deep_2018}
Lukas Ruff, Robert Vandermeulen, Nico Goernitz, Lucas Deecke, Shoaib~Ahmed
  Siddiqui, Alexander Binder, Emmanuel Müller, and Marius Kloft.
\newblock Deep {One}-{Class} {Classification}.
\newblock In Jennifer Dy and Andreas Krause, editors, \emph{Proceedings of the
  35th {International} {Conference} on {Machine} {Learning}}, volume~80 of
  \emph{Proceedings of {Machine} {Learning} {Research}}, pages 4393--4402.
  PMLR, July 2018.

\bibitem[Schmon et~al.(2021)Schmon, Cannon, and
  Knoblauch]{schmon_generalized_2021}
Sebastian~M. Schmon, Patrick~W. Cannon, and Jeremias Knoblauch.
\newblock Generalized {Posteriors} in {Approximate} {Bayesian} {Computation},
  February 2021.
\newblock arXiv:2011.08644 [stat].

\bibitem[Sch{\"o}niger et~al.(2015)Sch{\"o}niger, Illman, W{\"o}hling, and
  Nowak]{Schoniger2015}
Anneli Sch{\"o}niger, Walter~A Illman, Thomas W{\"o}hling, and Wolfgang Nowak.
\newblock {Finding the right balance between groundwater model complexity and
  experimental effort via Bayesian model selection}.
\newblock \emph{Journal of Hydrology}, 531:\penalty0 96--110, 2015.

\bibitem[Shiono(2021)]{bayesflow_agent}
Takashi Shiono.
\newblock {Estimation of agent-based models using Bayesian deep learning
  approach of BayesFlow}.
\newblock \emph{Journal of Economic Dynamics and Control}, 125:\penalty0
  104082, 2021.

\bibitem[{Stan Development Team}(2022)]{Stan2022}
{Stan Development Team}.
\newblock {The Stan Core Library}, 2022.
\newblock Version 2.30.0.

\bibitem[Stine(1989)]{Stine1989}
Robert Stine.
\newblock {An Introduction to Bootstrap Methods}.
\newblock \emph{Sociological Methods {\&} Research}, 18\penalty0
  (2-3):\penalty0 243--291, November 1989.
\newblock \doi{10.1177/0049124189018002003}.

\bibitem[Säilynoja et~al.(2021)Säilynoja, Bürkner, and
  Vehtari]{sailynoja_graphical_2021}
Teemu Säilynoja, Paul-Christian Bürkner, and Aki Vehtari.
\newblock Graphical {Test} for {Discrete} {Uniformity} and its {Applications}
  in {Goodness} of {Fit} {Evaluation} and {Multiple} {Sample} {Comparison},
  November 2021.
\newblock arXiv:2103.10522 [stat].

\bibitem[Talts et~al.(2020)Talts, Betancourt, Simpson, Vehtari, and
  Gelman]{talts_validating_2020}
Sean Talts, Michael Betancourt, Daniel Simpson, Aki Vehtari, and Andrew Gelman.
\newblock Validating {Bayesian} {Inference} {Algorithms} with
  {Simulation}-{Based} {Calibration}, October 2020.
\newblock arXiv:1804.06788 [stat].

\bibitem[Tejero-Cantero et~al.(2020)Tejero-Cantero, Boelts, Deistler,
  Lueckmann, Durkan, Gon{\c{c}}alves, Greenberg, and Macke]{tejero2020sbi}
Alvaro Tejero-Cantero, Jan Boelts, Michael Deistler, Jan-Matthis Lueckmann,
  Conor Durkan, Pedro~J Gon{\c{c}}alves, David~S Greenberg, and Jakob~H Macke.
\newblock Sbi--a toolkit for simulation-based inference.
\newblock \emph{arXiv preprint arXiv:2007.09114}, 2020.

\bibitem[Thomas and Corander(2019)]{mms_genbayes}
Owen Thomas and Jukka Corander.
\newblock {Diagnosing model misspecification and performing generalized Bayes'
  updates via probabilistic classifiers}.
\newblock \emph{arXiv preprint arXiv:1912.05810}, 2019.

\bibitem[Tolstikhin et~al.(2017)Tolstikhin, Bousquet, Gelly, and
  Schoelkopf]{Tolstikhin2017}
Ilya Tolstikhin, Olivier Bousquet, Sylvain Gelly, and Bernhard Schoelkopf.
\newblock Wasserstein auto-encoders, 2017.

\bibitem[van~der Vaart(2000)]{vandervaart2000asymptotic}
Aad~W van~der Vaart.
\newblock \emph{Asymptotic statistics}, volume~3.
\newblock Cambridge university press, 2000.

\bibitem[Vehtari and Ojanen(2012)]{vehtari_survey_2012}
Aki Vehtari and Janne Ojanen.
\newblock A survey of {Bayesian} predictive methods for model assessment,
  selection and comparison.
\newblock \emph{Statistics Surveys}, 6\penalty0 (none), January 2012.
\newblock \doi{10.1214/12-SS102}.

\bibitem[von Krause et~al.(2022)von Krause, Radev, and
  Voss]{von_krause_mental_2022}
Mischa von Krause, Stefan~T. Radev, and Andreas Voss.
\newblock Mental speed is high until age 60 as revealed by analysis of over a
  million participants.
\newblock \emph{Nature Human Behaviour}, 6\penalty0 (5):\penalty0 700--708, May
  2022.
\newblock \doi{10.1038/s41562-021-01282-7}.

\bibitem[Ward et~al.(2022)Ward, Cannon, Beaumont, Fasiolo, and
  Schmon]{ward_robust_2022}
Daniel Ward, Patrick Cannon, Mark Beaumont, Matteo Fasiolo, and Sebastian~M.
  Schmon.
\newblock Robust {Neural} {Posterior} {Estimation} and {Statistical} {Model}
  {Criticism}, October 2022.
\newblock arXiv:2210.06564 [cs, stat].

\bibitem[Wiqvist et~al.(2021)Wiqvist, Frellsen, and Picchini]{snpla}
Samuel Wiqvist, Jes Frellsen, and Umberto Picchini.
\newblock {Sequential Neural Posterior and Likelihood Approximation}.
\newblock \emph{arXiv preprint arXiv:2102.06522}, 2021.

\bibitem[Zhang and Gao(2020)]{zhang_convergence_2020}
Fengshuo Zhang and Chao Gao.
\newblock Convergence rates of variational posterior distributions.
\newblock \emph{The Annals of Statistics}, 48\penalty0 (4), August 2020.
\newblock \doi{10.1214/19-AOS1883}.

\end{thebibliography}
